%
%
%
%
%


\documentclass[bibtotoc,                                      
    a4paper,                                               
    twoside,                                              
    12pt,                                                  
    numbers  = noendperiod,
    headings = normal,
    listof   = leveldown,
    pagesize=auto                                          
    version=3.03
]{scrartcl}

\usepackage[english, intoc]{nomencl}
\makenomenclature
\usepackage[english]{babel}
\usepackage{import}

%
%
%
%
%


\NeedsTeXFormat{LaTeX2e}[1996/12/01]
\newcommand{\classname}{MastersDoctoralThesis}
\ProvidesClass{\classname}[2016/11/22 v1.5 LaTeXTemplates.com]

\RequirePackage{etoolbox}
\RequirePackage{xparse}
\newbool{nolistspace}
\newbool{chapteroneline}
\newbool{listtoc}
\newbool{toctoc}
\newbool{parskip}
\newbool{hyperrefsupport}
\booltrue{hyperrefsupport}
\newbool{headsepline}
\newbool{consistentlayout}
\DeclareOption{nohyperref}{\boolfalse{hyperrefsupport}}
\DeclareOption{nolistspacing}{\booltrue{nolistspace}}
\DeclareOption{liststotoc}{\booltrue{listtoc}}
\DeclareOption{toctotoc}{\booltrue{toctoc}}
\DeclareOption{parskip}{\booltrue{parskip}}
\DeclareOption{headsepline}{\booltrue{headsepline}}
\DeclareOption{consistentlayout}{\booltrue{consistentlayout}}

\ProcessOptions\relax


\patchcmd{\listoffigures}{\MakeUppercase}{\MakeMarkcase}{}{}
\patchcmd{\listoffigures}{\MakeUppercase}{\MakeMarkcase}{}{}
\patchcmd{\listoftables}{\MakeUppercase}{\MakeMarkcase}{}{}
\patchcmd{\listoftables}{\MakeUppercase}{\MakeMarkcase}{}{}

\ifbool{nolistspace}{
	\patchcmd{\listoffigures}{%
		\@starttoc{lof}
	}{%
		\begingroup%
		\singlespace\@starttoc{lof}\endgroup%
	}{}{}%
	\patchcmd{\listoftables}{%
		\@starttoc{lot}
	}{%
		\begingroup%
		\singlespace\@starttoc{lot}\endgroup%
	}{}{}%
	{%
		\begingroup%
		\singlespace\@starttoc{toc}\endgroup%
	}{}{}%
}{}


\RequirePackage{babel} 

\RequirePackage{scrbase} 

\RequirePackage{scrhack} 

\RequirePackage{setspace} 

\RequirePackage{longtable} 

\RequirePackage{siunitx} 

\RequirePackage{graphicx} 
\graphicspath{{Figures/}{./}} 

\RequirePackage{booktabs} 

\RequirePackage{caption} 
\captionsetup{justification=centerlast,font=small,labelfont=sc,margin=50pt}


\NewDocumentCommand{\thesistitle} { o m }{%
 \IfValueTF{#1}{\def\shorttitle{#1}}{\def\shorttitle{#2}}%
 \def\@title{#2}%
 \def\ttitle{#2}%
}
\DeclareDocumentCommand{\author}{m}{\newcommand{\authorname}{#1}}
\NewDocumentCommand{\supervisor}{m}{\newcommand{\supname}{#1}}
\NewDocumentCommand{\examiner}{m}{\newcommand{\examname}{#1}}
\NewDocumentCommand{\degree}{m}{\newcommand{\degreename}{#1}}
\NewDocumentCommand{\addresses}{m}{\newcommand{\addressname}{#1}}
\NewDocumentCommand{\university}{m}{\newcommand{\univname}{#1}}
\NewDocumentCommand{\department}{m}{\newcommand{\deptname}{#1}}
\NewDocumentCommand{\group}{m}{\newcommand{\groupname}{#1}}
\NewDocumentCommand{\faculty}{m}{\newcommand{\facname}{#1}}
\NewDocumentCommand{\keywords}{m}{\newcommand{\keywordnames}{#1}}

\NewDocumentCommand{\bhrule}{}{\typeout{--------------------}}
\NewDocumentCommand{\tttypeout}{m}{\bhrule\typeout{\space #1}\bhrule}

\newcommand{\HRule}{\rule{.9\linewidth}{.6pt}} 

\setcounter{tocdepth}{2} 


\usepackage{xcolor} 

\colorlet{mdtRed}{red!50!black}


\RequirePackage{geometry}

\raggedbottom


\doublehyphendemerits=10000 
\brokenpenalty=10000 
\widowpenalty=9999 
\clubpenalty=9999 
\interfootnotelinepenalty=9999 



\DeclareDocumentCommand\cleardoublepage{}{\clearpage\if@twoside \ifodd\c@page\else
	\hbox{}
	\thispagestyle{\blank@p@gestyle}
	\newpage
	\if@twocolumn\hbox{}\newpage\fi\fi\fi%
}


\newcommand{\abbrevname}{List of Abbreviations}
\providecaptionname{english,british,american}{\abbrevname}{List of Abbreviations}
\providecaptionname{ngerman,german,austrian,naustrian}{\abbrevname}{Abk\"urzungsverzeichnis}
\NewDocumentEnvironment{abbreviations}{ m }{%
	\ifbool{nolistspace}{\begingroup\singlespacing}{}
	\ifbool{listtoc}{\section{\abbrevname}}{\section*{\abbrevname}}
	\begin{longtable}{#1}
	}{%
	\end{longtable}
	\addtocounter{table}{0}
	\ifbool{nolistspace}{\endgroup}{}
}


\DeclareDocumentCommand{\abstractauthorfont}{}{}
\DeclareDocumentCommand{\abstracttitlefont}{}{}
\newcommand{\byname}{by}

\providecaptionname{german,ngerman,austrian,naustrian}{\byname}{von}
\providecaptionname{american,australian,british,canadian,english,newzealand,UKenglish,USenglish}{\byname}{by}
\ifbool{consistentlayout}{
	\DeclareDocumentEnvironment{abstract}{ O{} }{%
		\addsec*{\abstractname}%
		{\chapteralign\normalsize\abstractauthorfont \authorname \par}
		\vspace{\baselineskip}
		{\chapteralign\parbox{.7\linewidth}{\chapteralign\normalsize\itshape\abstracttitlefont\@title}\par}
		\bigskip\noindent\ignorespaces
	}%
	{}
}{%
	\DeclareDocumentEnvironment{abstract}{ O{\null\vfill} }{
		\tttypeout{\abstractname}
		#1
		\thispagestyle{plain}
		\begin{center}
			{\normalsize \MakeUppercase{\univname} \par}
			\bigskip
			{\huge\textit{\abstractname} \par}
			\bigskip
			{\normalsize \facname \par}
			{\normalsize \deptname \par}
			\bigskip
			{\normalsize \degreename\par}
			\bigskip
			{\normalsize\bfseries \@title \par}
			\medskip
			{\normalsize \byname{} \authorname \par}
			\bigskip
		\end{center}
	}
	{
		\vfill\null
	}
}

\DeclareDocumentEnvironment{extraAbstract}{ O{\null\vfill} }{
	\tttypeout{\abstractname}
	#1
	\thispage{empty}
	\begin{center}
		{\normalsize \MakeUppercase{\univname} \par}
		\bigskip
		{\huge\textit{\abstractname} \par}
		\bigskip
		{\normalsize \facname \par}
		{\normalsize \deptname \par}
		\bigskip
		{\normalsize \degreename\par}
		\bigskip
		{\normalsize\bfseries \@title \par}
		\medskip
		{\normalsize \byname{} \authorname \par}
		\bigskip
	\end{center}
}
{
	\vfill\null
}


\usepackage{xcolor}
\colorlet{mdtRed}{red!50!black}
\newcommand{\acknowledgementname}{Acknowledgements}
\providecaptionname{american,australian,british,canadian,english,newzealand,UKenglish,USenglish} {\acknowledgementname}{Acknowledgements} 
\providecaptionname{german,ngerman,austrian,naustrian}{\acknowledgementname}{Danksagung} 

\ifbool{consistentlayout}{
	\DeclareDocumentEnvironment{acknowledgements}{}{%
		\tttypeout{\acknowledgementname}
		\section*{\acknowledgementname}
	}
}
{
	\DeclareDocumentEnvironment{acknowledgements}{}{%
		\tttypeout{\acknowledgementname}
		\thispagestyle{plain}
		\begin{center}{\LARGE\textsc{\acknowledgementname}\par}\end{center}
		\vspace{2cm}
	}
	{
		\vfil\vfil\null
	}
}


\newcommand{\authorshipname}{Declaration of Authorship}
\providecaptionname{american,australian,british,canadian,english,newzealand,UKenglish,USenglish}{\authorshipname}{Declaration of Authorship} 
\providecaptionname{german,ngerman,austrian,naustrian}{\authorshipname}{Eidesstattliche Erkl\"arung} 

\ifbool{consistentlayout}{
	\DeclareDocumentEnvironment{declaration}{}{
		\section*{\authorshipname}
		}{}%
}{
	\DeclareDocumentEnvironment{declaration}{}{
		\tttypeout{\authorshipname}
		\thispagestyle{plain}
		\null\vfil
		{\noindent\huge\bfseries\authorshipname\par\vspace{10pt}}
	}{}
}


\ifbool{consistentlayout}{
	\DeclareDocumentCommand{\dedicatory}{
		m O{\vspace*{.7\textheight} }  }{
			\tttypeout{Dedicatory}
			\markboth{}{}
			#2
			{\hfill\parbox{.4\textwidth}{\flushright#1\par}}
		}
}{
	\newcommand\dedicatory[1]{
		\tttypeout{Dedicatory}
		\null\vfil
		\thispagestyle{plain}
		\begin{center}{\Large\slshape #1}\end{center}
		\vfil\null
	}
}


\newcommand{\symbolsname}{List of Symbols}
\providecaptionname{english,british,american}{\symbolsname}{List of Symbols}
\providecaptionname{ngerman,german,austrian,naustrian}{\symbolsname}{Symbolverzeichnis} 

\NewDocumentEnvironment{symbols}{ m }{%
	\ifbool{nolistspace}{\begingroup\singlespacing}{}
	\ifbool{listtoc}{\section{\symbolsname}}{\section*{\symbolsname}}
	\begin{longtable}{#1}
	}{%
	\end{longtable}
	\addtocounter{table}{-1}
	\ifbool{nolistspace}{\endgroup}{}
}


\ifbool{hyperrefsupport}{
\AtEndPreamble{\RequirePackage{hyperref}
\hypersetup{pdfpagemode={UseOutlines},
bookmarksopen=true,
bookmarksopenlevel=0,
hypertexnames=false,
colorlinks=true,
citecolor=magenta,
linkcolor=darkblue,
urlcolor=mdtRed,
pdfstartview={FitV},
unicode,
breaklinks=true,
}

\pdfstringdefDisableCommands{
	\let\\\space%
}
	}
}{
}


\usepackage{fancyhdr}

\usepackage[utf8]{inputenc} 
\usepackage[T1]{fontenc} 
\usepackage{tocbasic}
\usepackage{amsmath}
\usepackage{amsthm}
\usepackage{amssymb}
\usepackage{dsfont}
\usepackage{mathrsfs}
\usepackage{mathtools}
\usepackage{enumerate}
\usepackage{feynmp-auto}
\usepackage{datetime}
\usepackage[justification=RaggedRight, singlelinecheck=false]{caption} 
\usepackage{subfig}

\usepackage{multirow}
\usepackage{xcolor,colortbl}

\usepackage{tikz}
\usetikzlibrary{calc,decorations,decorations.pathreplacing, shapes.arrows, snakes}  
\usetikzlibrary{decorations.text}

\usepackage{wrapfig}
\usepackage{placeins}
\usepackage{siunitx}

\usepackage{empheq}
\usepackage[most]{tcolorbox}

\usepackage{dblfloatfix}


\def \BColor {blue!7.5}

\newtcbox{\mymath}[1][]{%
	nobeforeafter, math upper, tcbox raise base,
	enhanced, colframe=black, sharp corners, 
	colback=\BColor, boxrule=1pt,
	#1}

\usepackage[backend=bibtex,style=numeric,natbib=true,sorting=none]{biblatex} 

\addbibresource{Layout/bibliography-1.bib} 
\setcounter{biburllcpenalty}{7000}
\setcounter{biburlucpenalty}{8000}

\usepackage[autostyle=true]{csquotes} 
\usepackage{afterpage}

\usepackage[english, intoc]{nomencl}
\makenomenclature

\usepackage{sectsty}
\allsectionsfont{\mdseries}
\sectionfont{\mdseries\scshape}

\numberwithin{equation}{section}

\newcommand{\dif}{\mathrm{d}}
\newcommand{\A}{\mathcal{A}}
\newcommand{\B}{\mathcal{B}}
\newcommand{\K}{\mathcal{K}}
\newcommand{\M}{\mathcal{M}}
\newcommand{\N}{\mathcal{N}}
\newcommand{\V}{\mathcal{V}}
\newcommand{\RE}{\operatorname{Re}}
\newcommand{\IM}{\operatorname{Im}}
\newcommand{\nl}{\nonumber\\}
\newcommand{\I}{\mathcal{I}}
\newcommand{\R}{\mathcal{R}}
\newcommand{\F}{\mathcal{F}}
\newcommand{\Q}{\mathcal{Q}}
\newcommand{\1}{\mathds{1}}
\newcommand{\ha}{{\hat{a}}}

\newcommand{\J}{\mathcal{J}}
\newcommand{\Z}{\mathcal{Z}}

\hyphenation{man-i-fold man-i-folds}
\hyphenation{su-per-sym-me-try su-per-sym-me-tries su-per-sym-me-tric}
\hyphenation{in-stan-ton}

\newtheoremstyle{Def}
{0.5cm}                 
{0.5cm}                 
{}                         
{}                         
{\normalfont\bfseries}  
{} {\newline}
{}

\newtheoremstyle{Theorem}   
{0.5cm}                 
{0.5cm}                 
{\it}                         
{}                         
{\normalfont\bfseries}  
{} {\newline}
{}
\theoremstyle{Def}
\newtheorem{definition}{Definition}[section]
\theoremstyle{Theorem}
\newtheorem{satz}[definition]{Theorem}

\makeatletter
\renewcommand*\thefootnote{\@arabic{\c@footnote})}
\makeatother


\begin{document}

\thesistitle{The Axion-Instanton Weak Gravity Conjecture and Scalar Fields} 
\supervisor{PD Dr. Eran \textsc{Palti}} 
\examiner{} 
\degree{Master of Science} 
\author{Clemens \textsc{Vittmann}} 
\addresses{} 

\subject{Physics} 
\keywords{} 
\university{\href{http://www.uni-heidelberg.de}{University of Heidelberg}} 
\department{\href{https://www.thphys.uni-heidelberg.de}{Institute for Theoretical Physics}} 
\group{\href{https://www.thphys.uni-heidelberg.de/index.php?lang=e&n1=bsm_group}{String Theory and Beyond the Standard Model Group}} 
\faculty{\href{http://www.thphys.uni-heidelberg.de/index.php?lang=e}{Faculty Name}} 


\hypersetup{pdftitle=\ttitle} 
\hypersetup{pdfauthor=\authorname} 
\hypersetup{pdfkeywords=\keywordnames} 
\definecolor{darkblue}{rgb}{0.0,0.0,0.4}
\definecolor{darkgreen}{rgb}{0.0,0.4,0.0}
\hypersetup{
	colorlinks,
	citecolor=darkgreen,
	urlcolor=darkblue
}
\setcounter{page}{1}
\pagenumbering{roman}


\begin{titlepage}

\begin{center}
{\scshape\LARGE \univname\par}\vspace{0.5cm} 
\Large\sffamily Department of Physics and Astronomy\\
\vspace{1.4cm}
\textsc{\LARGE Master Thesis}\\[0.7cm] 

\HRule \\[0.3cm] 
{\Large \bfseries \ttitle\par}\vspace{0.3cm} 
\HRule \\[1.4cm] 
 
\begin{minipage}[t]{0.4\textwidth}
\begin{flushleft} \large
\emph{Submitted by:}\\
\href{}{\authorname} 
\end{flushleft}
\end{minipage}
\begin{minipage}[t]{0.5\textwidth}
\begin{flushright} \large
\emph{Supervised by:} \\
\href{https://www.thphys.uni-heidelberg.de/~palti/teaching.html}{\supname}\\[.2cm]
\emph{Referees:}\\
\href{https://www.thphys.uni-heidelberg.de/~hebecker/}{Prof. Dr. Arthur \textsc{Hebecker}} \\
\href{https://www.thphys.uni-heidelberg.de/~weigand/}{Prof. Dr. Timo \textsc{Weigand}} \\
\end{flushright}
\end{minipage}\\[3.5cm]
 
\vfill

\large \textit{This thesis has been carried out}\\[0.2cm] 
\textit{at the}\\[0.2cm]
\deptname\\[1.3cm] 

\vfill

{\large \formatdate{4}{8}{2018}}
\vfill
\end{center}
\end{titlepage}
\leavevmode\thispagestyle{empty}
\newpage

\begin{acknowledgements}
Foremost, I thank Eran Palti for giving me the opportunity to write this thesis, for guiding me through the process and for the hours of discussing and explaining even after he left Heidelberg University for a position in Munich. 
\\\\
Second, I thank Arthur Hebecker, whose lectures introduced me to quantum field theory, and Timo Weigand, who kept me busy with string theory at the same time, for being willing to be the referees of my thesis.\\\\
I am grateful to Marc Merstorf and Clemens Frub\"ose as well as Sascha Leonhardt and Thomas Mikhail for tons of helpful and entertaining discussions (not only) about physics and for providing a very pleasant work atmosphere at the Philosophenweg 19.\\\\
My last thanks goes to Mikio Nakahara for writing his excellent textbook on differential geometry, topology and physics which taught me most of the mathematics necessary for dealing with complex manifolds.
\end{acknowledgements}
\newpage

\leavevmode\thispagestyle{empty}
\newpage


\begin{center}
	{\LARGE\textsc{Abstract}}
	\vspace{1cm}
\end{center}
We study the Weak Gravity Conjecture in the presence of scalar fields. The Weak Gravity Conjecture is a consistency condition for a theory of quantum gravity asserting that for a $U(1)$ gauge field, there is a particle charged under this field whose mass is bounded by its charge. It was extended to a statement about any canonical pair of $(p-1)$-dimensional object and $p$-form coupling to it, in particular to axion-instanton pairs. The gauge-scalar Weak Gravity Conjecture is a modification of this bound that includes scalar interactions. We propose a similar extension to cases where scalar fields are present for the axion-instanton Weak Gravity Conjecture and provide evidence from Type IIA supergravity.\\[2cm]

\begin{center}
	{\LARGE\textsc{Zusammenfassung}}
	\vspace{1cm}
\end{center}
Wir untersuchen die \emph{Weak Gravity Conjecture} in Anwesenheit skalarer Felder. Die \emph{Weak Gravity Conjecture} ist eine Konsistenzbedingung f\"ur Theorien der Quantengravitation und behauptet, dass es zu einem $U(1)$-Eichfeld ein Teilchen gibt, welches Ladung unter diesem Feld tr\"agt und dessen Masse von der Ladung be\-schr\"ankt ist. Sie wurde zu einer Aussage \"uber jedes kanonische Paar bestehend aus $(p-1)$-dimensionalem Objekt und daran koppelnder $p$-Form verallgemeinert, insbesondere zu einer Bedingung f\"ur Axion-Instanton-Paare. Die \emph{Gauge-Scalar Weak Gravity Conjecture} ist eine Erweiterung dieser Bedingung, die Wechselwirkungen mit skalaren Feldern beinhaltet. Wir schlagen eine \"ahnliche Verallgemeinerung zu F\"allen, in denen skalare Felder vorhanden sind, f\"ur die Axion-Instanton-Variante der \emph{Weak Gravity Conjecture} vor und untermauern diese mit Indizien von Typ-IIA-Supergravitation.
\newpage

\leavevmode\thispagestyle{empty}
\newpage

{\hypersetup{linkcolor=black}
\tableofcontents 
\newpage
\listoffigures 
\vspace{1cm}
\emph{All figures were created by the author.}\\[1cm]
\listoftables}
\newpage


\pagestyle{fancy}
\fancyhf{}

\fancyhead[RE]{\mdseries\scshape\nouppercase{\leftmark}}
\fancyhead[LO]{\textsl{\rightmark}}
\fancyfoot[C]{\fancyfoot[C]{--~\thepage~--}}

\setcounter{page}{1}
\pagenumbering{arabic}
\section{Introduction}\thispagestyle{empty}
When attempts to find a description of the \emph{strong interaction} were made fifty years ago - the force binding protons and neutrons together - young Gabriele Veneziano found that the Euler beta function had certain features one would expect from the scattering amplitude of strongly interacting particles. No theory was known at that time, though, that would produce such a scattering amplitude. It was about two years later that Yoichiro Nambu, Holger Bech Nielsen and Leonard Susskind independently discovered that it was not a theory of point particles but one of \emph{vibrating strings} that would give rise to Veneziano's amplitude.  Soon, those working on the physics of such strings realized two things: First, this model was making predictions about the strong force which were not in accordance with the experimental findings and the theory of quarks and gluons developed at that time - \emph{quantum chromodynamics} - turned out to be superior as a description of the strong interaction. Second - and quite surprisingly - this theory of strings seemed to offer a solution to a very different but fundamental problem of theoretical physics: Bringing together Einstein's general relativity and the Standard Model in a single unifying theory.

After the two revolutions of theoretical physics in the first half of the twentieth century - the discoveries of general relativity and quantum mechanics - these two fields developed essentially separate from each other and appeared to be drastically different: For example, in quantum field theory - the relativistic quantum framework now underlying our description of the electromagnetic, weak and strong interactions in what is known as the Standard Model - fields at two points in spacetime whose separation is space-like should (anti-)commute. Meanwhile, in general relativity - the theory describing the remaining fundamental interaction, namely gravity - the metric is dynamical and one does \emph{a priori} not even know whether a distance is space-like \cite{BeckerBecker}. Above all, general relativity is \emph{classical} and one runs into the following difficulties when attempting to perform the same procedure of perturbative quantization which proved so successful for the other fundamental interactions: Unlike the Standard Model, gravity is non-renormalizable meaning that the infinities one encounters in the quantization process cannot be removed by a finite number of counter-terms. Although the resulting theory \emph{is} predictive at low energies - i.e. below the Planck mass - it breaks down at shorter distances \cite{Books:QFT}. While this does not play a role for most practical purposes in physics, things do become problematic in the quantum description of small black holes and early universe cosmology.

Without elaborating on the idea of unification as a driving force in the history of physics, we want to briefly review why it is plausible to assume that the Standard Model and gravity themselves are - in spite of their success in describing basically all experimentally observed phenomena - not \emph{fundamental} descriptions but rather limits of such a complete theory. We already mentioned the problem of infinities in quantum field theories. While it is well-known that the Standard Model is a \emph{renormalizable} QFT, which means that these infinities can be absorbed in a finite number of counter terms, their very appearance suggests that this description is merely an effective one, a low-energy limit of a more fundamental theory. The same is true with regard to the spacetime singularities in general relativity which are contained in the centers of black holes. The circumstance that the Standard Model has about twenty free parameters - like electron mass or mixing angles which have to be taken from experiment - renders the theory arbitrary. Even worse, some of these parameters appear to be fine-tuned. We will not elaborate on this any further but go back to the UV infinities of quantum field theory.
\begin{figure}[h]
	\centering
	\includegraphics[width=.75\linewidth]{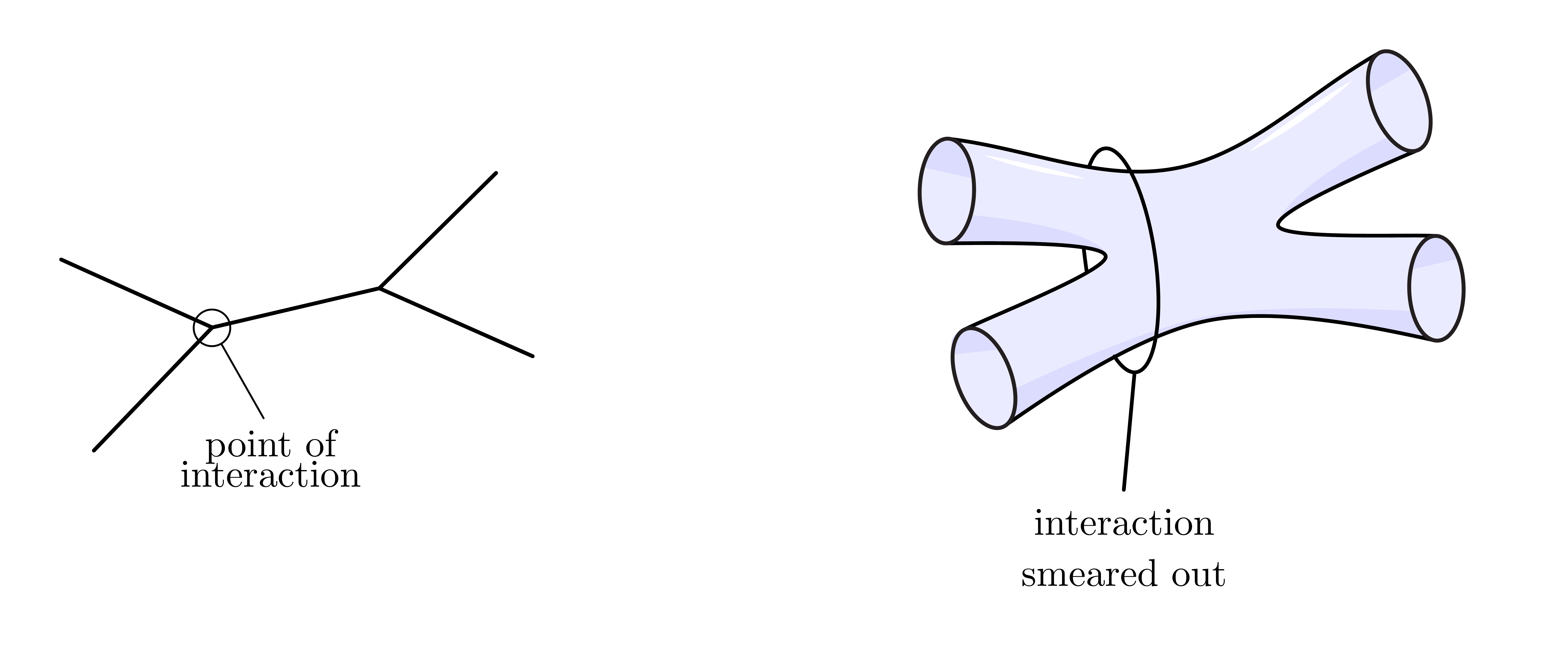}
	\caption[Feynman diagram and string worldsheet]{Feynman diagram of point particles (l.h.s.) and closed strings scattering diagram (r.h.s.).\label{Fig:Intro:Scattering}}
\end{figure}
It was realized that if the fundamental entities of nature were not point-like but rather of finite length, then these UV divergences would not occur. This was proved for the one- and two-loop diagrams analogous to the Feynman diagrams of point particles and there is no reason to expect anything different at higher orders.
Even without performing the actual calculation, this behavior is intuitively accessible when looking at such stringy diagrams like the one in fig. \ref{Fig:Intro:Scattering} on the right-hand side: Unlike the case of point particles, there is no single point in spacetime where scattering strings interact. Instead, the worldsheet in fig. \ref{Fig:Intro:Scattering} always looks \emph{locally} like that of a single freely propagating string and it is only the \emph{topology} of this worldsheet which encodes any interaction.

Even more surprisingly, the theory included a particle in its spectrum that had precisely the numbers of freedom one would expect from a particle that carries the gravitational force, the graviton. At the same time, some problems like the ambiguity arising from free parameters would not appear: String theory has no free dimensionless parameters, all couplings are expectation values of fields and the theory seems to be an up to dualities unique and consistent theory of quantum gravity.

While string theory is unique in \emph{ten} spacetime dimensions, there are many consistent ways to obtain a \emph{four}-dimensional effective theory corresponding to the choice of compactification manifold. The set of these solutions to string theory is known as the \emph{string landscape} and one is led to the question if, conversely, any consistent effective field theory can be coupled to gravity\footnote{If that were the case, one would not try to construct four-dimensional theories as compactifications of ten-dimensional string theory but rather take some effective theory that fits best to experiment \cite{WGC:Vafa}.}. About thirteen years
\begin{wrapfigure}{r}{.45\linewidth}
	\vspace{-20pt}
	\begin{center}
		\includegraphics[width=\linewidth]{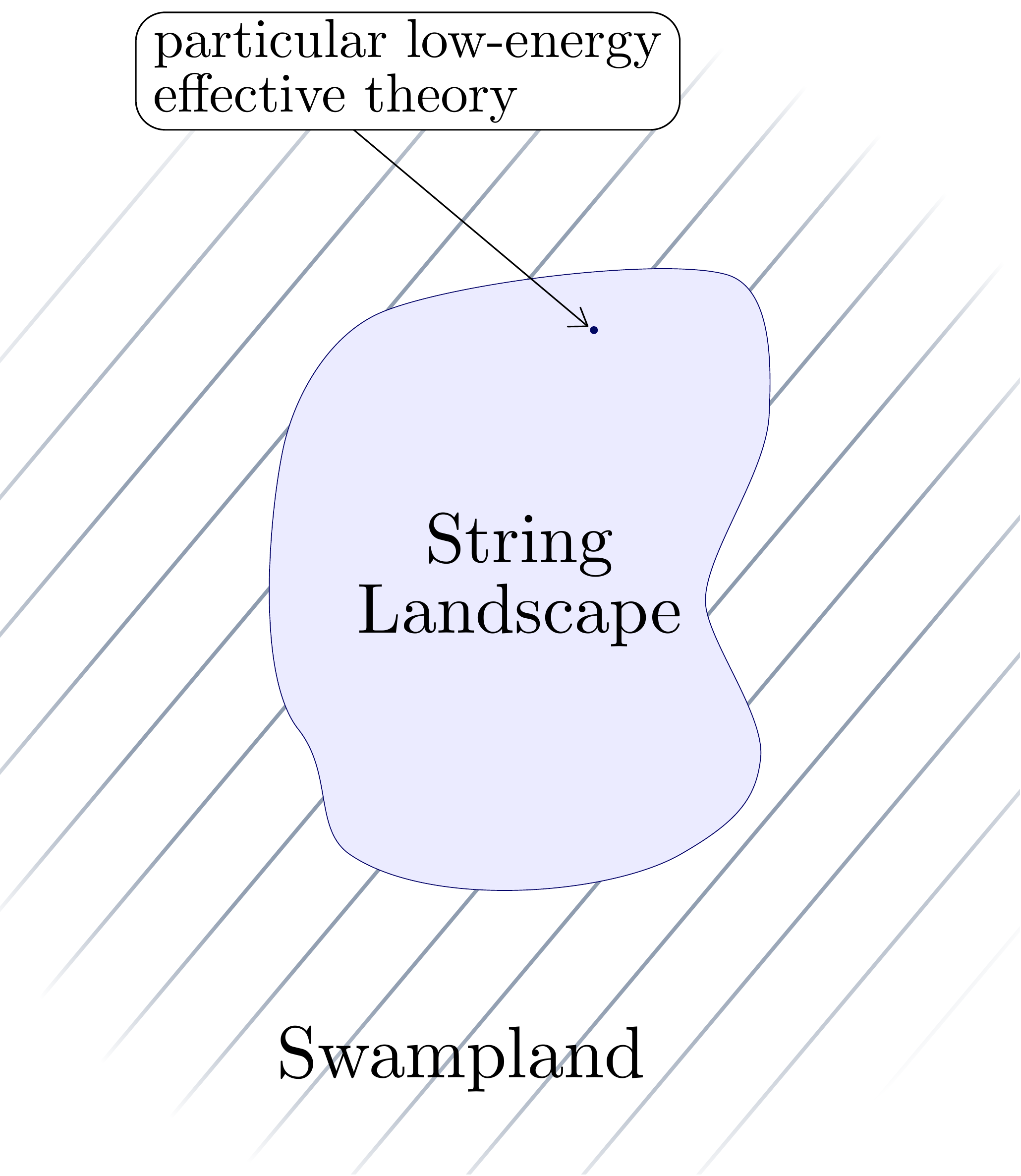}
	\end{center}
	\vspace{-15pt}
	\caption[Swampland surrounding string landscape]{Swampland of theories that lack quantum consistency surrounding string landscape.}
	\vspace{-5pt}
\end{wrapfigure}
ago, \citeauthor{WGC:Vafa} argued \cite{WGC:Vafa} that this was not the case and termed the set of those theories which lack quantum consistency the \emph{swampland}. 

Several criteria were proposed \cite{WGC:Vafa, WGC:Vafa2} to distinguish effective theories that can be UV-completed to a consistent theory of quantum gravity from those that merely lie in the ``surrounding'' swampland. The arguably best known criterion to exclude a theory from the landscape is the ``folk theorem'' that quantum gravity does not allow for global symmetries. This is turned into a quantitative statement by the conjecture proposed by \citeauthor{WGC} in \cite{WGC}  that ``gravity must be the weakest force''. More precisely: For a consistent quantum theory containing gravity and a $U(1)$ gauge field there must exist a particle charged under this gauge field with a mass bounded from above by its charge. This \emph{Weak Gravity Conjecture (WGC)} has attracted much interest in the past decade and found application e.g. in cosmology, where it is used to constrain models of large field inflation \cite{WGC:Inflation,WGC:Inflation2,WGC:Fencing}. It was generalized to settings with several $U(1)$s \cite{WGC:Naturalness}, $p$-forms \cite{Heidenreich} and quite recently to situations where scalar fields are present \cite{WGC:Palti1, WGC:Palti2, WGC:Palti3}. 

\begin{figure}
	\vspace{10pt}
	\begin{center}
		\includegraphics[width=\linewidth]{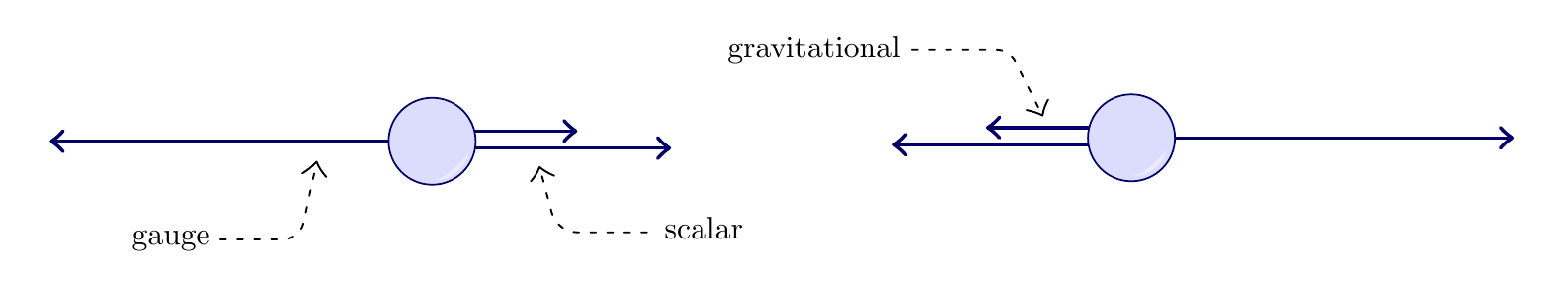}
		\vspace{2.5pt}
		\caption[Particles satisfying gauge-salar WGC]{Two particles satisfying the Gauge-Scalar WGC.\label{Fig:WGC:GaugeScalarWGC}}
	\end{center}
\end{figure}

This last generalization is particularly important for us. It proposes bounds (henceforth called \emph{Scalar} and \emph{Gauge-Scalar WGC})
\begin{align}
m^2 < \mu^2M_p^2,\quad g^2 M_p^2 \ge m^2+\mu^2 M_p^2
\end{align}
for a particle $m$ coupled to a scalar field with coupling $\mu$. The second inequality is saturated by BPS states and for two such particles, the combined scalar, vector and gravitational forces cancel as sketched in fig. \ref{Fig:WGC:GaugeScalarWGC}. We will elaborate on this in section \ref{Section:GaugeScalarWGC}.
In the main part of this thesis, we will eventually suggest a similar condition for \emph{axion-instanton} pairs in the presence of scalar fields. Before doing so, we need to lay the basis. Section \ref{Section:WGC} begins with a review of the no-global-symmetries conjecture and the black hole based argument supporting it. The remaining part of the section contains a review of the Weak Gravity Conjecture, a discussion of motivation as well as evidence for it and we will present some of the extensions to more general settings. 

Evidence for our conjecture about axions, instantons and scalar fields will come from compactification of Type IIA string theory where the scalars will be the moduli of the compactification manifold. In order to lay the ground for sections \ref{Section:ScalarWGC} and \ref{Section:ScalarIIAWGC} and make the thesis self-contained  - we review the idea and procedure of Calabi-Yau compactification in section \ref{Section:Compactification}, focusing on the moduli spaces of Calabi-Yau manifolds. In particular, we discuss the \emph{special K\"ahler} structure, which the complex structure and K\"ahler moduli spaces carry. While a basic knowledge of differential geometry is of course indispensable, appendix \ref{Appendix:Maths} gives a recap of complex manifolds while \ref{Appendix:String:SpecialGeometry} introduces the notion of \emph{special geometry}. 

\begin{figure} 
	\centering                 
	\includegraphics[width=.8\linewidth]{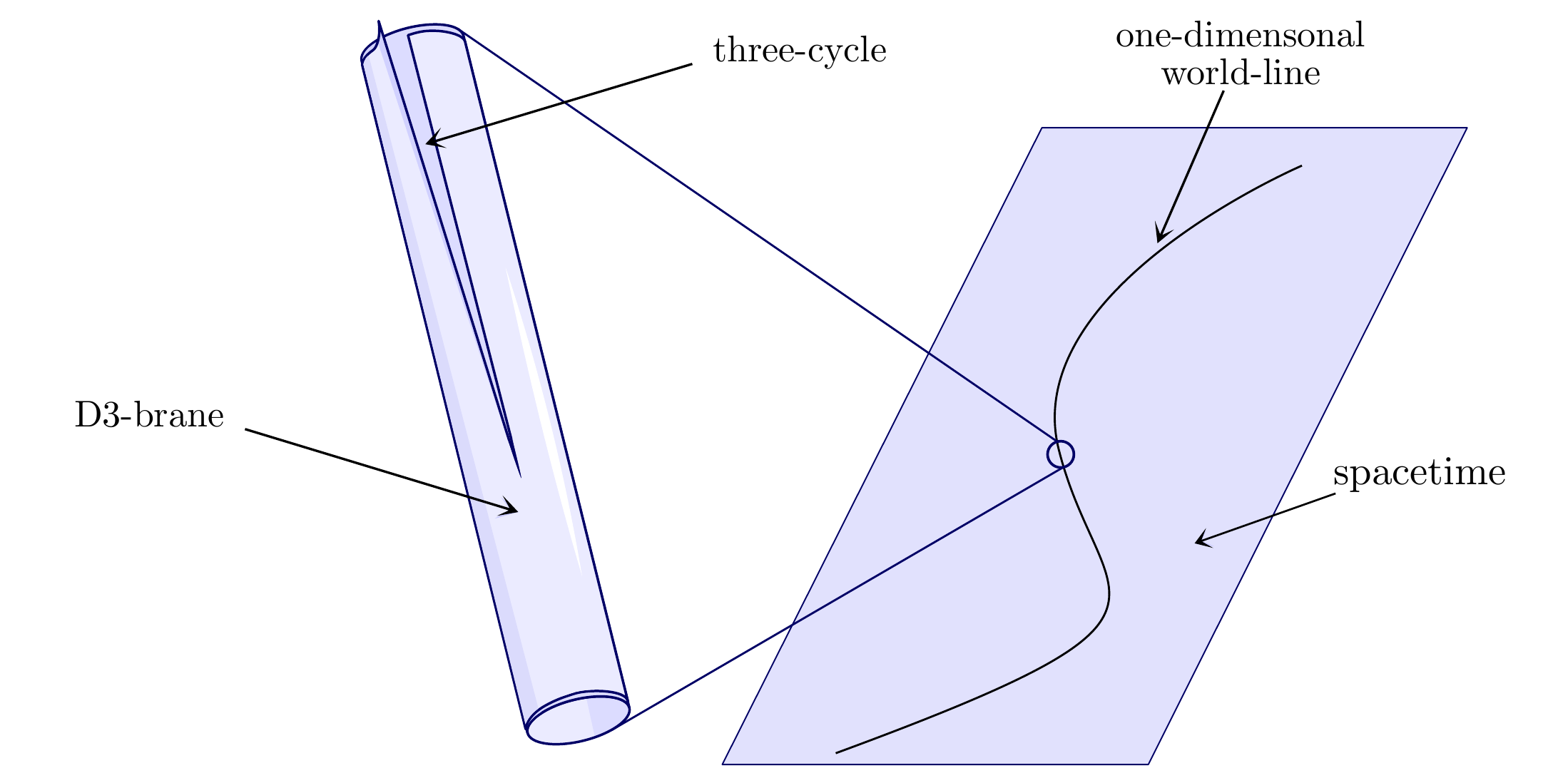}  
	\caption[D3-brane wrapping three-cycle]{D3-brane wrapping three-cycle looks like particle from low-energy perspective. Surface symbolizing brane is cut open to expose cycle underneath.\label{Fig:D3BraneCompactification}}
\end{figure}

In section \ref{Section:ScalarWGC} which marks the beginning of the main part of this thesis, we will see that compactification of Type IIB supergravity on a Calabi-Yau three-fold gives rise to several vector fields in spacetime while D3-branes wrapping supersymmetric three-cycles look like particles charged under these gauge fields from the perspective of the four-dimensional theory. The latter is due to the fact that we can perform the spatial integrations in the brane action which results in a one-dimensional path in spacetime as illustrated in fig. \ref{Fig:D3BraneCompactification}. 
We will show that these particles satisfy the condition
\begin{align}
q_AG^{AB}q_B=8\V \left(m^2 +4G^{AB}\nabla_Am\bar\nabla_B m\right).
\end{align}
Here, $q_A$ are the charges corresponding to the vector fields, $G^{AB}$ is the metric on the moduli space, $\V$ the compactification manifold's volume, the particle mass $m$ is a function of the moduli and units are chosen such that $M_p=1$.

The same can be done in Type IIA supergravity, where Euclidean E2-branes couple to the axions $\xi$ and $\tilde\xi$ arising from one of the forms upon compactification to four dimensions. In section \ref{Section:ScalarIIAWGC}, we will find that wrapping supersymmetric three-cycles, these E2-branes look like points in space-time and can therefore be interpreted as \emph{instantons}. We will show that these satisfy a condition which is similar to the Gauge-Scalar WGC and relates instanton action, axion decay constants and the scalars:
\begin{align}
Q^2 = S^2 + G^{ab}\nabla_a S\bar\nabla_b S
\label{IntroQ2}
\end{align} 
with $S$ the instanton action and $\Q^2$ defined as
\begin{align}
\Q^2 = \frac{1}{2}\begin{pmatrix}
p && q
\end{pmatrix}\begin{pmatrix}
-(\I + \R\I^{-1}R) && \R\I^{-1}\\
\I^{-1}\R && -\I^{-1}
\end{pmatrix}\begin{pmatrix}
p \\ q
\end{pmatrix}.
\end{align} 
In this expression, the charges $q$ and $p$ are the couplings to $\xi$ and $\tilde\xi$ respectively while the matrix entries come from the kinetic terms for the axions in the four-dimensional action. The matrices $\I$ and $\R$ are the imaginary and real parts of the gauge-coupling matrix $\M$ which is defined in terms of the prepotential for the Calabi-Yau moduli. We will establish \eqref{IntroQ2} for the axions with kinetic term
\begin{align}
	\int\left[ (\I)^{-1}\right]^{\ha\hat b}\dif\tilde{\xi}_\ha\wedge *\dif\tilde{\xi}_{\hat b}
\end{align}
which corresponds to having purely $p$-charges but expect it to hold also for charges $q$ or both, $q$ and $p$.

The Gauge-Scalar WGC can be phrased as the statement that for two particles, the gauge repulsion exceeds the combined gravitational and scalar attraction while all forces cancel in case of two BPS states. It is therefore natural to propose that analogously, the equality \eqref{IntroQ2} becomes an \emph{inequality} 
\begin{align}
Q^2 \ge S^2 + G^{ab}\nabla_a S\bar\nabla_b S
\end{align} 
in the absence of supersymmetry. For a single axion with decay constant $f$ and only one scalar field $\phi$, this translates to
\begin{align}
S^2 + \partial_\phi S\bar \partial_\phi S \le 1/f^2
\end{align}
Further, we conjecture that this relation is a general extension of the axion-instanton WGC to situations with scalar fields.

Finally, some (especially lengthy) calculations were put in appendix \ref{Appendix:Calculations} and are cited when needed in order to make the thesis clear and readable.\newpage

\section{Weak Gravity Conjecture}\thispagestyle{empty}
\label{Section:WGC}
\subsection{Absence of global symmetries in quantum gravity}
Since in the limit $g\rightarrow 0$ a gauge symmetry becomes a global one, the WGC is strongly motivated by the statement that quantum gravity does not allow for global symmetries.
\begin{figure}[!h]
		\centering
		\includegraphics[width=.9\textwidth]{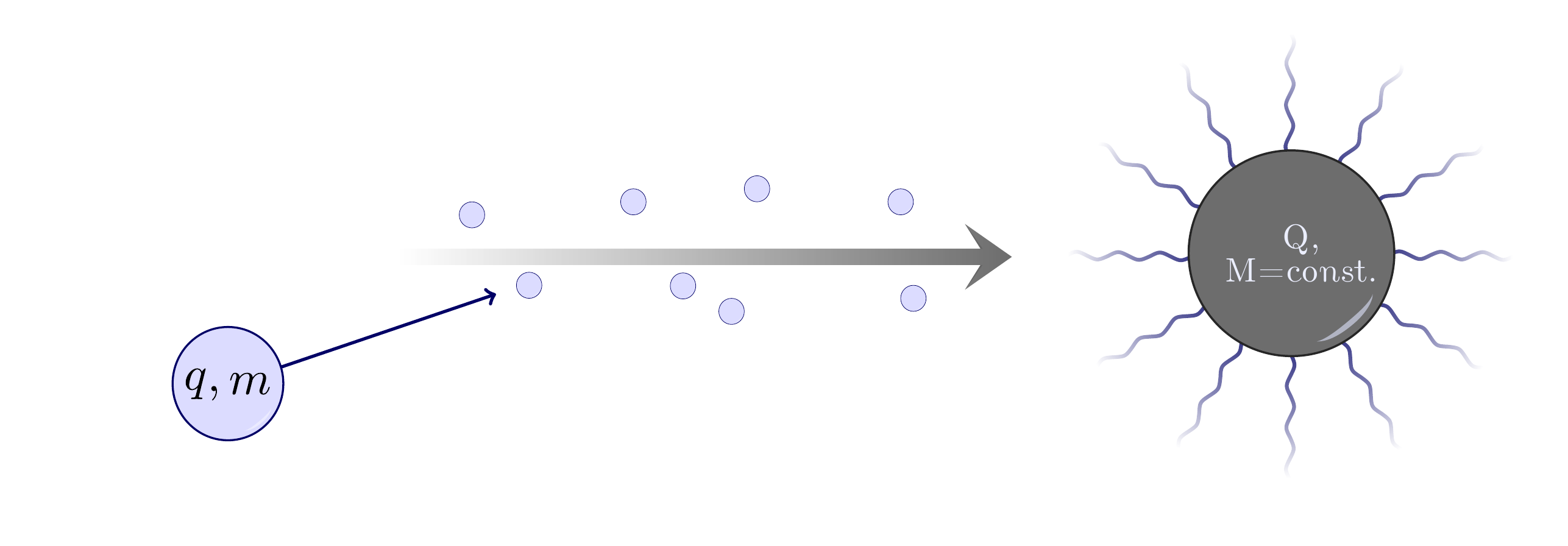}
		\caption[Increasing black hole charge]{Throwing particles into the black hole to increase its charge to $Q$ while radiating the excess mass away.}
		\label{Fig:WGC:GlobalSymmetries}
\end{figure}
Before giving a sharper formulation of the conjecture, we review the general black-hole based argument \cite{WGC:GaugeGravity, WGC:GlobalSymmetries} for the absence of global symmetries. While a black hole charged under a gauge theory needs to satisfy the extremality bound\footnote{See equation \eqref{WGC:ExtremalityBound} below.} \begin{align}M\ge QM_p\label{Eq:WGC:GlobalBound1},\end{align} for a global symmetry we can construct a black hole at a fixed mass with arbitrarily high charge $Q$: We can always increase the charge by throwing enough charged particles into the black hole while keeping the black-hole mass constant. The latter is achieved by waiting for the excess mass to be radiated away by emission of uncharged particles (e.g. photons) via Hawking radiation before throwing another particle into the black hole. This is depicted in fig. \ref{Fig:WGC:GlobalSymmetries}.

\begin{wrapfigure}{r}{.55\textwidth}
	\vspace{-15pt}
	\begin{center}
		\includegraphics[width=.5\textwidth]{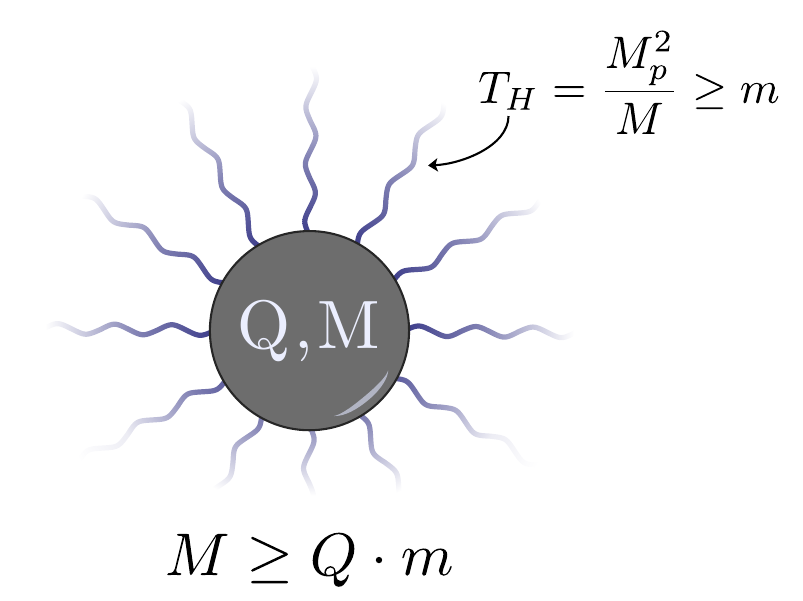}
		\caption[Energy conservation for decaying black hole]{Bound demanded by energy conservation for decaying BH.}
		\vspace{-20pt}
		\label{Fig:WGC:GlobalSymmetries2}
	\end{center}
\end{wrapfigure}

Having established that we can consider a black hole with any global charge $Q$, we now see what happens if we let such a black hole evaporate its mass. At some point, its Hawking temperature $T_H =M_p^2/M$ will eventually exceed  the mass $m$ of the lightest charged particle. In order to get rid of its charge, the black hole needs to have a mass that is equal to at least $Q$ times $m$ at this point,
\begin{align}
	M \ge Qm.
	\label{Eq:WGC:GlobalBound2}
\end{align}This situation is sketched in fig. \ref{Fig:WGC:GlobalSymmetries2} on the right-hand side. From the two
bounds \eqref{Eq:WGC:GlobalBound1} and \eqref{Eq:WGC:GlobalBound2} we conclude
\begin{align}
Q\le \left(\frac{M_p}{m}\right)^2,
\label{WGC:BlackHoleChargeBound}
\end{align}
which is violated if we only take $Q$ big enough.

While a similar argument can also be applied for gauge theories with tiny couplings \cite{WGC},
 in the case of a global symmetry, Hawking radiation produces equal numbers of particles with charge $+q$ and $-q$ since there is no electric field outside the black hole to produce a chemical potential term that favors discharge \cite{WGC:GaugeGravity}. Thus, the black hole cannot decay completely and what remains is a stable black hole \emph{remnant}, an object with size and mass of Planck order. Since there is no upper bound on the global charge, this leads to an infinite number of such remnants, which are argued \cite{Susskind} to let the entropy per unit area go to infinity and Newton's constant to zero. This strongly suggests that global symmetries should be absent from a quantum theory of gravity which is further supported by string theory, where in fact all symmetries are gauged \cite{GlobalSymmetriesInST}.
\subsection{Weak Gravity Conjecture}
\label{Section:WGC:WGC}
In view of the above claim, something should prevent us from taking the limit $g\rightarrow 0$ for the coupling of a \emph{local} symmetry where it becomes indistinguishable from a \emph{global} one. Naturally, the question arises if small but non-zero couplings are also problematic and if so, what the lower bound for allowed couplings is. 
In their paper \cite{WGC}, \citeauthor{WGC} proposed the following:
\newline
\begin{tcolorbox}[title=Weak Gravity Conjecture,colback=\BColor]
In a theory containing a $U(1)$ gauge field with coupling $g$ and gravity,
\begin{enumerate}[i)]
	\item there must be a particle carrying charge $Q$ under the gauge field with its mass satisfying the bound 
	\begin{align}
		m \le qM_p,
		\label{Eq:WGC}
	\end{align}
	where we defined $q = \sqrt{2}Qg$,
	\item and the effective theory breaks down at a scale $\Lambda \lesssim gM_p$.
\end{enumerate}
\end{tcolorbox}
\vspace{.5em}
These statements are called \emph{electric} and \emph{magnetic WGC}. The latter actually follows from the former by considering a magnetic monopole: The Dirac-quantization condition demands $g_{\text{mag}}\sim 1/g_{\text{el}}$. Since the monopole mass $m$ needs to account for the energy stored in its magnetic field, we have 
\begin{align}
m\ge g_{\text{mag}}^2\Lambda,
\end{align}
where $\Lambda$ is the cutoff of the effective theory. Applying the electric WGC, 
\begin{align}
m\lesssim g_{\text{mag}}M_p,
\end{align}
it follows that 
\begin{equation}
g_{\text{mag}}\Lambda\lesssim M_p,
\end{equation}
which is the magnetic WGC for $g_{\text{mag}}\sim 1/g_{\text{el}}$. Phrased differently, the WGC thus suggests that there is a lower bound on the strengths of interactions associated with the gauge boson and hence turns the merely qualitative argument that global symmetries should be absent from a quantum theory of gravity into a quantitative criterion. Before discussing the conjecture in more detail, we want to point out that while generally thought to be true, the WGC still is somewhat speculative and the best evidence so far is that all models obtained from string theory seem to satisfy it \cite{WGC:Pablo}. We will test the conjecture (or rather an extension of it) explicitly in section \ref{Section:ScalarWGC}.

The argument we made to exclude global symmetries relied on the fact that the charge of the black hole was not observable from outside, which is no longer true if we consider non-zero gauge couplings. Nevertheless, we can adapt the above argument as follows: Consider a black hole in four dimensions with mass $M$ electrically charged under a $U(1)$ field with coupling $g$. Such a solution of the Einstein action
\begin{align}
S=\frac{1}{2\kappa^2}\int \R*\1 -\frac{1}{2g^2}\int F\wedge *F
\end{align}
is called  a \emph{Reissner-Nordstr\"om black hole}. We chose convention such that
\begin{align}
 F = \dif A,\quad A=\frac{g^2Q}{4\pi r},
\end{align}
i.e. the charges are defined as
\begin{align}
	Q = \frac{1}{g^2}\int_{S^2}*F.
	\label{Eq:WGC:BHCharges}
\end{align}
Its metric is \cite{Books:BeckerBecker}
	\begin{align}
		\dif s^2 = -\Delta \dif t^2 + \Delta^{-1}\dif r^2 + r^2\dif\Omega_2^2,
	\end{align}
	where
	\begin{align}
		\Delta &= 1 - \frac{2MG}{r}+\frac{(gQ)^2G}{4\pi r^2}\nl
		&=1 - \frac{M\kappa^2}{4\pi r}+\frac{(gQ)^2\kappa^2}{4\pi (8\pi)r^2}
	\end{align}
	with $\kappa = M_p^{-1}$. Looking for the roots of $\Delta(r)$, 
	\begin{align}
	r = MG\pm\sqrt{\left(\frac{M\kappa^2}{8\pi}\right)^2-\frac{g^2Q^2\kappa^2}{(4\pi)(8\pi)}},
	\end{align}
	we can see that the $r=0$ singularity is only shielded by an event horizon if the \emph{extremality bound}
	\begin{align}
		M \ge \sqrt{2}gQM_p
		\label{WGC:ExtremalityBound}
	\end{align}
is satisfied. According to the cosmic censorship conjecture \cite{Books:BeckerBecker, CosmicCensor}, naked singularities should not appear in physical situations and therefore, \eqref{WGC:ExtremalityBound} needs to be fulfilled. 
\begin{figure}[h]
\centering
\includegraphics[width=.85\textwidth]{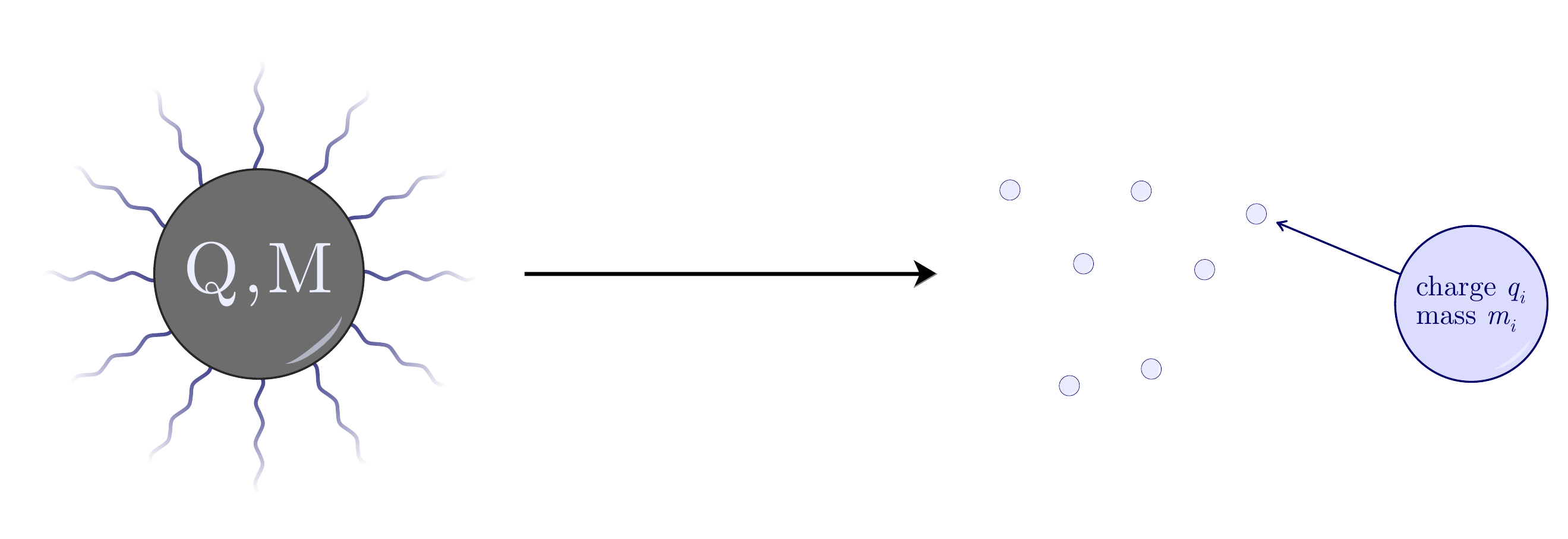}
\caption[Extremal black hole decaying]{Extremal black hole decaying to collection of states of masses $m_i$ and charges $q_i$.}
\label{Fig:WGC:BHDecay}
\end{figure}
We demand that an extremal black hole is still able to decay. If we label the final states by an index $i$ and write $m_i,q_i$ for their masses and charges respectively as depicted in fig. \ref{Fig:WGC:BHDecay}, energy conservation demands that the black hole mass $M=\sqrt{2}gQM_p$ amounts to at least the sum of the masses of these states,
\begin{align}
M\ge \sum_i m_i.
\end{align}
At the same time, we have
\begin{align}
Q = \sum_i q_i
\end{align}
due to charge conservation. Until now, we did not give any specification regarding the WGC particle. We will now argue that the weak gravity bound is satisfied by the particle whose charge-to-mass ratio is maximal. Let $z_i\coloneqq q_i/m_i$. Then
\begin{align}
	Q=\sum_i z_i m_i \le z_{\text{max}}\sum_i M = z_{\text{max}}M,
\end{align}
where $z_{\text{max}}=q/m$ denotes the particle with maximal charge-to-mass ratio. Using $  M=\sqrt{2}gQM_p $, we conclude
	\begin{align}
		1\le \sqrt{2} \frac{q}{m}gM_p,
		\end{align}
which is precisely the electric Weak Gravity Conjecture \eqref{Eq:WGC}. 
\subsection{Generalized Conjectures}
If we take a look at the weak gravity bound \eqref{Eq:WGC} again, we find that there are different possible ways in which it could be modified. First, one can ask if an analogous bound holds also for higher-dimensional objects charged under some $p$-form. Indeed, the argument we just gave is not restricted to zero-dimensional objects, i.e. particles. Second, we can stick to particles but consider a setup where these are charged under several gauge fields and see how this restricts the particle mass. At last, we can take additional - namely scalar - forces into account. We will now discuss these modifications in turn.
\subsubsection{The WGC for $p$-form gauge fields}
As mentioned, it seems only natural to extend the Weak Gravity Conjecture to general $p$-form gauge fields, as was argued in \cite{WGC}. We assume that $p$-forms appear as
\begin{align}
	\frac{1}{2g}\int F_{p+1}\wedge *F_{p+1}.
\end{align}

The statement as given in \cite{Heidenreich} is:
\vspace{.5em}
\begin{tcolorbox}[title=$p$-Form Weak Gravity Conjecture,colback=\BColor]
	In $d$ dimensions, for each Abelian $p$-form gauge field with coupling $g$, there must be a $(p-1)$-dimensional object (a $(p-1)$-brane) with tension $T_p$ that carries integer charge $Q$ under this gauge field and satisfies
	\begin{align}
	\frac{p(d-p-2)}{d-2}T_p^2\le g^2Q^2M_d^{d-2},
	\label{Eq:WGC:General}
	\end{align}
	where $M_d$ is the $d$-dimensional Planck mass.
	\label{Conjectures:GeneralWGC}
\end{tcolorbox}
\vspace{.5em}
We will not discuss the precise nature of the prefactor and are content with seeing that \eqref{Eq:WGC:General} reduces to the precise form of the electric WGC given in \eqref{Eq:WGC} if we set $p=1$ and $d=4$, which corresponds to the case of a point-particle with mass $m=T_1$ in four-dimensional spacetime that couples to a $U(1)$-gauge field. In that case, the inequality \eqref{Eq:WGC:General} reads
\begin{align}
	\frac{1}{2}m^2\le g^2Q^2M_4^2.
\end{align}
With $M_p=M_4$, we recover \eqref{Eq:WGC}.

The conjecture is supported by the same argument as the one for the Weak Gravity Conjecture for a single $U(1)$: We demand that extremal black branes should be able do decay. There are some values of $p$ where this fails though, namely $p=0$ and $p\ge d-2$, and we need to address them separately:
\begin{enumerate}[i)]
	\item A $d-1$ form is non-dynamical and a $d$-form couples to a $(d-1)$-brane which is space-time filling.
	\item While the $p=0$ case is particularly interesting since it corresponds to axions and could yield a falsifiable prediction for the QCD axion. The obvious candidates for $(-1)$-dimensional objects are instantons that couple to these axions with the inverse decay constant playing the role of the gauge coupling. Evidence for this relation comes from string theory and we will come back to this issue in section \ref{Section:ScalarIIAWGC}.
\end{enumerate}
\subsubsection{The WGC for axions}
It was suggested in \cite{WGC} that the WGC can be extended to axions and instantons that couple to them in the following way:
\vspace{.5em}
\begin{tcolorbox}[title=Weak Gravity Conjecture for Axions,colback=\BColor]
		For any axion with decay constant $f$ there must be an instanton with action $S_E$ coupling to the axion such that
		\begin{align}
		S_E\lesssim \frac{1}{f}M_p
		\label{Eq:WGC:Axions}
		\end{align}
		is satisfied.
\end{tcolorbox}
\vspace{.5em}
We see that the euclidean instanton action $S_E$ is analogue to the mass ``$m$'' while the inverse of the axion decay constant $f$ plays the role of the coupling ``$g$''. That this statement is a consequence of the standard WGC can be derived from T-dualities in string theory, where particles are mapped to instantons and vice versa \cite{WGC:Fencing}.

One needs $S_E > 1$ in order to have $e^{-S_E}<1$ for an instanton and hence, we conclude that the axion decay constant is at most of order of the Planck mass,
\begin{align}
	f \lesssim M_p.
\end{align}
This is very interesting inasmuch as it provides a potentially falsifiable prediction for the QCD axion.
\subsubsection{The WGC for multiple $U(1)$s}
In \cite{WGC:Naturalness}, the Weak Gravity Conjecture was extended from $U(1)$ to a product gauge group of several $U(1)$s. Before we give the statement, note that the WGC can be rephrased in a slightly different way:
Consider a $U(1)$ gauge theory with coupling $g$ and particles of mass $m_i$ and charge $q_i$\footnote{That is, the index $i$ labels the species, each consisting of particles and antiparticles of mass $m_i$ and charge $q_i$ and $-q_i$ respectively.}. We define the (dimensionless) ratios
\begin{align}
	z_i\coloneqq Q_i\frac{M_p}{m_i},\quad\text{where }Q_i\coloneqq\sqrt{2}gq_i.
\end{align}
Then, the electric WGC \eqref{Eq:WGC} is the statement that there exists a species $k$ such that
\begin{align}
	1 \le z_k.
\end{align}
\begin{wrapfigure}{r}{.5\textwidth}
	\vspace{-10pt}
	\includegraphics[width=.5\textwidth]{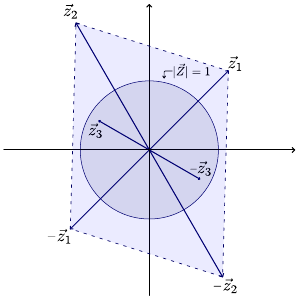}
	\caption[Convex hull condition]{Example with $n=2$ that satisfies the convex hull condition.}
	\vspace{-20pt}
	\label{Fig:WGC:ConvexHull}
\end{wrapfigure}
Now, let us extend the gauge group to a product $\prod_{a=1}^{n}U(1)_a$ with couplings $g^a$ such that each particle $m_i$ carries charges $q_i^a$. Similar to before, we define the ratios
\begin{align}
z_i^a\coloneqq Q_i^a\frac{M_p}{m_i},\quad\text{where } Q_i^a\coloneqq\sqrt{2}q_i^ag^a.
\end{align}
For $i$ fixed, $q_i^a$ and $z_i^a$ are vectors of $SO(n)$ and we write $\vec{q}_i$ and $\vec{z}_i$ respectively.

It turns out that the ``most obvious'' generalization of the WGC to the present case, namely to conjecture that there is a species $k$ such that $\left|\vec{z}_k\right|> 1$, is not sufficient. However, demanding that all vectors $\vec z_i$ have $|\vec z_i|> 1$ is too strict.
To determine the proper condition, we consider a black hole with charge $\vec{Q}$ and mass $M$ and - following the same line of reasoning as in \ref{Section:WGC:WGC} -  demand that it be able to decay into a state with $n_i$ particles of species $i$ for $i=1,...,n$. Due to charge conservation,
\begin{align}
\vec Q = \sum_i \vec Q_i
\end{align}
and with $\vec{Z}=\vec{Q}M_p/M$ and the vectors $\vec{z}_i$ as defined above, it follows that
\begin{align}
	\vec Z &= \sum_i\vec Q_i\frac{M_p}{M}= \frac{1}{M}\sum_in_i m_i\vec{z}_i.
\end{align}
Thus, we can interpret $\vec Z$ as weighted average of the charge vectors $\vec z_i$. From energy conservation it follows that the mass $M$ has to account for at least the masses of the decay particles, i.e.
\begin{align}
1 > \frac{1}{M} \sum_i n_im_i.
\end{align}
Hence, the black hole charge vector $\vec{Z}$ lies in the convex hull spanned by the vectors $\pm\vec{z}_i$. Since for an extremal black hole $|\vec{Z}|=1$, this translates to the requirement that the unit ball must be enclosed by the complex hull. This motivates the following conjecture:
\FloatBarrier
\vspace{.5em}
\begin{tcolorbox}[title=Weak Gravity Conjecture for Several $U(1)$s, colback=\BColor, width=\linewidth]
	Let $\{m_i\}$ be a number of particles carrying charge $\vec q_i = (q_i^a)$ under a product gauge group $\prod_{a=1}^{N}U(1)_a$ with couplings $g^a$ and vectors $\vec Q_i$ be defined via $Q^a_i = \sqrt{2}q_i^ag^a$. Then, the  convex hull spanned by the (dimensionless) charge-to-mass ratios $\vec z_i=\vec Q_iM_p/m_i$ encloses the unit ball.
\end{tcolorbox}
\vspace{.5em}
Fig. \ref{Fig:WGC:ConvexHull} illustrates a situation with two $U(1)$s where the conjecture is satisfied. Note that although $|\vec z_3|<1$, the unit ball is enclosed in the convex hull. Likewise, one can easily think of a situation where all charge vectors $\vec z_i$ have $|\vec z_i|>1$ that would still violate the conjecture: Take for example $n=2$ and assume there are two species with $\vec z_{1,2}$ orthogonal to each other. Then, we can take their lengths slightly bigger than one but such that their convex hull intersects the unit disk. Since the boundary of the unit disk consists of extremal black hole solutions, this would correspond to a situation where stable black hole remnants exist that render the theory unphysical.
\subsubsection{Gauge-Scalar Weak Gravity Conjecture}
\label{Section:GaugeScalarWGC}
Finally, we come to an extension of the weak gravity conjecture that takes not only gauge and gravitational but also scalar forces into consideration. This is going to play a major role in the main part of the thesis where it is elaborated in more detail. Here - for the sake of simplicity - we will restrict to a situation where the WGC particle couples to a single scalar field. In that case, it was proposed in \cite{WGC:Palti1} that the WGC bound \eqref{Eq:WGC} needs to be altered in the following way:
\vspace{.5em}
\begin{tcolorbox}[title=Gauge-Scalar Weak Gravity Conjecture, colback=\BColor, width=\linewidth]
	If the WGC particle is coupled to a scalar field with coupling $\mu$, the WGC bound is modified to
	\begin{align}
		m^2+\mu^2M_p^2\le g^2M_p^2
		\label{Eq:WGC:GaugeScalar}
	\end{align}
	where absorbed the charge $q$ in the definition of $g$.
\end{tcolorbox}
It is easy to give a physical interpretation to \eqref{Eq:WGC:GaugeScalar}. We consider a fermion $\psi$ as WGC particle whose mass $m=m(\phi)$ is parameterized by a scalar $\phi$ as well as a single gauge-field $A_\mu$ under which the WGC particle is charged. Expanding the scalar field about its VEV, $\phi = \langle\phi\rangle+\delta\phi$ introduces a coupling to the WGC particle: We have $m(\phi)=m(\langle\phi\rangle)+\partial_\phi m(\langle\phi\rangle)\delta\phi$ and the Lagrangian contains a term
\begin{align}
\mathscr L\supset\partial_\phi m\delta\phi\bar\psi\psi.
\end{align}
Similarly, if the WGC particle is itself a (complex) scalar $\varphi$, the term $m^2(\phi)\varphi\varphi^*$ gives rise to a coupling
\begin{align}
\mathscr L \supset 2m\partial_\phi m\delta\phi\varphi\varphi^*
\end{align}
in the Lagrangian. We stick with the fermion case where
\begin{align}
\mathscr L\supset m\bar\psi\psi + QA_\mu\gamma^\mu\bar\psi\psi + \partial_\phi m\delta\phi\bar\psi\psi.
\end{align}\newpage
Clearly, this gives rise to the following tree-level interactions

\begin{figure}[h!]
	\begin{center}
		\subfloat{{\includegraphics[width=.25\linewidth,page=1]{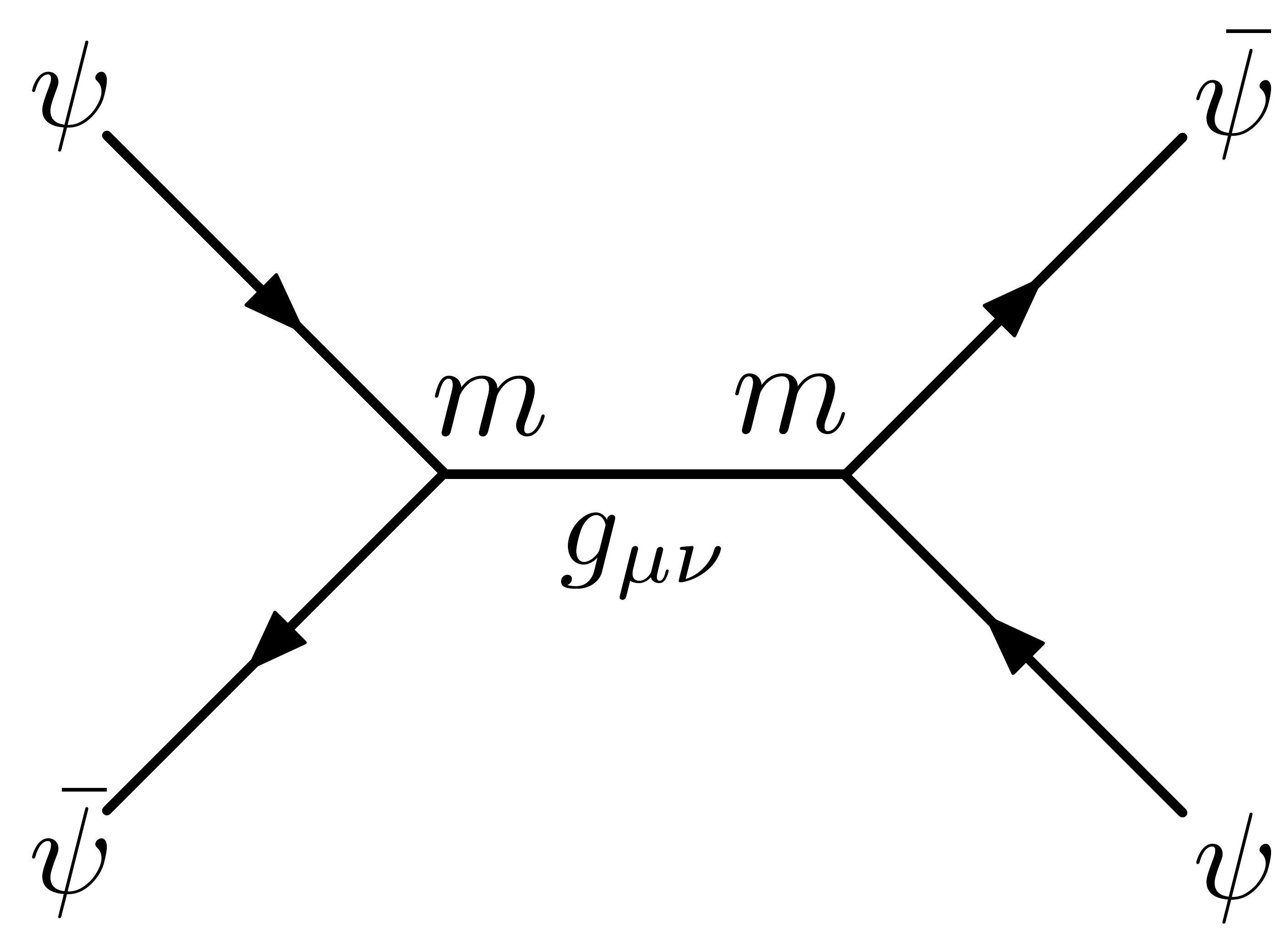}}}%
		\qquad
		\subfloat{{	\includegraphics[width=.25\linewidth,page=2]{Diagrams/Feynman.pdf}}}
		\qquad
		\subfloat{{	\includegraphics[width=.25\linewidth,page=3]{Diagrams/Feynman.pdf}}}%
	\end{center}
\end{figure}
between two WGC particles via gravity, the gauge field and the scalar. Associated with these are forces of the form
\begin{align}
F\sim \frac{g^2}{r^2},
\end{align}
where $g$ is the respective coupling. These are attractive for gravity and the scalar and repulsive for the gauge field. Thus, \eqref{Eq:WGC:GaugeScalar} can be read as the statement that the gauge repulsion exceeds the combined forces of gravitational and scalar attraction. This was already sketched in figure \ref{Fig:WGC:GaugeScalarWGC}.
\\
We will establish in section \ref{Section:D3Branes} that equality holds for BPS states. Due to the different signs, in this case the sum over all forces vanishes: \emph{Two similar BPS particles put next to each other do not feel any force}.
For gravity to truly be the weakest force, we need that in addition to \eqref{Eq:WGC:GaugeScalar}, the scalar interaction must exceed the gravitational force:
\vspace{.5em}
\begin{tcolorbox}[title=Scalar Weak Gravity Conjecture, colback=\BColor, width=\linewidth]
	If the WGC particle is coupled to a scalar field with coupling $\mu$, then
	\begin{align}
	|\mu| M_p > m.
	\label{Eq:WGC:Scalar}
	\end{align}
\end{tcolorbox}
Note that with $\mu=\partial_\phi m$, this is a differential equation that is easily integrated:
\begin{align}
	m\sim e^{-\frac{\phi}{M_p}}.
\end{align}
Interestingly, this connects the WGC with another quantum-consistency condition proposed in \cite{WGC:Vafa2} which we discuss as last conjecture about quantum gravity before turning to Calabi-Yau compactification.
\subsubsection{Swampland Conjecture}It is well-known that the moduli space of a consistent quantum theory of gravity (this is true in string theory but assumed to be a general feature of quantum gravity) is parameterized by the expectation values of massless scalar fields. Hence, we can talk about the \emph{geometry} of the moduli space by defining a metric via the kinetic terms of these scalars.
\begin{figure}
	\begin{center}
	\includegraphics[width=1\linewidth]{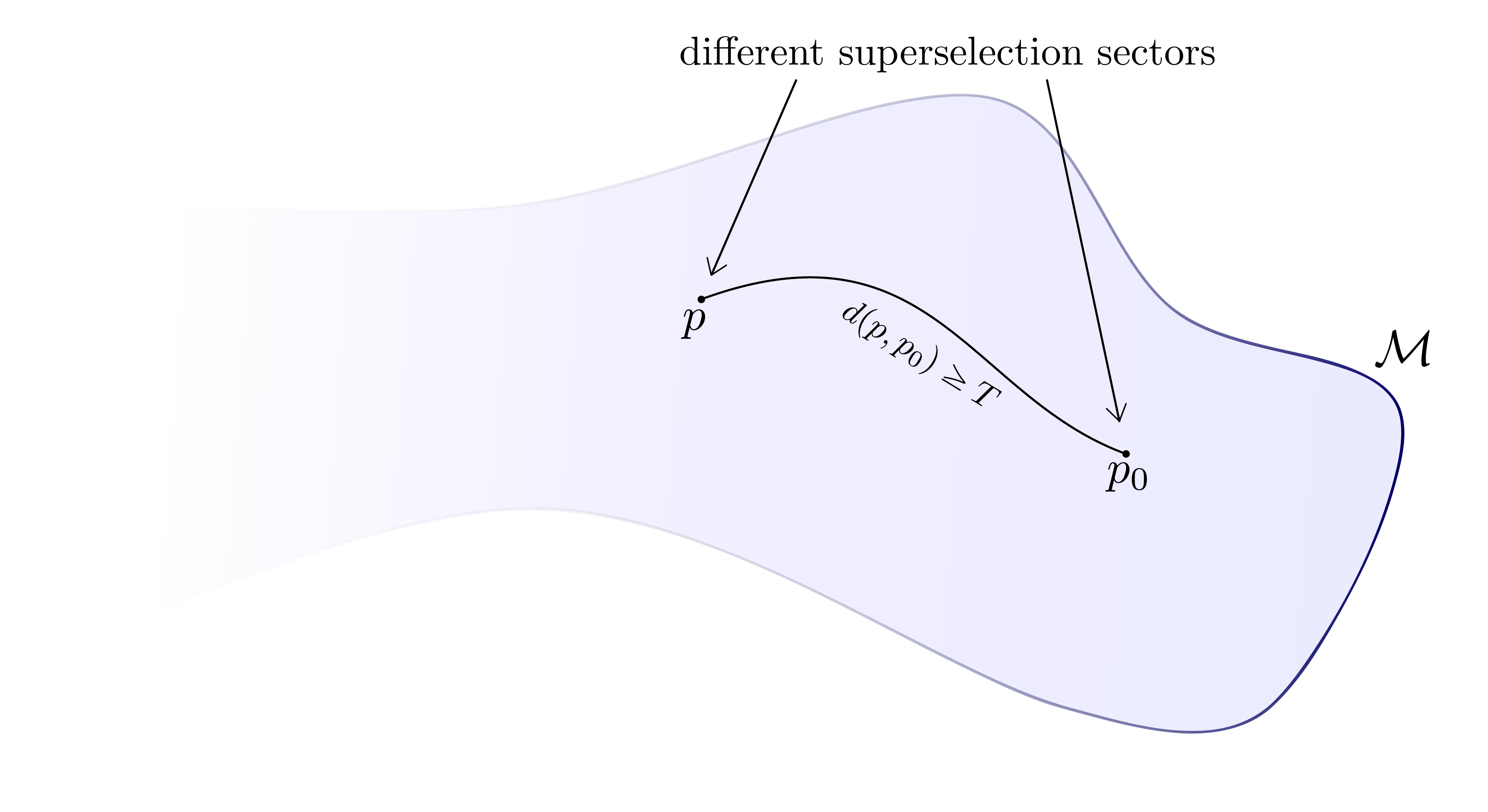}
	\caption[Moduli space of a quantum theory of gravity]{Moduli space of a quantum theory of gravity. Points correspond to distinct low-energy effective theories. For a fixed point $p_0$, it is conjectured that for any $T>0$, there is a point $p$ at a distance $d(p,p_0)\ge T$.\label{Fig:WGC:ModuliSpace}}
	\end{center}
\end{figure}
A single point (as depicted in fig. \ref{Fig:WGC:ModuliSpace}) corresponds to a certain low-energy effective action and \citeauthor{WGC:Vafa2} conjectured that displacements from such a point in the moduli space of a quantum theory of gravity lead to an infinite tower of states that become light exponentially fast. Distances in the moduli space can be defined as shortest geodesics with respect to the metric mentioned above. 
\vspace{.5em}
\begin{tcolorbox}[title=Swampland Conjecture, colback=\BColor, width=\linewidth]
	Let $\M$ denote the moduli space of a quantum theory of gravity and let the distance $d(p,q)$ between two points of $\M$ be defined as shortest geodesic between them. Then, for any $p_0\in\M$, the interval
	\begin{align}
		\{d(p,p_0)|p\in\M\}
	\end{align}
	is not bounded from above and the theory at $p$ has an infinite tower of states with mass of order
	\begin{align}
		m~\sim e^{-\alpha d(p,p_0)}
	\end{align}
	with some $\alpha > 0$.
\end{tcolorbox}
The conjecture implies that as the distance diverges, the low-energy effective theory breaks down due to the appearance of an infinite tower of light states. Hence, the theory corresponding to a particular point $p_0$ only makes sense in some finite region around this point.\newpage

\leavevmode\thispagestyle{empty}
\newpage
	
\section{Calabi-Yau Compactification}
\label{Section:Compactification}
\begin{figure}[h]            
	\begin{center}        
		\includegraphics[width=\linewidth]{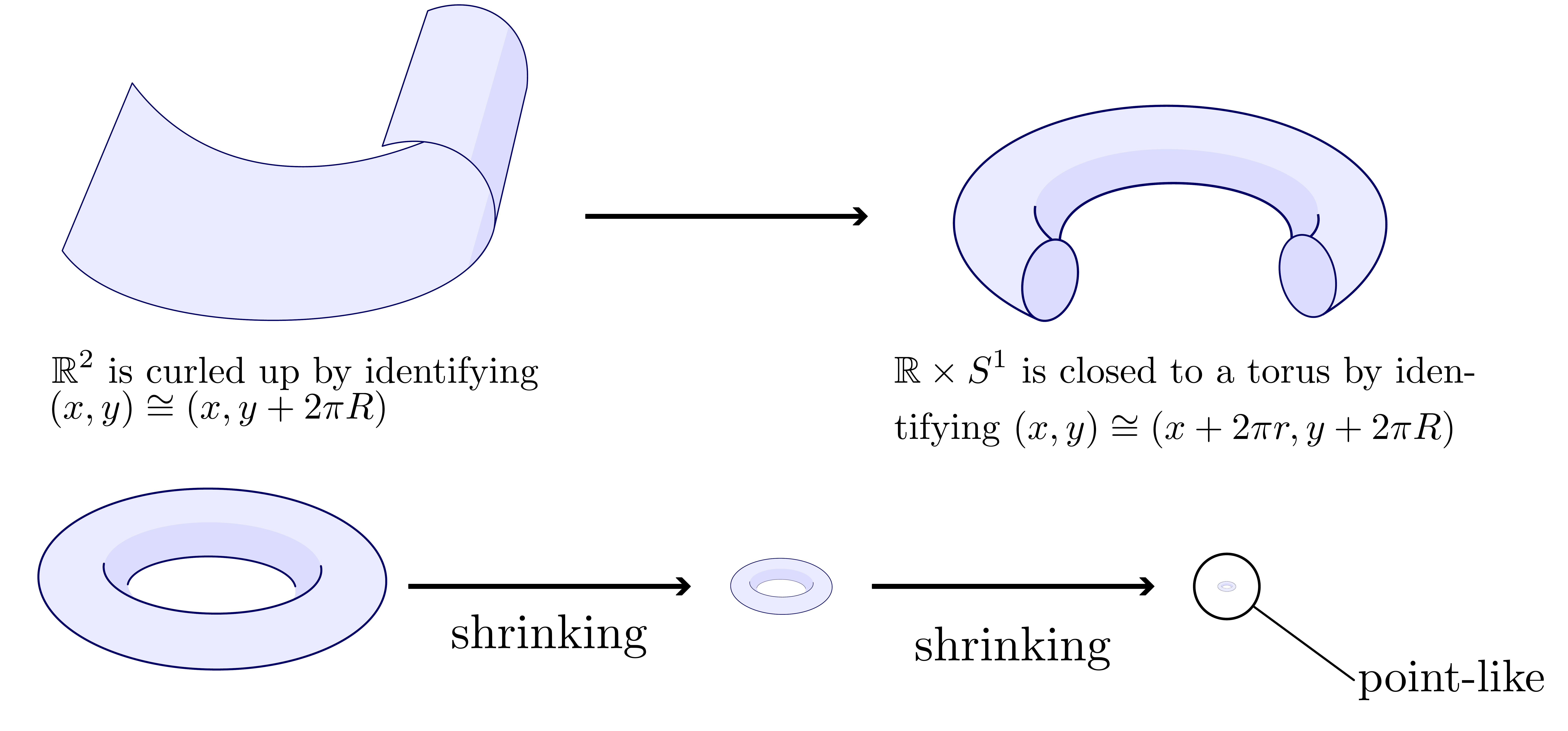}  
	\caption[Compactification of two-dimensional surface]{Compactified two-dimensional surface looks like a single point from low-dimensional perspective.\label{Fig:Compactification:Compactification}}
	\end{center}     
\end{figure}

In the thesis, we will mostly be dealing with the low-energy actions of Type IIA and IIB string theory. These are formulated in ten spacetime dimensions. Since this is in contradiction with our every-day experience, six of these dimensions must be such that they are not detectable in experiment. A possible way to resolve this issue is requiring the extra-dimensions to be small and compact such that they are invisible above a certain length scale. This is illustrated in fig. \ref{Fig:Compactification:Compactification}: Identifying two opposite sides of a rectangle yields a cylinder. Doing the same with the two remaining sides, one arrives at a torus which looks like a single point if we take its radii small enough. Before turning to the Type II theories, we will briefly review Kaluza-Klein reduction and discuss the moduli spaces of Calabi-Yau manifolds.
\subsection{Kaluza-Klein reduction}
In the Kaluza-Klein ansatz, one assumes that spacetime $\M$ has product structure\footnote{More generally, one can consider a so-called \emph{warped product} but we will assume $\M$ to be a direct product.}
\begin{align}
	\M = \M^{1,3}\times Y,
	\label{Compactification:KaluzaKleinAnsatz}
\end{align}
where $\M^{1,3}$ is some maximally symmetric space with four non-compact dimensions that represents our observed world  - e.g. Minkowski space - and $Y$ is a compact manifold called \emph{internal} or \emph{compactification manifold}. We denote the coordinates on $\M^{1,3}$ by $x$ and those on the internal manifold by $y$. The ansatz \eqref{Compactification:KaluzaKleinAnsatz} corresponds to having a metric of the form
\begin{align}
	G_{MN}(x,y) = \begin{pmatrix}
	g_{\mu\nu}^{(4)}(x) && 0 \\
	0 && g_{mn}^{(6)}(y)
	\end{pmatrix}.
\end{align}
The field content of IIA/IIB supergravity consists - apart from the metric - of $p$-forms. The equation of motion of a $p$-form $\hat C_p$ is
\begin{align}
\dif *\dif \hat C_p = 0
\end{align}
and together with the gauge condition \begin{equation}
\dif * \hat C_p=0,
\end{equation} this can be written as
\begin{align}
\Delta^{(10)}\hat C_p=0,
\end{align}
where the Laplacian is defined as $\Delta^{(10)} = (\dif + \dif^\dagger)^2$. In order to obtain a four-dimensional effective theory, one expands such a $p$-form $\hat C_p$ into a sum
\begin{align}
\hat C_p(x,y) = C^k(x)\varphi_k(y),
\end{align}
where the $C^k$ and $\varphi_k$ are fields on $\M^{(1,3)}$ and $Y$ respectively. With the compactification ansatz \eqref{Compactification:KaluzaKleinAnsatz}, the ten-dimensional Laplacian decomposes as \begin{equation}
\Delta^{(10)}=\Delta^{(4)}+\Delta^{(6)}.
\end{equation}In order for the four dimensional field to remain massless, we need \begin{equation}
\Delta^{(4)}\hat C_p = 0
\end{equation}
and consequently
\begin{align}
	\Delta^{(6)}\varphi_k=0.
\end{align}
This tells us that the expansion is one in terms of \emph{harmonic forms} on the internal manifold.
\subsection{Calabi-Yau requirement}
So far, it is not clear what to chose as a compactification manifold, since we posed no further restrictions. But surely, the obtained four-dimensional theory would depend on this choice in a crucial way. The easiest ansatz $Y= \mathbb{T}^6$, i.e. taking the six-torus as compactification manifold, does not yield an appealing theory from a phenomenological point of view, as it leaves all ($\N=4$ or $\N=8$) supersymmetry unbroken in four dimensions. Still - despite the fact that so far, no traces of supersymmetry have been observed in experiment - it seems reasonable \cite{Polchinski1} to expect that at least some supersymmetry survives compactification to a four-dimensional theory. In particular, one can look for a model that possesses $\N=1$ supersymmetry in four dimensions at high energies which are low
\begin{figure}[t]
	\centering
	\includegraphics[width=.7\linewidth]{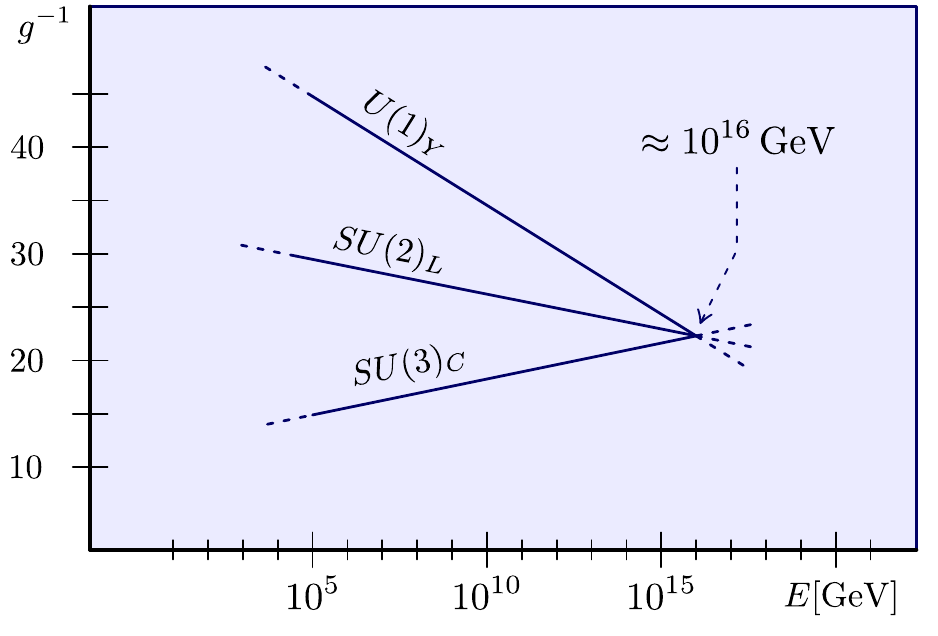}
	\caption{Running couplings in minimal supersymmetric extension of standard model.\label{Fig:GaugeCouplings}}
\end{figure}
compared to the compactification scale. A striking piece of evidence for this assumption is the resulting unification of the three gauge-couplings at about $\SI{E16}{\giga\electronvolt}$ in such a supersymmetric extension of the standard model as illustrated in fig. \ref{Fig:GaugeCouplings}.
What exactly does the requirement of unbroken supersymmetry imply for the structure of the compactification manifold $Y$? Spontaneous symmetry breaking is associated with non-vanishing vacuum expectation values: Consider a SUSY charge $\mathcal Q$ with SUSY parameter $\varepsilon$. Having unbroken supersymmetry means (see e.g. \cite{Polchinski2} on this) that
\begin{align}
	\bar\varepsilon \Q|0\rangle =0.
\end{align}
In order to have SUSY preserved by the vacuum, we therefore need all supersymmetry variations to vanish in the vacuum, 
\begin{align}
	\langle \delta_\varepsilon\Phi\rangle = \langle 0|[\bar\varepsilon \Q,\Phi]|0\rangle = 0\qquad\text{for all fields }\Phi.
\end{align}
Note that the variation $\delta b$ of a boson is fermionic and thus does not have a non-zero vacuum expectation value. Hence, we need only demand $\langle \delta f\rangle=0$ for fermions.

In Type II string theory, the SUSY charges form two independent superalgebras. Associated with each of these is a ten-dimensional Majorana-Weyl spinor $\hat \varepsilon^i$ with $i=1,2$. The gravitino which is present in both Type IIA and IIB transforms as \cite{Books:Supergravity}
\begin{align}
	\delta\hat\psi_A^i =\nabla_A\hat\varepsilon^i + \cdots
\end{align}
where $\langle\cdots\rangle = 0$ in the absence of fluxes which we do not consider. Thus, the condition for unbroken SUSY is
\begin{align}
	\langle \nabla_A\hat \varepsilon^i\rangle=0,
\end{align}
i.e. $\hat \varepsilon^i$ must be \emph{covariantly constant} with respect to the background metric. Clearly, this restricts the allowed compactification manifolds\footnote{Consider e.g. the two-sphere $S^2$. It is well-known that one cannot define a vector field that vanishes nowhere on $S^2$. But a covariantly constant vector field that vanishes at one point vanishes everywhere.}. We will examine what this implies. In Type IIA the two spinors transform in the $\textbf{16}$ and $\textbf{{16}'}$ of $SO(1,9)$, while in IIB both transform in the chiral representation $\textbf{16}$. With the ansatz \eqref{Compactification:KaluzaKleinAnsatz}, $SO(1,9)$ decomposes as
\begin{align}
	SO(1,9)\rightarrow SO(1,3)\times SO(6)
\end{align}
and we see that on $Y$, the spinors have two pieces that transform\footnote{See e.g. \cite{ZeeGroups}.} as $\textbf{4}$ and $\bar{\textbf{4}}$ of $SU(4)\cong SO(6)$.  If the spinors are covariantly constant on $Y$, they are left invariant upon parallel transport around any closed loop. But this is just another way to say that they transform as a singlet under the holonomy group\footnote{We assume $Y$ to be orientable. Otherwise, $\operatorname{Hol(Y)}\subset O(6)$.}  \begin{equation}
\operatorname{Hol(Y)}\subset SO(6)\cong SU(4).
\end{equation}There are different possibilities for $\operatorname{Hol(Y)}$. Let us consider $\operatorname{Hol(Y)}=SU(3)$ where $\textbf{4}$ decomposes into $\textbf{3}+\textbf{1}$ and - in a consequence - we arrive at one covariantly constant spinor of each chirality. To make this clear, we write (for Type IIA)
\begin{align}
	\varepsilon^1 &= \varepsilon^1_+\otimes\eta_+ + \varepsilon^1_-\otimes\eta_-,\nl
	\varepsilon^2 &= \varepsilon^2_+\otimes\eta_- + \varepsilon^2_-\otimes\eta_-.
\end{align}
The spinors $\varepsilon_+^1$ and $\varepsilon_+^2$ are two independent Weyl-spinors in four dimensions and for each covariantly constant $\eta_+$, we get $4+4$ supercharges in four dimensions - one for each component of $\varepsilon_+^i$. Since we found one $\eta_+$, this corresponds to $\N=2$ SUSY in four dimensions.

We saw that the existence of exactly one covariantly constant spinor means reducing the holonomy of our compactification manifold to $SU(3)$. For K\"ahler manifolds, this is the requirement to be \emph{Calabi-Yau}, which we discuss in detail in appendix \ref{Appendix:Maths}. See in particular definition \eqref{Maths:CalabiYau:Definition} from where we quote that a compact K\"ahler $n$-fold is Calabi-Yau iff it has holonomy $\operatorname{Hol}\subset SU(n)$. We did not yet establish that the internal manifold is K\"ahler, though. As discussed in the appendix, we need to show that the manifold is complex and has a closed K\"ahler form. For the former, it suffices to construct an almost complex structure with vanishing Nijenhuis tensor field. By means of the covariantly constant spinor, we build a bilinear
\begin{align}
	{\J_{m}}^n\coloneqq i\eta_+^\dagger{\gamma_{mp}}g^{pn}\eta_+ = -i\eta_-^\dagger\gamma_{mp}g^{pn}\eta_-,
\end{align}
where $\gamma_{mn}=\frac{1}{2}[\gamma_m,\gamma_n]$ and $\gamma^m$ are the internal gamma matrices. With help of the Fierz transformation formula (see e.g. \cite{BeckerBecker}), one finds
\begin{align}
	{\J_m}^p{\J_p}^n=-\delta_m^n,
\end{align}
that is, the compactification manifold is almost complex with
\begin{align}
 \mathcal{J} = {\mathcal{J}_m}^n \dif x^m\otimes \frac{\partial}{\partial x^n}
\end{align}
the almost complex structure. Since $\eta_+$, $\eta_-$ and the metric are covariantly constant, $\J$ is also covariantly constant and consequently, the Nijenhuis tensor vanishes:
\begin{align}
	{N^m}_{np} = {\J_n}^{l}\partial_{[l}{\J_{p]}}^{m}-{\J_p}^{l}\partial_{[l}{\J_{n]}}^m=0.
\end{align}
This implies that $Y$ is complex\footnote{See \eqref{Maths:Nihenuis} and below.}. Finally, we can - according to \eqref{Maths:ComplexCoordinates} - define complex coordinates $z^i,\bar z^{\bar j}$ such that the metric is Hermitian. Bearing in mind that the metric is also covariantly constant, one sees that the K\"ahler form
\begin{align}
	J \coloneqq ig_{i\bar j}\dif z^i\wedge \dif \bar z^{\bar j}
\end{align}
is closed. We briefly recapitulate: In order to maintain a minimal amount of supersymmetry, the six-dimensional compactification space is required to be a K\"ahler manifold with closed K\"ahler form and $SU(3)$ holonomy - in short: A \emph{Calabi-Yau threefold}.
\subsection{Reduction on the Calabi-Yau}
\label{Section:Compactification:ReductionOnCY}
We found that in Kaluza-Klein reduction, ten-dimensional fields are expanded in harmonic forms on the internal manifold. As discussed in the appendix, these are in one-to-one correspondence with elements of the cohomology groups (see \eqref{HodgeIsomorphism}) and therefore counted by the Hodge numbers which - for a Calabi-Yau - take the form shown in fig. \ref{Maths:CalabiYau:HodgeDiamond}.

To make clear how this works in practice, consider as an example a three-form $\hat C_3$ with components $\hat C_{LMN}$ in $d=10$. Then, $\hat C_{\mu\nu\sigma}$ does not carry any Calabi-Yau indices and thus is a scalar from the perspective of the compactification manifold. Likewise, $\hat C_{\mu\nu i}$ is a $(1,0)$-form (which does not exist on the internal space), $\hat C_{ij\bar{k}}$ a $(2,1)$-form and so on. This leads us to the correspondence
\begin{align}
\hat C_{\mu\nu\sigma} \leftrightarrow H^{0,0}(Y),\quad\hat C_{ijk}\leftrightarrow H^{3,0}(Y),\quad \hat C_{\mu i\bar j}\leftrightarrow H^{1,1}(Y),\nl \hat C_{ij\bar k}\leftrightarrow H^{2,1}(Y),\quad \hat C_{i\bar j\bar k}\leftrightarrow H^{1,2}(Y)
\end{align}
and
\begin{align}
\hat C_{\mu\nu i} \leftrightarrow H^{1,0}(Y) =\emptyset,\quad\hat C_{\mu i j} \leftrightarrow H^{2,0}(Y) = \emptyset
\end{align}
since $h^{1,0}=0=h^{1,1}$. We will follow this scheme when performing the compactification of Type IIA and IIB.

In the appendix, a complex basis for $H^{2,1}(Y)$ is defined consisting of the $(2,1)$-forms $\{\eta_a\}$ with $a=1,...,h^{2,1}$. Another choice \cite{Micu} is to consider a basis for the whole space
\begin{align}
H^3(Y) = H^{3,0}\oplus H^{2,1}\oplus H^{1,2}\oplus H^{0,3}.
\end{align}
Note that $h^{3,0}=1=h^{0,3}$ (the corresponding cohomology groups only consist of $\Omega$ and $\bar{\Omega}$ respectively) and hence, we can chose a real basis  $\{\alpha_{\ha},\beta^{\ha}\}$ with $\ha = 0,1,...,h^{2,1}$ satisfying
\begin{align}
\int_Y \alpha_{\ha}\wedge \alpha_{\hat b}=0=\int_Y\beta^{\ha}\wedge\beta^{\hat b},\quad \frac{1}{\V_0}\int_Y\alpha_{\ha}\wedge\beta^{\hat b} = \delta_{\ha}^{\hat b}.
\label{Maths:CalabiYau:RealBasis}
\end{align}
The various basis forms of the different cohomology groups are listed in table \ref{Table:BasisForms}.

The relations \eqref{Maths:CalabiYau:RealBasis} are preserved under the symplectic group $\operatorname{Sp}(2h^{2,1}+2)$, i.e. under transformations
\begin{align}
	\binom{\beta}{\alpha}\mapsto \begin{pmatrix}
	A && B \\
	C && D
	\end{pmatrix} \binom{\beta}{\alpha}
	\label{Maths:CalabiYau:Symplectic}
\end{align}
with
\begin{align}
	A^TD-C^TB = \1 = (A^TD-C^TB)^T
\end{align}
and
\begin{align}
	A^TC = (A^TC)^T,\quad B^TD=(B^TD)^T.
\end{align}
We will make use of this structure in order to define a symplectic section that serves as projective coordinates on the Calabi-Yau moduli space.
\begin{table}
\begin{center}
\begin{tabular}{|c|c|c|}
	\hline 
	\rowcolor{\BColor}
	\rule[-1.5ex]{0pt}{4.5ex} \textbf{Cohomology} & \textbf{Basis}  & \textbf{Defined} \\
	\hline 
	\rule[-1.5ex]{0pt}{4.5ex} $H^{1,1}(Y)$ & $\omega_A$  & in \eqref{Maths:CalabiYau:H11Basis} \\
	\hline 
	\rule[-1.5ex]{0pt}{4.5ex} $H^{2,2}(Y)$ & $\tilde{\omega}^A$ &  in \eqref{Maths:CalabiYau:H11Basis}\\
	\hline 
	\rule[-1.5ex]{0pt}{4.5ex} $H^{2,1}(Y)$ &  $\eta_a$ & below \eqref{Maths:CalabiYau:H11Basis}\\
	\hline 
	\rule[-1.5ex]{0pt}{4.5ex} $H^{1,2}(Y)$ &  $\bar\eta_a$ &  below \eqref{Maths:CalabiYau:H11Basis}\\
	\hline 	
	\rule[-1.5ex]{0pt}{4.5ex} $H^3(Y)$ & $(\alpha_{\hat a},\beta^{\hat b}) $ & in \eqref{Maths:CalabiYau:RealBasis}\\
	\hline
\end{tabular} 
\caption[Bases for cohomologies]{Bases for cohomology groups of $Y$.\label{Table:BasisForms}}
\end{center}
\end{table}
\subsection{Moduli spaces of Calabi-Yau threefolds}
A Calabi-Yau three-fold $Y$ with given Hodge numbers is not uniquely determined. Instead, we can consider perturbations \begin{equation}
g\rightarrow g+\delta g
\end{equation} of its K\"ahler metric that leave it Ricci-flat. In the following discussion of this, we follow \cite{Candelas:ModuliSpaces} and \cite{Greene}. In order to still have $R_{mn}(g+\delta g)=0$ and thus maintain the Calabi-Yau property, the variations $\delta g$ need to satisfy the \emph{Lichnerowicz} equation
\begin{align}
\nabla ^k\nabla_k \delta g_{mn}+2{{{R_m}^p}_n}^q\delta g_{pq}= 0.
\label{Maths:Lichnerowicz}
\end{align}
In complex coordinates, this splits into two independent equations for perturbations $\delta g_{i\bar{j}}$ with \emph{mixed indices} and such with \emph{pure indices}, $\delta g_{ij}$ and $\delta g_{\bar{i}\bar{j}}$, respectively.

\textbf{1.} The \emph{mixed} variations correspond to a real $(1,1)$-form 
	\begin{align}
	i\delta g_{i\bar{j}}\dif y^i\wedge\dif \bar y^j \in H^{1,1}(Y)
	\end{align}
	which allows us to expand
	\begin{align}
	(g_{i\bar j}+\delta g_{i\bar j})(x,y) = -iv^A(x)(\omega_A)_{i\bar j}(y),\quad A = 1,...,h^{1,1},
	\end{align}
	where $\{\omega_A\}$ is the basis of $H^{1,1}(Y)\cong \operatorname{Harm}^{1,1}(Y)$. Since the metric is directly related to the K\"ahler form, \eqref{Maths:KählerComplexCoord}, we have $J=v^A\omega_A$ and call the real 
		\begin{wrapfigure}{r}{.4\linewidth}
		\begin{center}
			\includegraphics[width=.9\linewidth]{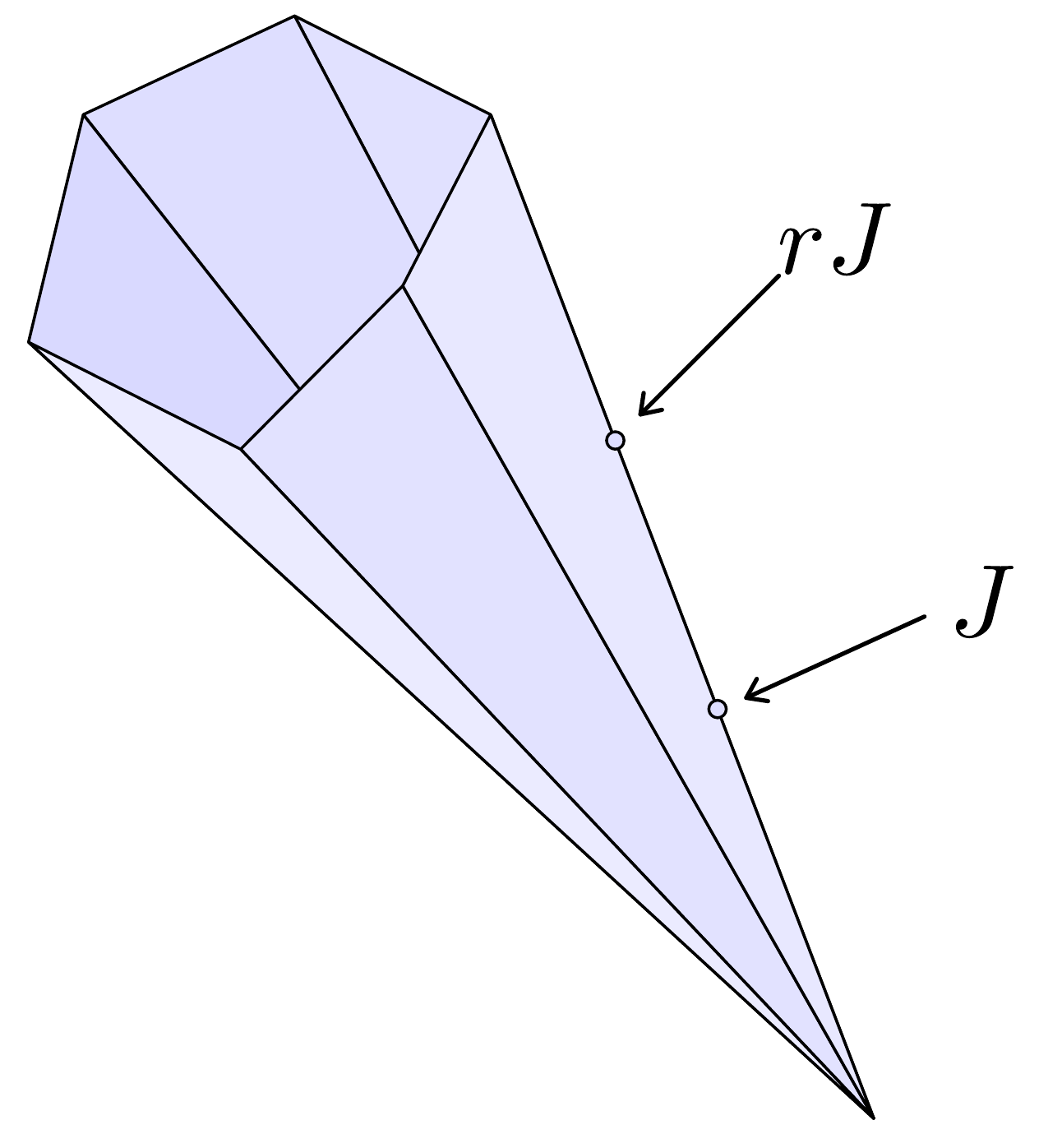}
		\caption{K\"ahler cone\label{Fig:Compactification:KählerCone}}
		\end{center}
	\end{wrapfigure}
scalars $v^A$ \emph{K\"ahler moduli}. Note that since $J\wedge J\wedge J$ is a volume form\footnote{See \eqref{Maths:VolumeForm}}, that is
\begin{align}
	\int_{\M_k}\bigwedge^k J > 0
	\label{Eq:Compactification:KählerConePositivity}
\end{align}
for $k=1,2,3$ and any complex $k$-dimensional submanifold of $Y$, these determine the \emph{volume} of the internal manifold. In general, the metric deformations which preserve \eqref{Eq:Compactification:KählerConePositivity} form a cone as illustrated in the figure on the right: If the equation holds for $J$, then it holds for any $rJ$ with $r>0$.
Together with the $h^{1,1}$ real scalars $b^A$ arising from the expansion of the two-form $\hat B_2$ which appears together with the metric in Type II string theory, $\hat B_2(x,y) = B_2(x)+b^A(x)\omega_A(y)$,
we define $h^{1,1}$ complex scalar fields
	\begin{align}
	t^A \coloneqq b^A + iv^A
	\end{align}
forming the so-called \emph{complexified K\"ahler cone} $\mathcal{M}^\text{ks}$.
	
	\textbf{2.} Since there are no $(2,0)$-forms on a Calabi-Yau, the \emph{pure} variations cannot be expanded directly. Instead, they correspond to a complex $(2,1)$-form
	\begin{align}
	\Omega_{ijk}g^{k\bar m}\delta g_{\bar m\bar l}\dif y^i\wedge\dif y^j\wedge \dif y^{\bar l} \in H^{2,1}(Y)
	\end{align}
	where $\Omega$ is the holomorphic $(3,0)$-form. We expand
	\begin{align}
	\Omega_{ijk}g^{k\bar m}\delta g_{\bar m\bar l}= \bar{z}^a(\bar{\eta}_a)_{ij\bar{l}},\quad a=1,...,h^{1,2}
	\end{align}
	or
	\begin{align}
	\delta g_{ij} = \frac{i}{\|\Omega\|^2} \bar{z}^a(\bar{\eta}_a)_{i\bar k\bar l}{\Omega^{\bar k\bar l}}_{j},\quad\|\Omega\|^2\coloneqq \frac{1}{3!}\Omega_{ijk}g^{i\bar l}g^{j\bar m}g^{k\bar n}\bar{\Omega}_{\bar l\bar m\bar n}
	\label{Maths:KählerModuli:MetricExpansion}
	\end{align}
in terms of $h^{2,1}$ complex scalar fields $\bar z^a$ and the basis $\bar\eta_a$ for $H^{1,2}(Y)$. 
There are two things that need to be noted at this point. First, we have
\begin{align}
	\dif z^1\wedge\dif z^2\wedge\dif z^3\wedge\dif\bar z^1\wedge\dif\bar z^2\wedge\dif\bar z^3 = i^3 3! \dif x^1\wedge\dif x^2\wedge\dif x^3\wedge\dif y^1\wedge\dif y^2\wedge\dif y^3
	\label{Eq:Compactification:DifferentialsComplexReal}
\end{align}
in coordinates $z^i=x^i+iy^i$\footnote{We write $\dif^6x$ for $\dif x^1\dif x^2\dif x^3\dif y^1\dif y^2\dif y^3$.} Second, since the components $\Omega_{ijk}$ are antisymmetric, we need to have
\begin{align}
	\Omega_{ijk}(z)=f(z)\varepsilon_{ijk}
\end{align}
for some holomorphic (and nowhere vanishing) function $f$ and $\varepsilon_{ijk}$ the epsilon tensor. Hence, 
\begin{align}
\|\Omega\|^2 = |f|^2(\sqrt{g})^{-1}
\end{align}
and with this
\begin{align}
\int\Omega\wedge\bar\Omega &= \int\frac{1}{3!}|f|^2\varepsilon_{ijk}\varepsilon_{lmn}\dif z^i\wedge \dif z^j\wedge \dif z^k\wedge \dif\bar z^l\wedge \dif\bar z^m\wedge \dif\bar z^n\nl
&= \frac{i^33!}{3!}\int \|\Omega\|^2 d^6 x\sqrt{q}\nl
&= -i\V\|\Omega\|^2
\end{align}
or
\begin{align}
\|\Omega\|^2 = \frac{i}{\V}\int_Y\Omega\wedge\bar\Omega.
\label{Maths:CSModuliSpace:K-Omega-identity}
\end{align}
Deformations \eqref{Maths:KählerModuli:MetricExpansion} usually yield a metric which does no longer satisfy \eqref{Maths:HermMetricPure} and thus fails to be Hermitian. By a suitable coordinate transformation, the metric can be put in a form where the mixed-index components again vanish. But such a transformation is not holomorphic and this metric is thus Hermitian which respect to new complex coordinates, i.e. another complex structure \cite{Candelas:ModuliSpaces}. The corresponding scalars $z^a$ are therefore called \emph{complex structure moduli} and we denote the space they span by $\mathcal{M}^{cs}$.

Together, these scalars span the geometric moduli space of the Calabi-Yau, which locally is a product
\begin{align}
\mathcal{M}=\mathcal{M}^\text{cs}\times\mathcal{M}^\text{ks}
\end{align}
of complex structure and K\"ahler structure moduli space respectively.

The most general metric \cite{Candelas:ModuliSpaces} one can write  for $\M$ is
\begin{align}
	\dif s^2 = -\frac{1}{2\V}\int \dif^6x \sqrt{g}g^{i\bar k}g^{j\bar l}\left(\delta g_{ij}\delta g_{\bar k\bar l}+\delta g_{i\bar l}\delta g_{j\bar k}-\delta B_{i\bar l}\delta B_{j\bar k}\right)
	\label{Eq:Compactification:ModuliSpaceMetric}
\end{align}
with $\V$ the volume of the Calabi-Yau.
\subsection{Complex structure moduli space}
\label{Section:Compactification:ComplexStructureModuliSpace}
The part of the metric \eqref{Eq:Compactification:ModuliSpaceMetric} corresponding to the complex structure moduli is
\begin{align}
2G_{a\bar{b}}z^a\bar{z}^{b}\coloneqq- \frac{1}{2\V}\int_Y \dif^6x \sqrt{g}g^{i\bar k}g^{j\bar l}\delta g_{ij}\delta g_{\bar k\bar l}.
\end{align}
and using the expansion \eqref{Maths:KählerModuli:MetricExpansion} as well as \eqref{Eq:Compactification:DifferentialsComplexReal} again, we have
\begin{align}
	2G_{a\bar b}z^a\bar z^b = -\frac{2i\|\Omega\|^2}{\V\|\Omega\|^4}\int\eta_a\wedge \bar\eta_b  z^a\bar z^b,
\end{align}
that is,
\begin{align}
G_{a\bar{b}} = \frac{-i}{\V\|\Omega\|^2}\int_Y\eta_a\wedge\bar\eta_b,
\label{ComplexStructureMetric}
\end{align}
We will work in the real basis defined in \eqref{Maths:CalabiYau:RealBasis} and denote the three-cycles dual to $\{\alpha_{\ha},\beta^{\ha}\}$ by $\{\mathcal A^{\ha},\mathcal B_{\ha}\}$, i.e. one has
\begin{align}
	\A^\ha\cap\B_{\hat b} = \delta_{\hat b}^\ha = -\B_{\hat b}\cap\A^\ha
\end{align}
and
\begin{align}
	\int_{\mathcal A^{\ha}}\alpha_{\hat b} ~=~ \sqrt{\V_0}\delta_{\hat b}^\ha ~=~ -\int_{\mathcal B^{\hat b}}\beta^\ha.
	\label{Eq:Compactification:CSModuli:DualCycles}
\end{align}
In terms of this basis, $\Omega$ can be expanded as
\begin{align}
\Omega=Z^{\ha}\alpha_{\ha} - \mathcal{F}_{\hat b}\beta^{\hat b}
\label{HolomThreeFormExp}
\end{align}
with the periods
\begin{align}
Z^{\ha} = \frac{1}{\sqrt{\V_0}}\int_{\mathcal A_{\ha}}\Omega,\qquad \mathcal{F}_{\ha}= \frac{1}{\sqrt{\V_0}}\int_{\mathcal B^\ha}\Omega.
\end{align}
The coordinates $Z^\ha$ are actually projective because $\Omega$ is homogeneous of degree one,
\begin{align}
(Z^0,Z^1,...,)\cong (\lambda Z^0,\lambda Z^1,...),
\end{align}
which allows us to chose
\begin{align}
z^a = \frac{Z^a}{Z^0}.
\end{align}
This is discussed in great detail in \ref{Appendix:String:SpecialGeometry}. The expansion
\begin{align}
\partial_{z^a}\Omega = k_a\Omega + i\eta_a
\label{Compactification:Kodaira}
\end{align}
which is derived in \ref{Calculations:Kodaira} can be used to define a K\"ahler potential $K^{cs}$ for the metric \eqref{ComplexStructureMetric} via
\begin{align}
e^{-K^{cs}} \coloneqq \frac{i}{\V_0}\int_Y\Omega\wedge\bar\Omega = \frac{\V}{\V_0}\|\Omega\|^2
\label{ComplexStructrueModuli:KählerPotential}
\end{align}
since
\begin{align}
\partial_{z^a}\partial_{\bar{z}_b} K^{cs} =&~-\partial_{z^a}\left(\frac{1}{\int_Y\Omega\wedge\bar\Omega}\int_Y\Omega\wedge(\bar k_b\bar{\Omega}-i\bar\eta_b)\right)\nonumber\\
=&~ \frac{1}{(-i\V\|\Omega\|^2)^2}\int_Y(k_a\Omega+i\eta_a)\wedge\bar\Omega\int_Y\Omega\wedge(\bar k_b\bar{\Omega}-i\bar\eta_b)\nl
&-\frac{1}{-i\V\|\Omega\|^2}\int_Y(k_a\Omega+i\eta_a)\wedge(\bar k_b\bar{\Omega}-i\bar\eta_b)\nonumber\\
=&~k_a\bar k_b - k_a\bar{k}_b - \frac{i}{\V\|\Omega\|^2}\int_Y\eta_a\wedge\bar\eta_b\nonumber\\
=&~ G_{a\bar b}.
\end{align}
Plugging in the expansion \eqref{HolomThreeFormExp} for $\Omega$,
\begin{align}
\frac{i}{\V_0}\int_Y\Omega\wedge\bar\Omega &= \frac{i}{\V_0}\int_Y(Z^{\ha}\alpha_{\ha} - \mathcal{F}_{\hat b}\beta^{\hat b})\wedge(\bar Z^{\hat c}\alpha_{\hat c} - \bar{\mathcal F}_{\hat d}\beta^{\hat d})\nonumber\\
&= i(Z^{\ha}\mathcal F_{\hat{a}} -\bar Z^{\ha}\bar{\mathcal{F}}_{\ha}),
\end{align}
we find that this expression is equal to the symplectic product introduced in \ref{Appendix:String:SpecialGeometry},
\begin{align}
	e^{-K^{cs}} = -i\langle v,\bar v\rangle\quad\text{where}\quad v=\binom{Z^\ha}{\F_\ha}.
\end{align}
Hence, the K\"ahler metric can be written as the K\"ahler and symplectic covariant expression \eqref{String:SUGRA:KählerMetric}:
\begin{align}
	G_{a\bar b} = i\langle\nabla_a V,\bar\nabla_{\bar b}\bar V\rangle
\end{align}
Here, $V=e^{K^{cs}}v$ and the K\"ahler covariant derivatives are introduced in \eqref{Eq:String:SUGRA:KählerCovariant}.

The $k_a$ appearing in the expansion \eqref{Compactification:Kodaira} can be determined explicitly from
\begin{align}
	\partial_a \int_Y \Omega\wedge\bar\Omega = -i \V_0e^{-K^{cs}}\partial_a K^{cs}
\end{align}
and
\begin{align}
	\partial_a \int_Y\Omega\wedge \bar\Omega = \int_Y (k_a\Omega + i\eta_a)\wedge\bar\Omega = i\V_0e^{-K^{cs}}k_a.
\end{align}
From this we conclude
\begin{align}
	k_a &= -\partial_a K^{cs},\nl
	\partial_a\Omega &= -(\partial_a K^{cs}) +i\eta_a
	\label{Eq:Compactification:Kodaira3}
\end{align}
and can therefore use the K\"ahler covariant derivative to write
\begin{align}
i\eta_a = \nabla_a\Omega
\label{Compactification:Kodaira2}
\end{align}
and thus the metric on the complex structure moduli space as
\begin{align}
G_{a\bar b} = \frac{-i}{\V\|\Omega\|^2}\int\nabla_a\Omega\wedge\bar\nabla_{\bar b}\bar\Omega = -\frac{\int\nabla_a\Omega\wedge\bar \nabla_{\bar b}\bar\Omega}{\int \Omega\wedge\bar\Omega}
\end{align}

\subsection{K\"ahler moduli space}
\label{Section:KählerModuliSpace}
The metric on the Kähler moduli space is
\begin{align}
G_{AB} = \frac{1}{4\V}\int_Y\omega_A\wedge *\omega_B
\label{Maths:KählerModuli:Metric}
\end{align}
which we derive in \ref{Calculations:Compactification:KählerModuliMetric}. Using the volume form \eqref{Maths:VolumeForm}, we define
\begin{align}
\K\coloneqq\int_Y J\wedge J\wedge J = 6\V
\end{align}
and
\begin{align}
\K_{ABC}\coloneqq\int_Y\omega_A\wedge\omega_B\wedge\omega_C,~\K_{AB}\coloneqq\K_{ABC}v^C,~\K_A\coloneqq\K_{AB}v^B.
\end{align}
Note that with this notation, $\K=\K_Av^A$.
Using the fact \cite{StromingerIdentity} that
\begin{align}
*\omega_A = -J\wedge\omega_A+\frac{3\mathcal{K}_A}{2\mathcal{K}}J\wedge J,
\end{align}
the metric $G_{AB}$ on $\mathcal{M}^{ks}$ can then be written as
\begin{align}
G_{AB}
&=\frac{3}{2}\left(\frac{\K_{AB}}{\K}-\frac{3}{2}\frac{\K_A\K_B}{\K^2}\right).
\label{KählerModuliMetric}
\end{align}
It has a \emph{K\"ahler potential}\footnote{We included the prefactor of $\frac{4}{3}$ for later convenience.}
\begin{align}
K^{ks}\coloneqq -\ln\frac{4}{3}\K.
\label{KählerModuliKählerPotential}
\end{align}
To see this, note that we have $\partial_{\bar t^A} = -\partial_{t^A}=-\frac{1}{2i}\partial_{v^A}$ on a function $f(t^A)=f(v^A)$. Thus,
\begin{align}
\partial_{\bar t^B} \K = \frac{i}{2}\partial_{v^B}\K_{CDE}v^Cv^Dv^E=\frac{3i}{2}\K_B,\quad \partial_{t^A}\K_B = \frac{1}{2i}\partial_{t^A}\K_{BCD}v^Cv^D=\frac{1}{i}\K_{AB}.
\end{align}
Using this, we have
\begin{align}
\partial_{t^A}\partial_{\bar t^B}\left(-\ln\frac{4}{3}\K\right) &= -\partial_{t^A}\left(\frac{3i}{2\K}\K_B\right)\nl
&=-\frac{3i}{2}\left(\frac{-1}{\K^2}\frac{3}{2i}\K_A\K_B + \frac{1}{\K}\frac{1}{i}\K_{AB}\right)\nl
&= -\frac{3}{2}\left(\frac{\K_{AB}}{\K}-\frac{3}{2\K}\K_A\K_B\right),
\end{align}
which shows that $K^{ks}$ indeed is a K\"ahler potential for the metric. We define the inverse $\K^{AB}$ via
\begin{align}
\K_{AB}\K^{BC}=\delta_A^C
\end{align}
which implies
\begin{align}
\K_B\K^{BC} &= \K_{BD}v^D\K^{BC} =  v^D \delta_D^C= v^C.
\end{align}
The \emph{inverse metric} $G^{AB}$ can be written as
\begin{align}
G^{AB} = -\frac{2\K}{3}\left(\K^{AB}-3\frac{v^Av^B}{\K}\right)
\label{KählerModuliMetricInverse}
\end{align}
as can be seen by direct computation:
\begin{align}
G_{AB}G^{BC} &= \left(\K_{AB}-\frac{3}{2}\frac{\K_A\K_B}{\K}\right)\left(\K^{BC}-3\frac{v^Bv^C}{\K}\right)\nonumber\\
&= \delta_A^C + \frac{K_Av^C}{\K}\left(-\frac{3}{2} -3 + \frac{9}{2}\right)\nonumber\\
&= \delta_A^C.
\end{align}
The K\"ahler manifold actually is of a certain kind type which is discussed in detail in appendix \ref{Appendix:String:SpecialGeometry}. Such $K^{ks}$ can be expressed in terms of a prepotential $\F$,
\begin{align}
e^{-K^{ks}} = i(\bar{X}^{\hat A}\mathcal{F}_{\hat{A}}-X^{\hat A} \bar{\mathcal F}_{\hat A}),
\label{Eq:Compactification:SpecialKähler}
\end{align}
which is defined as
\begin{align}
\mathcal{F} \coloneqq -\frac{1}{3!}\K_{ABC}\frac{X^AX^BX^C}{X^0},\qquad \mathcal{F}_{\hat A}\coloneqq\partial_{X^{\hat A}}\mathcal F,
\end{align}
with coordinates $X^{\hat A} \coloneqq (1,t^A)$. For explanations on the projective nature of the coordinates and a more general discussion of special K\"ahler manifolds we again refer to the appendix on special geometry.
In \ref{Calculations:Compactification:KählerModuliPrepotential}, we show that \eqref{Eq:Compactification:SpecialKähler} indeed is a prepotential, i.e.
\begin{align}
i(\bar{X}^{\hat A}\mathcal{F}_{\hat{A}}-X^{\hat A}\bar{\mathcal F}_{\hat A})=\frac{4}{3}\K = 8\V=e^{-K^{ks}}.
\end{align}
We sum up metric and K\"ahler potential for the two moduli spaces in table \ref{Table:ModuliSpaces} for later reference.
\begin{table}
	\begin{center}
		\includegraphics[width=.8\linewidth]{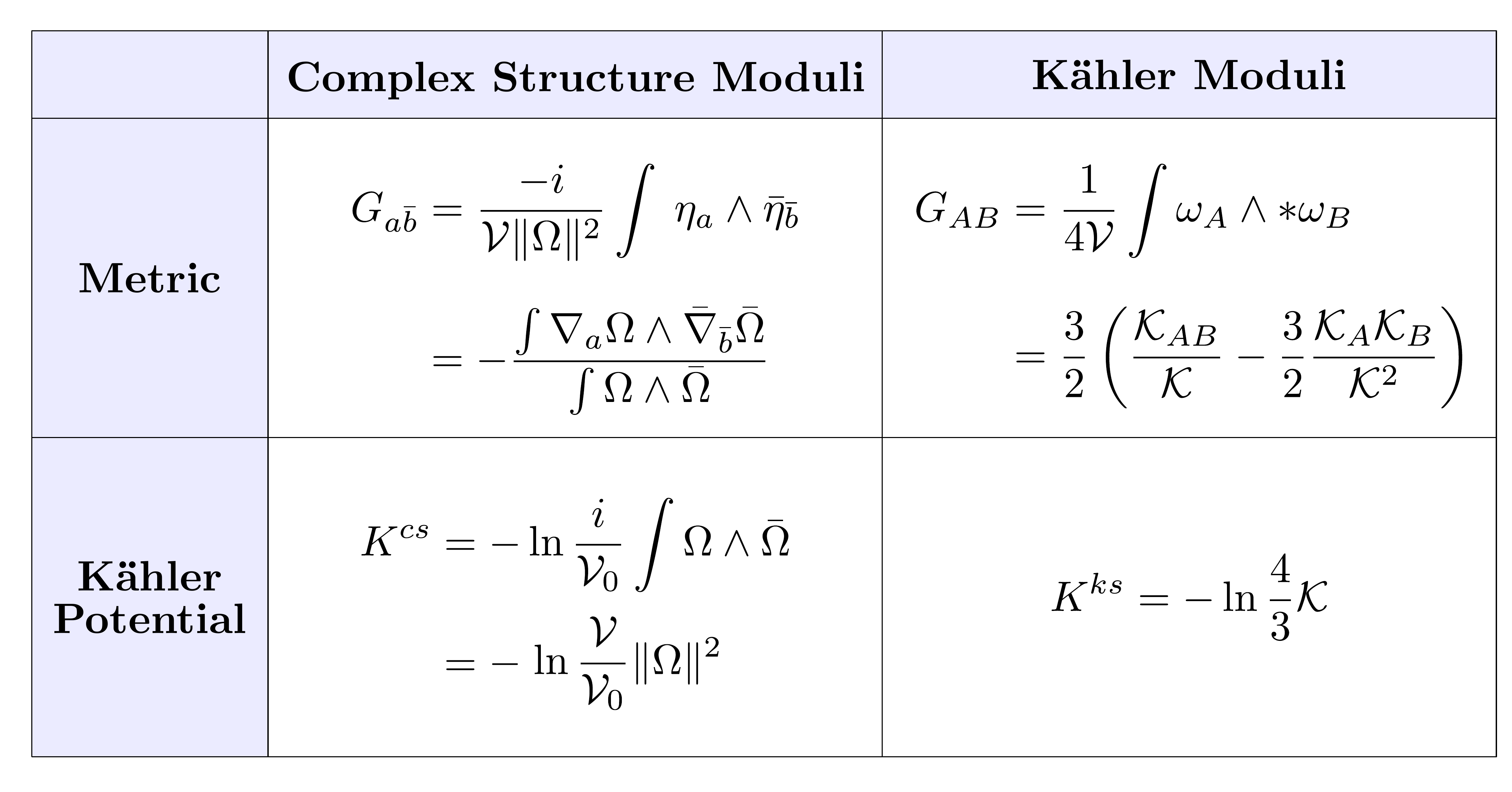}
		\caption[Metrics and K\"ahler potentials]{Metrics and K\"ahler potentials for moduli spaces.\label{Table:ModuliSpaces}}
		\end{center}
\end{table}
There is an interesting connection between the moduli spaces which we will exploit later. Hence, there is one topic to be discussed before turning to the main part.
\subsection{Mirror symmetry}
\label{Section:MirrorSymmetry}
For a Calabi-Yau manifold $Y$ (we only consider threefolds), \emph{mirror symmetry} states the following:
\begin{enumerate}[-]
	\item  there is a so-called \emph{mirror Calabi-Yau} $\tilde Y$ with even and odd cohomologies identified. That is, their Hodge numbers are related via
\begin{align}
	h^{1,1}\tilde (Y) = h^{2,1}(Y),\quad h^{2,1}(\tilde Y) = h^{1,1}(Y)
\end{align}
which amounts to reflecting the Hodge numbers in the Hodge diamond (see fig. \ref{Maths:CalabiYau:HodgeDiamond}) along the diagonal.

\item The complex structure and K\"ahler moduli spaces of the mirror Calabi-Yau $\tilde Y$ are identified with the K\"ahler and complex structure moduli spaces of $Y$, 
\begin{align}
	\M^{ks}(\tilde Y) = \M^{cs}(Y),\quad \M^{cs}(\tilde Y)=\M^{ks}(Y),
\end{align}
\end{enumerate} and the action resulting from compactification of Type IIA supergravity on some Calabi-Yau $Y$ is identical to the action of Type IIB supergravity compactified on the mirror manifold $\tilde Y$. In particular, one  can also assume a cubic prepotential
 \begin{align}
\mathcal{F} \coloneqq -\frac{1}{3!}\K_{ABC}\frac{X^AX^BX^C}{X^0},\qquad \mathcal{F}_{\hat A}\coloneqq\partial_{X^{\hat A}}\mathcal F,
\end{align}
for the \emph{complex structure} moduli space by switching to the mirror Calabi-Yau. We will not go into any more detail about mirror symmetry but refer to \cite{MirrorSymmetry} for more information.

Having acquired the tools necessary for dealing with the moduli spaces of Calabi-Yaus, we can now begin the main part of the thesis.\newpage

\section{A Scalar WGC for Type IIB Particles}
\label{Section:ScalarWGC}
\subsection{Supersymmetric black holes}
\label{Section:BlackHoles}
The argument we gave in support of the Weak Gravity Conjecture relied on the requirement that charged extremal black holes should be able to decay. Since compactification of Type II supergravity gives rise to $\N=2$ SUGRA in $d=4$ and in SUGRA, gravity is coupled to various scalars $\phi^a$, we will consider extremal black hole solutions of $\N=2$ supergravity in presence of scalar fields \cite{WGC:Palti1}. The latter enter the action of a theory containing $U(1)$ gauge fields $V^A$ via symmetric real functions $\I_{AB}(\phi)$ and $\R_{AB}(\phi)$. The  bosonic part of the action takes the generic form \cite{BlackHoles}
\begin{align}
	S= \int *\R - G_{a\bar b}\dif \phi^a \wedge *\dif\phi^{\bar b} + \frac{1}{2}\I_{AB}F^A\wedge *F^{B} +\frac{1}{2}\R_{AB} F^A\wedge F^B
	\label{Eq:BPS:Action}
\end{align}
where $F^A=\dif V^A$ and the metric $g$ - entering implicitly via the Hodge-$*$ - is a function of the fields $\phi$.
\subsubsection{Electromagnetic duality}
The theory enjoys a symmetry that is analogous to the well-known\footnote{See e.g. \cite{Books:GaugeKnotsGravity} for an extensive treatment.} duality of Maxwell's electromagnetism, i.e. the interchange of $F\leftrightarrow *F$ . Obviously, this preserves the set of vacuum Maxwell equations
\begin{align}
	\dif F=0,\quad\dif G =0
	\label{Eq:BPS:Maxwell}
\end{align}
where the dual field strength $G$ is related to $F$ via
\begin{align}
	G = \frac{\delta}{\delta F}\int\frac{1}{2} F\wedge *F = *F.
\end{align}
In the theory described in \eqref{Eq:BPS:Action}, the dual field strengths $G_A$ are defined similarly but mix the forms $F^A$ and their Hodge duals:
\begin{align}
	G_A \coloneqq \frac{\delta S}{\delta F^A} = \I_{AB}*F^A +\R_{AB}F^A
\end{align}
with Hodge-dual\footnote{\label{Footnote:HodgeSquare}Note that $*^2=-1$ when acting on even forms on the Calabi-Yau-threefold.}
\begin{align}
	*G_A = \I_{AB}F^A-\R_{AB}*F^B.
	\label{Eq:BPS:HodgeDualG}
\end{align}
With this definition, we can write the Bianchi identities and equation of motion similar to \eqref{Eq:BPS:Maxwell} as
\begin{align}
	\dif F^A=0,\quad \dif G_A = 0
\end{align}
and look for a symmetry that preserves this set of equations as well as the definition $G_A = \delta S/\delta F^A$. The latter requirement restricts \cite{Books:Supergravity} the allowed transformations to elements of the symplectic group,
\begin{align}
	\binom{F^A}{G_A}\rightarrow\mathcal{S}\binom{F^A}{G_A},
\end{align}
with $\mathcal{S}\in \operatorname{Sp}(2h,\mathbb{R})$. Here, $h$ denotes the number of vector fields $V^A$, i.e. $A=1,...,h$. Since we are considering charged, spherically symmetric and asymptotically flat black-hole solutions, we define the magnetic and electric charges $(p^A,q_A)$ by the integrals
\begin{align}
	p^A\coloneqq\frac{1}{4\pi}\int_{S^2_\infty}F^A,\qquad q_A\coloneqq\frac{1}{4\pi}\int_{S^2_\infty}G_A
\end{align}
performed over a sphere at infinity.\footnote{Note that the convention differs slightly from the one used in \eqref{Eq:WGC}.} By definition,
\begin{align}
\Gamma\coloneqq \binom{p^A}{q_A}
\label{Eq:BPS:SymplecticCharge}
\end{align}
transforms as symplectic vector.

A discussion of the structure underlying $\N=2$ supergravity is given in appendix \ref{Appendix:String:Sugra}. From there, we quote that the scalars parameterizing $\I_{AB}$ and $\R_{AB}$ are the $h$ complex scalars $z^A$ from the gauge multiplets to which the one-forms $V^A$ belong. As discussed in the appendix, the manifold spanned by these scalars is endowed with \emph{special K\"ahler geometry}. Hence, there is a symplectic section $(Z^{\hat A},\F_{\hat A})$ with the $h+1$ fields $Z^{\hat A}(z^A)$ serving as projective coordinates on the manifold and $\F_{\hat A}$ are functions of these coordinates. The K\"ahler potential is given by the symplectic invariant expression
\begin{align}
	K = i\langle v,\bar v\rangle
\end{align}
with the symplectic vector $v$ defined in \eqref{Eq:String:SUGRA:Vectorv}. The metric can be written as the K\"ahler and symplectic invariant expression
\begin{align}
	G_{A\bar B} = i\langle\nabla_A V,\bar\nabla_{\bar B}\bar V\rangle
\end{align}
in terms of the K\"ahler covariant derivative \eqref{Eq:String:SUGRA:KählerCovariant} and
\begin{align}
	V= e^{\frac{1}{2}K} v= 	e^{\frac{1}{2}K} \binom{Z^{\hat A}(z^A)}{\F_{\hat A}(z^A)}
\end{align}
as shown in \eqref{Eq:String:SUGRA:KählerMetricSymplectic}.
\subsubsection{Superalgebras with central charge and BPS bound}
\label{Appendix:String:BPS}
Remember that for massive representations of the $\N=1, d=4$ SUSY-algebra, the anti-commutator of the supercharges in the rest frame takes the form \cite{Polchinski2}
\begin{align}
\lbrace Q_{\alpha},Q_\beta^\dagger\rbrace = 2m\delta_{\alpha\beta}
\end{align}
where $m$ is the mass of the states and $\alpha,\beta=1,...,4$ label the Majorana spinor components. For \emph{extended} supersymmetry, i.e. $\N>1$, there are additional terms that are allowed by Lorentz invariance and which we thus include:
\begin{align}
\lbrace Q^I_\alpha,Q^{\dagger J}_\beta\rbrace = 2m\delta^{IJ}\delta_{\alpha\beta}+2i\Z^{IJ}\Gamma^0_{\alpha\beta}
\label{Eq:String:SUGRA:Anticommutator}
\end{align}
where $I,J=1,...,\N$ with conserved quantities $\Z^{IJ}$. These are called \emph{central charges} since they commute with all other generators in the superalgebra. Note that the matrix $\Z^{IJ}$ is necessarily antisymmetric which is why central charges can only appear for extended supersymmetries. We will only consider $\N=2$ where the central charge matrix can be brought\footnote{To do so, perform unitary transformations $\Z\mapsto U^TZU$.} to the form
\begin{align}
\Z^{IJ} = \begin{pmatrix}
0 && \Z \\
-\Z && 0
\end{pmatrix}.
\end{align}
Since the left-hand side of \eqref{Eq:String:SUGRA:Anticommutator} is non-negative, the eigenvalues of the right-hand side, namely $2m+2\Z$ and $2m-2\Z$, must also be non-negative. The resulting inequality
\begin{align}
m \ge |\Z|
\label{Eq:String:SUGRA:BPSBound}
\end{align}
is called \emph{BPS bound}. The central charges are electric and magnetic charges coupling to the gauge fields \cite{Books:BeckerBecker}. States that saturate \eqref{Eq:String:SUGRA:BPSBound} are called \emph{BPS states}.
\subsubsection{Relation for the central charge}
We will now derive a relation for the central charge of the superalgebra that is going to play an important role in the next section. 

First, note that the name graviphoton of the vector in the $\N=2$ supergravity multiplet (see table \ref{Table:Strings:SugraMultiplets}) stems from the fact that its field strength appears in the transformation of the gravitino \cite{Books:Supergravity}. It is given by the (scalar field dependent) combination of field strengths
\begin{align}
e^{\frac{1}{2}K}( Z^{\hat A}G_{\hat A} - \F_{\hat A} F^{\hat A})
\end{align} and therefore, we have the relation
\begin{align}
	\Z &= \int e^{\frac{1}{2}K}( Z^{\hat A}G_{\hat A} - \F_{\hat A} F^{\hat A})\nl
		&= e^{\frac{1}{2}K}Z^{\hat A}q_{\hat A} - \F_{\hat A}p^{\hat A}
\end{align}
which identifies the central charge with the graviphoton charge at infinity. We write this in the concise form
\begin{align}
	\Z = \langle V,\Gamma\rangle \qquad\text{with }\quad\Gamma = \binom{p^A}{q_A},\quad V=e^{\frac{1}{2}K}\binom{X^A}{\F_A}.
	\label{Eq:BPS:CentralChargeSymplectic}
\end{align}
The sections $\F_A$ and $X^B$ are related (see eq. \eqref{Eq:Calculations:MZ}) via the gauge-coupling matrix $\M$ as
\begin{align}
	\F_A= \M_{AB}X^B
\end{align}
and one finds
\begin{align}
 \Z &= X^A\left(q_A- \M_{AB}p^B\right),\quad \nabla_a\Z\ = \nabla_a X^A(q_A - \M_{AB}p^B)
\end{align}
where we used that $\M$ is symmetric. Hence,
\begin{align}
	|\Z|^2 + \nabla_a\Z\bar\nabla_{\bar b}\bar\Z G^{a\bar b}&=(q_A-\M_{AC}p^C)(q_B-\bar\M_{BD}p^D)\left(X^A\bar X^B+\nabla_a X^A\bar\nabla_{\bar b} \bar X^B G^{a\bar b}\right)
	\label{Eq:SUSYBlackHoles:CentralChargeRelation}
\end{align}
and we identify the third factor on the right-hand side as $-\frac{1}{2}\mathcal{I}^{-1}$. To simplify this equation further, we use the matrix introduced in \cite{Ceresole}:
\begin{align}
	\mathscr M = \begin{pmatrix}
	-\left(\I+\R\I^{-1}\R\right)_{AB} && {\left(\R\I^{-1}\right)_A}^B\\
	{\left(\I^{-1}\R\right)^A}_B && -{\left(\I^{-1}\right)}^{AB}
	\end{pmatrix}
	\label{Eq:BPS:MMatrix}
\end{align}
and we define
\begin{align}
	\Q^2 \coloneqq \frac{1}{2}\Gamma^T\mathscr M\Gamma.
	\label{Eq:SUSYBlackHoles:Q^2Definition}
\end{align}
In terms of these, \eqref{Eq:SUSYBlackHoles:CentralChargeRelation} takes the form
\begin{empheq}[box=\mymath]{equation}
	\Q^2 = |\Z|^2 + \nabla_a\Z\bar\nabla_{\bar b}\bar\Z G^{a\bar b}
\label{Eq:SUSYBlackHoles:CentralChargeRelation2}
\end{empheq}
which we derive in \ref{Calculations:SUSYBlackHoles:MatrixM}. Note that \eqref{Eq:SUSYBlackHoles:CentralChargeRelation2} is a statement about BPS states which are extremal with respect to \eqref{Eq:String:SUGRA:BPSBound}, $m=|\Z|$. For a complex function $f$, one has $\partial|f| = \frac{1}{2|f|}(\partial f)\bar f$ and therefore, $\partial|f|\bar\partial|f| = \frac{1}{4}\partial f\bar\partial\bar f$. Hence, we can write this as
\begin{align}
\mathcal{Q}^2=m^2 +4G^{a\bar b}\nabla_am\bar\nabla_{\bar b} m.
\end{align}
\subsection{Gauge fields from reduction of Type IIB supergravity}
Compactification of Type IIB supergravity on a Calabi-Yau three-fold gives rise to a number of $U(1)$s in the four-dimensional theory. As we will see, D3-branes wrapping three-cycles can be viewed as particles carrying charge under these gauge fields and thus should satisfy the weak gravity conjecture. The compactification of Type II supergravity follows mainly the discussions in \cite{Micu, Grimm, Gurrieri, Grimm2}. For an overview of compactification in the presence of background fluxes see \cite{Grana}.
A summary of results on Calabi-Yau manifolds and their moduli spaces and a short discussion of Type II SUGRA is given in the appendices \ref{Appendix:Maths} and \ref{Appendix:String}.
\subsubsection{Action for Type IIB SUGRA}
In appendix \ref{Appendix:String}, we discuss the ten-dimensional supergravities Type IIA and IIB which are the low-energy limits of the Type II string theories. From there we cite the action \eqref{Appendix:IIBAction} where we now put hats on the ten-dimensional quantities:
\begin{align}
S_{\text{IIB}}^{(10)}=& \frac{1}{2\hat\kappa^2}\int e^{-2\hat\phi}\left(*\hat{\mathcal R}+ 4\dif\hat\phi\wedge*\dif\hat \phi- \frac{1}{2}\hat H_3\wedge *\hat H_3\right)\nl
&-\frac{1}{4\hat \kappa^2}\int \left(\hat F_1\wedge* \hat F_1 + \hat F_3\wedge* \hat F_3 + \hat F_5\wedge*\hat F_5\right)-\frac{1}{4\hat \kappa^2}\int \hat C_4\wedge \hat H_3\wedge \hat F_3
\label{IIB:10dAction}
\end{align}
with the Kalb-Ramond $2$-form $B_2$, the dilaton $\hat\phi$ and metric $\hat g$ from the NS-NS sector as well as the axion $\hat C_0$, $2$-form $\hat C_2$ and $4$-form $\hat C_4$ from the R-R sector and
\begin{align}
\hat H_3 &= \dif \hat B_2,\nonumber\\
\hat F_1 &= \dif \hat C_0,\nonumber\\
\hat F_3 &= \dif \hat C_2 - \hat C_0 \hat H_3,\quad \hat F_5 = \dif\hat C_4-\hat C_2\wedge\hat H_3,
\label{IIB:10dFieldStrengths}
\end{align}
\subsubsection{Expanding the fields}
We make the Kaluza-Klein reduction ansatz discussed in section \ref{Compactification:KaluzaKleinAnsatz} and expand the fields \eqref{IIB:10dFieldStrengths} in terms of harmonic forms on the Calabi-Yau, i.e. in terms of the basis forms listed in table \ref{Table:BasisForms}. We will discuss these fields in turn.
\begin{enumerate}[i)]
	\item The axion $\hat C_0$ has no internal index and corresponds to $H^{0,0}(Y)$.
	\item The two-form $\hat B_2$ (and likewise $\hat C_2$) decomposes in $\hat B_{\mu\nu}$ corresponding to $H^{0,0}(Y)$ and $\hat B_{\mu i}$ corresponding to $H^{1,1}(Y)$. There are no further terms, see our discussion in section \ref{Section:Compactification:ReductionOnCY}.
	\item The $\hat C_{\mu\nu}$ part of $\hat C_4$ is a four-form in $d=4$ and therefore closed. Hence, it does not contribute to the action. The pieces $\hat C_{ijk},\hat C_{ij\bar k}, \hat C_{i\bar j\bar k}$ and $\hat C_{\bar i\bar j\bar k}$ correspond to $H^3(Y) = H^{3,0}(Y)\oplus H^{2,1}(Y)\oplus H^{1,2}(Y)\oplus H^{0,3}(Y)$ and the remaining $\hat C_{\mu\nu i\bar j}$ and $\hat C_{ij\bar k\bar l}$ to $H^{1,1}(Y)$ and $H^{2,2}(Y)$ respectively.
\end{enumerate}
Accordingly, we expand
\begin{align}
B_2(x,y) &= B_2(x)+b^A(x)\omega_A(y),\nl
\hat C_2(x,y) &= C_2(x)+c^A(x)\omega_A(y),\nl
\hat C_4(x,y) &= D_2^A(x)\wedge\omega_A(y) +\varrho_A(x)\tilde{\omega}^A(y) + V^{\hat a}(x)\wedge\alpha_{\hat a}(y) - U_{\hat a}(x)\wedge\beta^{\hat a}(y)
\label{IIB:FieldExpansion}
\end{align}
with
\begin{align}
\hat a &= 0,...,h^{2,1}, & a=1,...,h^{2,1},\nl
\hat A &= 0,...,h^{1,1}, & A=1,...,h^{1,1}.
\end{align}
From now on, we will not write the coordinates explicitly. The vectors $V^\ha$ and $U_\ha$ (as well as $D_2^{\hat A}$ and $\varrho_{\hat A}$) are actually not independent but related by the self-duality $F_5=*F_5$  of the field strength $F_5$. We will keep the one-forms $V^\ha$ and scalars $\varrho_{\hat A}$.

Defining  $H_3 \coloneqq \dif B_2$, $F^{\hat a}\coloneqq\dif V^{\hat a}$ and $G_\ha \coloneqq \dif U_\ha$  we have
\begin{align}
\hat H_3 &= H_3 + \dif b^A\wedge \omega_A,\nl
\dif \hat C_2 &= H_3 + \dif c^A\wedge \omega_A,\nl
\dif\hat C_4 &= \dif D_2^A\wedge\omega_A + F^{\hat a}\wedge\alpha_{\hat a}-G_{\hat a}\wedge\beta^{\hat a}+\dif\varrho_A\wedge\tilde\omega^A.
\end{align}
We are only interested in the action for the gauge fields $V^\ha$ and $G_\ha$ which comes from the $F_5\wedge * F_5$-term in the action \eqref{IIB:10dAction}. Plugging in the above-mentioned expansion for the expression of the field strength $F_5$ in \eqref{IIB:10dFieldStrengths}, we obtain
\begin{align}
	\hat F_5 =&~\dif D_2^A\wedge\omega_A + F^{\hat a}\wedge\alpha_{\hat a}-G_{\hat a}\wedge\beta^{\hat a}+\dif\varrho_A\wedge\tilde\omega^A - (C_2+c^A\omega_A)\wedge(H_3+\dif b^A\omega_A)\nl
	=&~F^{\hat a}\wedge\alpha_{\hat a}-G_{\hat a}\wedge\beta^{\hat a} + \dif\varrho_A\wedge\tilde\omega^A+(\dif D_2^A -C_2\wedge\dif b^A - c^A\wedge H_3)\wedge\omega_A \nl
	&- c^A\dif b^B\wedge\omega_A\wedge\omega_B.
\end{align}
\subsubsection{Field content of Type IIB in four dimensions}
\begin{table}
	\begin{center}
		\begin{tabular}{|c|c|c|}
			\hline 
			\rowcolor{\BColor}
			\rule[-1.5ex]{0pt}{4.5ex} \textbf{Multiplet}& \textbf{(Massless) Field Content}  & \textbf{Number} \\
			\hline 
			\rule[-1.5ex]{0pt}{4.5ex} Gravity multiplet& $(g_{\mu\nu},V^0)$  & 1 \\
			\hline 
			\rule[-1.5ex]{0pt}{4.5ex} Gauge multiplets & $(V^a,z^a)$ & $h^{2,1}$ \\
			\hline
			\rule[-1.5ex]{0pt}{4.5ex} \multirow{2}{*}{Hypermultiplets }&$(b^A,c^A,v^A,\varrho_A)$ & $h^{1,1}$\\
			\cline{2-3}
			\rule[-1.5ex]{0pt}{4.5ex} &$(h_B,h_C,\phi,C_0)$ & 1\\
			\hline
		\end{tabular} 
	\end{center}
	\caption[Type IIB multiplets]{Type IIB multiplets in $d=4$.\label{Table:IB:FieldContent}}
\end{table}
As discussed in section \ref{Section:Compactification}, the four-dimensional theory possesses $\N=2$ SUSY. Hence, the massless fields form three kinds of irreducible representations, namely a gravity multiplet, gauge multiplets and hypermultiplets \cite{Books:Supergravity}. Again, we will discuss these in turn.
\begin{enumerate}[i)]
	\item The gravity multiplet contains the graviton $g_{\mu\nu}$ and a one-form $V^0$.
	\item A gauge multiplet in $\N=2,d=4$ contains a one-form and a complex scalar. The only (remaining) one-forms are $V^a$ which combine with the complex structure moduli $z^a$.
	\item Hypermultiplets contain four real scalars. Hence, $b^A$ and the K\"ahler moduli $v^A$ combine with $c^A$ and $\varrho_A$ to $h^{1,1}$ multiplets.
	\item The two-forms $B_2$ and $C_2$ are actually Poincaré dual to scalars $h_B,h_c$ \cite{Micu} which form another hypermultiplet with the dilaton $\phi$ and $C_0$.
\end{enumerate}
The various multiplets are collected in table \ref{Table:IB:FieldContent}.
\subsubsection{Integrating on the Calabi-Yau}
The non-vanishing terms in the integral over $\hat F_5\wedge*\hat F_5$ containing the  field strengths $F^{\hat a}$ and $G_{\hat a}$ are
\begin{align}
I_1= -\frac{1}{4\hat \kappa^2}\int_Y (F^{\hat a}\wedge \alpha_{\hat a})\wedge*(F^{\hat b}\wedge\alpha_{\hat b}-G_{\hat b}\wedge\beta^{\hat b})
\end{align}
and
\begin{align}
I_2=-\frac{1}{4\hat \kappa^2}\int_Y (- G_{\hat a}\wedge\beta^{\hat a})\wedge*(F^{\hat b}\wedge\alpha_{\hat b}-G_{\hat b}\wedge\beta^{\hat b}).
\end{align}
We use the integrals from \ref{Maths:IntegralsOnCalabiYauThreefolds} to get
\begin{align}
4\kappa^2 I_1 =& -F^{\hat a}\wedge *F^{\hat b}[\IM \M+(\RE\M)(\IM\M)^{-1}(\RE\M)]_{\hat a\hat b}\nl 
&+ F^{\hat a}\wedge*G_{\hat b} [(\RE \M)(\IM\M)^{-1}]_{\hat a}^{\hat b},\nl
4\kappa^2 I_2=&~G_{\hat a}\wedge*F^{\hat b}[(\RE\M)(\IM\M)^{-1}]_{\hat a}^{\hat b}- G_{\hat a}\wedge*G_{\hat b}[(\IM\M)^{-1}]^{\hat a\hat b}
\end{align}
where we defined the four-dimensional coupling $\kappa^2=\hat\kappa^2/\V_0$.
This can be written in a more compact way. For example,
\begin{align}
	I \coloneqq \IM\M^{-1}(\M F)\wedge * (\bar{\M}F) 
\end{align} is the $F\wedge F$-term form above,
\begin{align}
I=~& [(\IM M)^{-1}]^{\hat a\hat b}(\RE\M+i\IM\M)_{a\hat c}F^{\hat c}(\RE\M - i\IM\M)_{b\hat d}\wedge*F^{\hat d}\nl
=~& [(\IM\M)^{-1}]^{\hat a\hat b}(\RE\M)_{a\hat c}(\RE \M)_{b\hat d}F^{\hat c}\wedge *F^{\hat d}\nl
&+ [(\IM\M)^{-1}]^{\hat a\hat b}(\IM\M)_{a\hat c}(\IM\M)_{b\hat d}F^{\hat c}\wedge*F^{\hat d}\nl
=~& (\RE\M)_{a\hat c}[(\IM\M)^{-1}]^{\hat c\hat d}(\RE \M)_{\hat db}F^{\hat a}\wedge *F^{\hat b} + (\IM M)_{\hat a\hat b}F^{\hat a}\wedge*F^{\hat b}.
\end{align}
One can see that together with the remaining terms, this takes the form
\begin{align}
\IM \M^{-1}(G-\M F)\wedge*(G-\bar{\M}F).
\end{align}
There are, of course, other non-vanishing terms arising from the $F_5\wedge*F_5$ integral but we are not interested in these. 
\subsubsection{Action for the gauge fields}
We have not yet imposed the self-duality condition $F_5=*F_5$. First, note that $F_5=*F_5$ implies
\begin{align}
*F^{\hat a}\wedge *\alpha_{\hat a} - *G_{\hat a}\wedge *\beta^{\hat a} =&~ *F^{\hat a}\wedge\left([(\RE\M)(\IM M)^{-1}]_{\hat a}^{\hat b}\alpha_{\hat b} +\right.\nl&\left. [-\IM\M -(\RE\M)(\IM\M)^{-1}(\RE\M)]_{\hat a\hat b}\beta^{\hat b}\right)-\nl
&*G_{\hat a}\wedge\left([(\IM\M)^{-1}]^{\hat a\hat b}\alpha_{\hat b}\right) - [(\RE\M)(\IM M)^{-1}]_{\hat b}^{\hat a}\beta^{\hat b}\nl
=&~F^{\hat a}\wedge\alpha_{\hat a} - G_{\hat a}\wedge\beta^{\hat a}
\end{align}
and by equating coefficients\footnote{See footnote \ref{Footnote:HodgeSquare} to eq. \eqref{Eq:BPS:HodgeDualG}.}
\begin{align}
*G &= \RE\M *F - \IM\M F\nl
G &= \RE\M F + \IM\M*F.
\label{IIB:EQforG}
\end{align}
The self-duality can then be imposed via the equation of motion for $G_\ha$ by adding the term
\begin{align}
\frac{1}{2}F^{\hat a}\wedge G_{\hat a}
\end{align}
as a Lagrange multiplier to the Lagrangian for the fields strengths $F^{\hat a}$ and $G_{\hat a}$:
\begin{align}
\mathcal L_{F^{\hat a}} = \frac{1}{4}(\IM\M)^{-1}(G-\M F)\wedge*(G-\bar{\M}F) + \frac{1}{2}F^{\hat a}\wedge G_{\hat a}.
\end{align}
Variation with respect to $G_{\hat a}$ shows that \eqref{IIB:EQforG} is now implemented:
\begin{align}
\delta_G \mathcal{L}_{F^{\hat a}} &= \frac{1}{4}(\IM\M)^{-1}(*(G-\bar{\M}F) + *(G - \M F)) + \frac{1}{2} F \nl
&= \frac{1}{2}(\IM \M)^{-1}(*G-\RE\M*F)+\frac{1}{2}F = 0,
\end{align}
i.e.
\begin{align}
*G &= \RE\M *F - \IM\M F.
\end{align}
Eliminating $G_{\hat a}$ in $\mathcal{L}_{F^{\hat a}}$ via its equation of motion yields
\begin{align}
\mathcal{L}_{F^{\hat a}} =& \frac{1}{4}(\IM M)^{-1}(\RE\M F + \IM M*F - \M F)\wedge(\RE \M*F-\IM\M F-\bar{\M}*F)\nl
&+\frac{1}{2}F\wedge(\RE\M F + \IM \M*F)\nl
=&\frac{1}{4}(\IM\M)^{-1}(\IM\M *F -i\IM\M F)\wedge(-\IM\M F + i\IM *F)\nl
&+\frac{1}{2}F\wedge(\RE\M F + \IM M*F)\nl
=&\frac{1}{2}\IM\M_{\hat a\hat b}F^{\hat a}\wedge*F^{\hat b} + \frac{1}{2}\RE\M_{\hat a\hat b}F^{\hat a}\wedge F^{\hat b}
\label{IIB:GaugeFields:Action}
\end{align}
which is a kinetic and a topological term for the gauge fields $V^{\hat a}$. We note:
\begin{empheq}[box={\mymath}]{equation}
S_{\text{gauge}} = \int\frac{1}{2}\IM\M_{\hat a\hat b}F^{\hat a}\wedge *F^{\hat b} + \frac{1}{2}\RE\M_{\hat a\hat b}F^{\hat a}\wedge F^{\hat b}.
\end{empheq}
This is precisely what we had expected for the action of the gauge fields in $\N=2$ supergravity - see section \ref{Section:BlackHoles}. By comparison with the action \eqref{Eq:BPS:Action}, we see that the roles of $\I$ and $\R$ are played by the imaginary and real part of the matrix $\M$. 
\subsection{Particles from D3-branes}
\label{Section:D3Branes}
As already mentioned, D3-branes wrapping three-cycles in Type IIB look like particles in the four-dimensional theory and it will turn out that these satisfy a modified version of the WGC.

We begin with the action \cite{Polchinski2,PaltiPhD}
\begin{align}
	S^{(10)}_{D3} = -\mu_3\int_{D3}\dif^4\xi e^{-\hat\phi}\sqrt{-\det(G)} + \mu_3\int_{D3}\hat{C}_4
	\label{D3:10dAction}
\end{align}
of a D3-brane coupled to the four-form $\hat{C}_4$ and wrapping a three-cycle $\mathcal{C}$. The brane tension $\mu_3$ is related\footnote{For a $Dp$-brane we have $\mu_p=(\sqrt{\pi}/\hat{\kappa})(2\pi\sqrt{\alpha'})^{3-p}$, see \cite{Polchinski2} eq. $(13.3.7)$.}  to the ten-dimensional gravitational coupling as $\mu_3=\sqrt{\pi}/\hat\kappa$ and $G$ denotes the pullback of the spacetime metric onto the brane's world-volume. Introducing the integer charges $q_\ha$ and $p^\ha$  by expanding $\mathcal{C}$ in terms of the three-cycles $\mathcal{A}^\ha$ and $\mathcal{B}_\ha$ dual to the three-forms $\alpha_\ha$ and $\beta^\ha$ (see \ref{Eq:Compactification:CSModuli:DualCycles}),
\begin{align}
	\mathcal{C} = q_\ha\mathcal{A}^\ha + p^\ha\mathcal{B}_\ha,
	\label{D3Brane:3Cycle}
\end{align}
and using the expansion \eqref{IIB:FieldExpansion}, the second integral in \eqref{D3:10dAction} reads
\begin{align}
	\mu_3\int_{D3}\hat{C}_4 & = \mu_3\sqrt{\V_0}\left( q_\ha \int V^\ha+p^\ha \int U_\ha\right).
\end{align}
To compute the first integral, we perform a Weyl rescaling \eqref{WeylRescaling} with $e^{-\frac{\hat\phi}{2}}$ to get rid of the dilaton factor and with $(\V_0/\V)^{-1/4}$ such that
\begin{align}
	-\mu_3\sqrt{\frac{\V_0}{\V}}\int_{D3}\dif^4\xi\sqrt{-\det(G)}.
\end{align} The volume of the cycle has a lower bound
\begin{align}
	\operatorname{Vol}(\mathcal{C}) &\ge \frac{1}{\sqrt{\|\Omega\|}}\left|\int_\mathcal{C}\Omega\right|
	=  \sqrt{\frac{\V}{\V_0}}e^{\frac{1}{2}K^{cs}}\left|\int_{\mathcal{C}}\Omega\right|
	\label{D3Brane:VolumeBound}
\end{align}
where we used \eqref{Maths:CSModuliSpace:K-Omega-identity} and the definition of the K\"ahler potential \eqref{ComplexStructrueModuli:KählerPotential}. Note that we did only choose a homology class in \eqref{D3Brane:3Cycle} and thus cannot give more than this bound for the volume. We can specify the cycle as to minimize this volume\footnote{Such a cycle is called \emph{supersymmetric special Lagrangian}.} in which case an equal sign holds. This is true for a BPS state.
We will come back to this issue later and assume for now that the volume is minimized. Using the expansion \eqref{HolomThreeFormExp} of the holomorphic three-form, we can perform the integration over $\mathcal{C}$,
\begin{align}
	\int_\mathcal{C}\Omega = \int_{\mathcal{C}}(Z^\ha\alpha_\ha -\mathcal F_{\hat b}\beta^{\hat b})=\sqrt{\V_0}\int \dif \tau (p_\ha Z^\ha-q^{\hat b}\mathcal F_{\hat b}),
\end{align}
to arrive at
\begin{align}
-\mu_3\sqrt{\frac{\V_0}{\V}}\operatorname{Vol}(\mathcal{C})=-\mu_3\sqrt \V_0e^{\frac{1}{2}K^{cs}}\left|q_\ha Z^\ha-p^{\hat b}\mathcal F_{\hat b}\right|\int\dif\tau.
\end{align}
We have
\begin{align}
\mu_3 = \frac{\sqrt{\pi}}{\hat\kappa}=\frac{\sqrt{\pi}}{\kappa\sqrt{\V_0}}
\end{align}and are working in units where $\sqrt{\pi}/\kappa = 1$. Thus, the four-dimensional action is
\begin{empheq}[box={\mymath}]{equation}
	S_{D3}^{(4)}= -e^{\frac{1}{2}K^{cs}}\left|q_\ha Z^\ha-p^{\hat b}\mathcal F_{\hat b}\right|\int\dif\tau+q_\ha \int V^\ha+p^\ha \int U_\ha.
	\label{D3Brane:4dAction}
\end{empheq}
Clearly, this is the action of a particle with mass 
 \begin{align}
 	m=e^{\frac{1}{2}K^{cs}}\left|q_\ha Z^\ha-p^{\hat b}\mathcal F_{\hat b}\right|
 	\label{Eq:D3Brane:ParticleMass}
 \end{align}
charged under the gauge fields $V^\ha$ and $U_\ha$. One recognizes $m$ as the central charge \eqref{Eq:BPS:CentralChargeSymplectic}. Hence, a brane wrapping a supersymmetric cycle gives rise to a particle that is extremal with respect to the BPS bound \eqref{Eq:String:SUGRA:BPSBound},
\begin{align}
	m = \left|\Z\right|.
\end{align}
\subsection{A scalar WGC for the particle}
\subsubsection{Applying the central charge relation}
We found out that the central charge is related to the symplectic charges and gauge couplings via \eqref{Eq:SUSYBlackHoles:CentralChargeRelation2} and that a D3-brane wrapping a supersymmetric Lagrangian three-cycle looks like a particle that is extremal with respect to the BPS bound upon compactification to four dimensional spacetime. Hence, we can apply the central charge relation derived in the last section,
\begin{empheq}[box={\mymath}]{equation}
\mathcal{Q}^2=m^2 +4G^{a\bar b}\nabla_am\bar\nabla_{\bar b} m.
\label{Eq:D3Brane:WGC}
\end{empheq}
If we consider only electric charges and set
\begin{align}
	\Gamma= \binom{0}{q_\ha},
\end{align} the quantity $\Q$ defined in \eqref{Eq:SUSYBlackHoles:Q^2Definition} becomes
\begin{align}
	\Q^2 &= -\frac{1}{2} q_\ha(\I^{-1})^{\ha\hat b}q_{\hat b}
\end{align}
where $\I = \IM\M$. In that case, $\Q$ is the analogue to the electric charge $Q$ appearing in the electric weak gravity conjecture \eqref{Eq:WGC} and we will make this clear by explicit calculation from the prepotential. Before doing so, we should point out the following: Recall that we assumed that the brane is wrapping the three-cycle such that it minimizes its volume and that only in this case the mass is extremal with respect to the BPS bound. Dropping this assumption would lead to $m\ge|\Z|$ but it is not obvious how the second term with derivatives of $m$ would be affected. We keep the volume minimized.

The equality \eqref{Eq:D3Brane:WGC} is in accordance with a generalization of the Gauge-Scalar WGC as we presented it in \eqref{Eq:WGC:GaugeScalar} to situations with several gauge and scalar fields. In particular - since the second term on the right-hand side is positive definite - it implies the Weak Gravity Conjecture
\begin{align}
	m^2 \le \mathcal{Q}^2.
\end{align}

\subsubsection{Explicit calculation from prepotential}
Now, to get better understanding of \eqref{Eq:D3Brane:WGC}, we explicitly carry out the calculation in the prepotential formalism. To do so, we need an explicit expression of the gauge-coupling matrix $\M$. Recall\footnote{See section \ref{Section:MirrorSymmetry}.} that mirror symmetry allows us to assume a cubic prepotential for the moduli in the following way: Let $\tilde Y$ be the mirror Calabi-Yau of $Y$. Since by mirror symmetry $\dim_{\mathbb C}H^{2,1}(Y)= \dim_{\mathbb C}H^{1,1}(\tilde Y)$ and $\dim_{\mathbb C}H^{1,1}(Y)= \dim_{\mathbb C}H^{2,1}(\tilde Y)$, we interpret $\tilde\K^{ks}\coloneqq \K^{cs}$ as K\"ahler potential for the \emph{K\"ahler} moduli on $\tilde Y$ and likewise, $\tilde X^{\hat A}\coloneqq Z^\ha, \tilde{\mathcal F}_{\hat A}\coloneqq\mathcal F_\ha$ as belonging to its prepotential. We will drop the tilde in the following. 

We only consider charges under the vectors in the gauge multiplet and hence set $q_{\hat A}=(0,q_{A}),~p^{\hat A}=0$, such that
\begin{align}
	m^2 = e^{K^{ks}}\left|q_At^A\right|^2.
\end{align}
We can now make use of an explicit expression for the gauge-coupling matrix $\M$. This can be derived directly from the prepotential but takes some time and effort while at the same time being not very illuminating. Hence, we perform the calculation in the appendix and from there cite eq. \eqref{Eq:MatrixMExplicit}:
\begin{align}
	(\IM \M^{-1})^{\hat A\hat B} = -\frac{6}{\K}\begin{pmatrix}
	1 & b^A\\
	b^A & \frac{1}{4}G^{AB}+b^Ab^B
	\end{pmatrix}.
\end{align}
Therefore,
\begin{align}
	\Q^2 &= -\frac{1}{2} q_A(\IM\M^{-1})^{AB}q_{ B}\nl
	&= e^{K^{ks}}\left(q_AG^{AB}q_B + 4(q_Ab^A)^2\right).
\end{align}
We can choose $t^A=b^A+iv^A$ to be purely imaginary\footnote{This is because the metric $G$ does not depend on the axions.} such that
\begin{align}
	\mathcal{Q}^2&= e^{K^{ks}}q_AG^{AB}q_B,\nl
	m^2 &= e^{K^{ks}}\left|q_Aiv^A\right|^2.
	\label{D3Brane:Q^2}
\end{align}
Next, we compute the K\"ahler covariant derivatives
\begin{align}
	\nabla_A m= \partial_{t^A}m +\frac{1}{2}\left(\partial_{t^A}K^{ks}\right)m
\end{align}
that appear on the right-hand side of \eqref{Eq:D3Brane:WGC}:
\begin{align}
\nabla_Am &= \frac{1}{2i}\left(\partial_{v^A}+\frac{1}{2}\partial_{v^A}K^{ks}\right)m\nl
&=\frac{1}{2i}e^{\frac{1}{2}K^{ks}}\left(\partial_{v^A}+\partial_{v^A}K^{ks}\right)(q_Av^A).
\end{align}
With
\begin{align}
\partial_{v^A}K^{ks} &= -\frac{1}{\K}\partial_{v^A}\K_{BCD}v^Bv^Cv^D = -\frac{3\K_A}{\K},
\end{align}
we find
\begin{align}
\nabla_Am = \frac{1}{2i}e^{\frac{1}{2}K^{ks}}\left(q_A - \frac{3}{2}\frac{\K_A}{\K} q_Bv^B \right).
\end{align}
We need the following expressions:
\begin{align}
G^{AB}\K_A &= -\frac{2\K}{3}\left(\K^{AB} - \frac{3v^Av^B}{\K}\right)\K_A\nl
&= -\frac{2\K}{3}(v^B-3 v^B)\nl
&= e^{-K^{ks}}v^B
\end{align}
and
\begin{align}
G^{AB}\K_A\K_B &= \frac{4}{3}\K v^A\K_B = \frac{4}{3}\K^2.
\end{align}
With these, the right-hand side of \eqref{Eq:D3Brane:WGC} takes the form
\begin{align}
m^2+4G^{AB}\nabla_Am\bar{\nabla}_Bm =&~\frac{1}{4}e^{K^{ks}} (q_Av^A)^2 + \nl&G^{AB}e^{K^{ks}}\left(q_A - \frac{3}{2}\frac{\K_A}{\K}q_Cv^C\right)\left(q_B-\frac{3}{2}\frac{\K_B}{\K}q_Dv^D\right)\nl
=&~e^{K^{ks}}\left[(q_Av^A)^2 + q_AG^{AB}q_B -2\frac{3}{2}\frac{4}{3}(q_Av^A)^2+ \frac{9}{4}\frac{4}{3}(q_Av^A)^2\right]\nl
=&~e^{K^{ks}} q_A G^{AB} q_B.
\end{align}
This is precisely what we found in \eqref{D3Brane:Q^2} and thus we have shown that
\begin{empheq}[box=\mymath]{equation}
q_AG^{AB}q_B=e^{-K^{ks}}\left(m^2 +4G^{AB}\nabla_Am\bar\nabla_B m\right),
\end{empheq}
confirming eq. \eqref{Eq:D3Brane:WGC}. In the next section, we will perform a similar calculation in order to establish a weak-gravity bound for an axion-instanton-pair.\newpage
\leavevmode\thispagestyle{empty}
\newpage
\section{A Scalar WGC for Type IIA Instantons}
\label{Section:ScalarIIAWGC}
\subsection{Axions from reduction of Type IIA supergravity}
We now turn to Type IIA supergravity where compactification gives rise to axions $\xi^{\hat a}$ and $\tilde\xi_{\hat a}$ in the four dimensional theory. Unlike before, there is no self-duality that relates these two sets of fields.

Our starting point is the ten-dimensional low-energy action \eqref{Appendix:IIAAction} that is obtained from eleven-dimensional supergravity as discussed in appendix \ref{Appendix:String:IIA}:
\begin{align}
S^{(10)}_{IIA} = -\frac{1}{4\hat \kappa^2}\int e^{-2\hat \phi} \hat H_3\wedge* \hat H_3 -\frac{1}{4\hat \kappa^2}\int \hat F_2\wedge* \hat F_2 - \frac{1}{4\hat\kappa^2}\int \hat F_4\wedge* \hat F_4 \nl
+ \frac{1}{4\hat\kappa^2}\int \hat H_3\wedge \hat C_3\wedge \dif \hat C_3
+ \frac{1}{2\hat\kappa^2}\int e^{-2\hat\phi}\left(*\mathcal R+ 4\dif\hat \phi\wedge*\hat \dif\phi\right)
\label{IIA:Action:10d}
\end{align}
We recall the definitions
\begin{align}
\hat H_3 = \dif\hat B_2,~~\hat F_2 = \dif \hat A_1,~\hat F_4 = \dif \hat C_3 - \hat A_1\wedge \hat H_3
\end{align}
and proceed similar to the discussion of Type IIB in the last section.
\subsubsection{Expanding the fields}
By now, it should be clear how the four-dimensional fields arise. The forms are expanded as
\begin{align}
\hat A_1 &= A^0\nonumber\\
\hat C_3 &= C_3 + A^A\wedge\omega_A + \xi^{\hat a}\alpha_{\hat a}-\tilde\xi_{\hat a}\beta^{\hat a}\nonumber\\
\hat B_2 &= B_2 + b^A\omega_A,
\end{align}
where
\begin{align}
a &= 1,...,h^{2,1}, &\hat a=0,1,...,h^{2,1},\nl
A &= 1,...,h^{1,1}, &\hat A=0,1,...,h^{1,1}.
\end{align}
The corresponding field strengths are
\begin{align}
\dif\hat A_1 &= \dif A^0,\nl
\dif\hat C_3 &= \dif C_3 + \dif A^A\wedge\omega_A + \dif\xi^{\hat a}\wedge\alpha_{\hat a}-\dif \tilde{\xi}_{\hat a}\wedge \beta^{\hat a},\nl
\hat H_3 &= H_3+\dif b^A\wedge\omega_A
\end{align}
with $H_3 := \dif B_2$.
Since we are eventually interested in the axions $\xi^\ha,\tilde\xi_\ha$, we take a closer look at the expansion of $\hat F_4$. Plugging in the above expressions this becomes
\begin{align}
\hat F_4= \dif C_3 + \dif A^A\wedge\omega_A + \dif\xi^{\hat a}\wedge\alpha_{\hat a}-\dif \tilde{\xi}_{\hat a}\wedge \beta^{\hat a} - A^0\wedge H_3 - A^0\wedge\dif b^A\wedge\omega_A.
\end{align}
\subsubsection{Field content of Type IIA in four dimensions}
\begin{table}
	\begin{center}
		\begin{tabular}{|c|c|c|}
			\hline 
			\rowcolor{\BColor}
			\rule[-1.5ex]{0pt}{4.5ex} \textbf{Multiplet}& \textbf{(Massless) Field Content}  & \textbf{Number} \\
			\hline 
			\rule[-1.5ex]{0pt}{4.5ex} Gravity multiplet& $(g_{\mu\nu},V^0)$  & 1 \\
			\hline 
			\rule[-1.5ex]{0pt}{4.5ex} Gauge multiplets & $(V^A,t^A)$ & $h^{1,1}$ \\
			\hline
			\rule[-1.5ex]{0pt}{4.5ex} \multirow{2}{*}{Hypermultiplets }&$(z^a,\xi^a,\tilde\xi_a)$ & $h^{2,1}$\\
			\cline{2-3}
			\rule[-1.5ex]{0pt}{4.5ex} &$(h_B,\phi,\xi^0,\tilde\xi_0)$ & 1\\
			\hline
		\end{tabular} 
		\caption[Type IIA multiplets]{Type IIA multiplets in $d=4$.\label{Table:IIAFieldContent}}
	\end{center}
\end{table}
Note that the three-form $C_3$ is dual to a constant in $d=4$ and thus does not carry any degree of freedom. The remaining fields form the following multiplets 
\begin{enumerate}[i)]
	\item The gravity multiplet consists of the graviton $g_{\mu\nu}$ and the one-form $V^0$.
	\item The one-forms $V^A$ combine with the complex $t^A=b^A+iv^A$ to $h^{1,1}$ gauge multiplets. 
	\item Complex structure moduli $z^a$ form $h^{2,1}$ hypermultiplets with the scalars $\xi^a$ and $\tilde\xi_a$.
	\item The two-form $B_2$ is dual to a scalar $h_B$ which combines in an additional hypermultiplet with the dilaton $\phi$, $\xi^0$ and $\tilde\xi_0$.
\end{enumerate}
For later reference, these multiplets are collected in table \ref{Table:IIAFieldContent}.

Now, we will derive the four-dimensional actions for the axions $\xi$ and $\tilde{\xi}$.
\subsubsection{Integrating on the Calabi-Yau}
The only non-vanishing terms in the integral of $\hat F_4\wedge*\hat F_4$ are
\begin{align}
 I_1&\coloneqq\int_Y (\dif C_3 - A^0\wedge H_3)\wedge*(\dif C_3 - A^0\wedge H_3),\nl
 I_2&\coloneqq\int_Y(\dif A^A-A^0\wedge\dif b^A)\wedge \omega_A\wedge*[(\dif A^B-A^0\wedge\dif b^B)\wedge \omega_B],\nl
 I_3&\coloneqq\int_Y (\dif \xi^{\hat a}\wedge\alpha_{\hat a}-\dif\tilde\xi_{\hat a}\wedge\beta^{\hat a})\wedge * (\dif\xi^{\hat b}\wedge\alpha_{\hat b}-\dif\tilde\xi_{\hat b}\wedge\beta^{\hat b})
\end{align}
and of these, only $I_3$ contributes a kinetic term for the axions:
\begin{align}
I_3=&~-\dif\xi^{\hat a}\wedge*\dif\xi^{\hat b}[\IM \M+(\RE\M)(\IM\M)^{-1}(\RE\M)]_{\hat a\hat b} \nl
&+2\dif\tilde\xi_{\hat a}\wedge*\dif\xi^{\hat b}[(\RE\M)(\IM\M)^{-1}]_{\hat b}^{\hat a}  - \dif\tilde\xi_{\hat a}\wedge *\dif\tilde\xi_{\hat b}[(\IM\M)^{-1}]^{\hat a\hat b}\nl
=&~{(\IM\M)^{-1}}^{\hat a\hat b}[\dif\tilde\xi_{\hat a}+\M_{\hat a\hat c}\dif\xi^{\hat c}]\wedge*[\dif\tilde\xi_{\hat b}+\bar\M_{\hat b\hat d}\dif\xi^{\hat d}].
\end{align}
The kinetic part in the action for the axions $\tilde\xi_\ha$ is therefore described by the inverse $\I^{-1}=(\IM\M)^{-1}$,
\begin{empheq}[box={\mymath}]{equation}
S_{\tilde{\xi}} =\int (\I^{-1})^{\hat a\hat b}\dif\tilde\xi_{\hat a}\wedge *\dif\tilde\xi_{\hat b}.
\end{empheq}
The one for the axions $\xi^\ha$ is
\begin{align}
S_{\xi} &=\int (\IM\M^{-1})^{\hat a\hat b}\M_{\hat a\hat c}\bar\M_{\hat b\hat d}\dif\xi^{\hat c}\wedge*\dif\xi^{\hat d}\nl
&=\int  (\IM\M^{-1})^{\hat a\hat b}(\RE\M+i\IM\M)_{\hat a\hat c}(\RE\M-i\IM\M)_{\hat b\hat d}\dif\xi^{\hat c}\wedge*\dif\xi^{\hat d}\nl
&= \int \left[(\IM\M^{-1})^{\hat a\hat b}(\RE\M_{\hat a\hat c}\RE\M_{\hat b\hat d})+\IM\M_{\hat c\hat d}\right]\dif\xi^{\hat c}\wedge*\dif\xi^{\hat d}
\end{align}
which we write as
\begin{empheq}[box={\mymath}]{equation}
S_{\xi} =\int \left(\R\I^{-1}\R +\I\right)_{\ha\hat b}\dif\xi^{\hat a}\wedge*\dif\xi^{\hat b}.
\label{Eq:IIA:AxionAction}
\end{empheq}
The next section will focus on the axions $\tilde\xi$ in order to make the calculation easier.
\subsection[Instantons from E2-Branes]{Instantons from E2-Branes}
\label{Section:E2Branes}
We saw in section \ref{Section:D3Branes} that a D3-brane wrapping a three-cycle in Type IIB looks like a zero-dimensional object - a point particle - in four dimensions that carries charge under the one-forms which arise from dimensional reduction of the four-form coupling to the D3-brane. Further, it was shown that this particle satisfies a refined form of the Weak Gravity Conjecture.

\begin{figure}[]             
	\centering                  
	\def\svgwidth{220pt}    
	\includegraphics[width=0.8\textwidth]{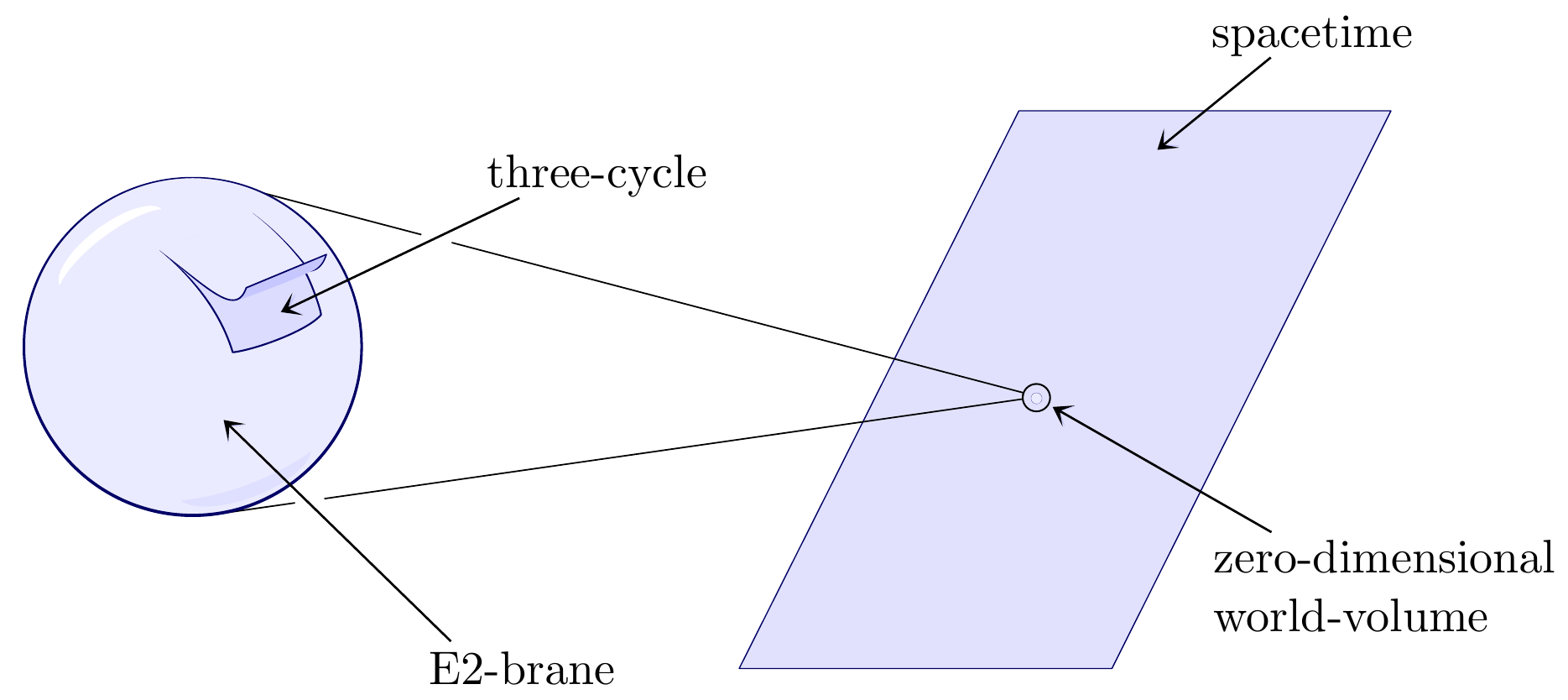}  
	\caption[E2-brane wrapping three-cycle]{E2-brane wrapping three-cycle looks like instanton from low-energy perspective. Surface symbolizing brane is cut open to expose cycle underneath.\label{Fig:E2BraneCompactification}   }     
\end{figure}

In the present section, we consider a setting with a Euclidean E2-brane\footnote{An E$p$-brane is a $p$-brane whose world-volume time is euclideanized. See e.g. \cite{BeckerBecker}.} that couples to the three-form $\hat C_3$. Wrapping such E2-branes around appropriate cycles gives rise to $(-1)$-dimensional objects upon compactification to four dimensions. This is symbolized by the sphere in fig. \ref{Fig:E2BraneCompactification} which from afar looks like a single point in spacetime.
We will now derive the four-dimensional action for such a brane. Our starting point is
\begin{align}
S^{(10)}_{E2} = -\mu_2\int_{E2}\dif_E^3\xi e^{-\hat\phi}\sqrt{-\det(G)} + \sqrt{2}\mu_2\int_{E2}\hat{C}_3,
\label{E2:10dAction}
\end{align}
the action of an E2-brane coupled to the three-form $\hat{C}_3$ which (like the D3-brane in the last section) wraps a supersymmetric three-cycle
\begin{align}
\mathcal{C} = q_\ha\mathcal{A}^\ha - p^{\hat b}\F_{\hat b}.
\end{align}

This time, the integral is Euclidean and therefore yields the brane volume. We find:
\begin{empheq}[box={\mymath}]{equation}
S_{E2}^{(4)}= -e^{\frac{1}{2}K^{ks}}\left|q_\ha Z^\ha-p^{\hat b}\mathcal F_{\hat b}\right|+q_\ha \xi^\ha+p^\ha \tilde\xi_\ha.
\label{E2Brane:4dAction}
\end{empheq}
This can be interpreted as an object localized (see the illustration in fig. \ref{Fig:E2BraneCompactification}) in spacetime - an \emph{instanton} - with action
\begin{align}
	S=e^{\frac{1}{2}K^{ks}}\left|q_\ha Z^\ha-p^{\hat b}\mathcal F_{\hat b}\right|
\end{align}
coupled to the axions $\xi^\ha$ and $\tilde{\xi}_\ha$.
\subsection{A scalar WGC for instantons and axions}
We will set $q_\ha=0$, i.e. consider charges
\begin{align}
\Gamma= \binom{p^\ha}{0},\qquad p^\ha = (0,p^a),
\end{align} 
implying that the instanton couples to $\tilde\xi$ only.
Similar to the last section, we compute the expression
\begin{align}
S^2 + \nabla_aS\bar\nabla_bS G^{ab}.
\end{align}
\subsubsection{Calculation}
The prepotential and coordinates are related via $\F_a = \M_{a\hat b}Z^{\hat b}$ (see eq. \eqref{Eq:Calculations:MZ}) and therefore,
\begin{align}
 S = e^{\frac{1}{2}K}\left|p^a\M_{ab}\right|.
\end{align}
Here, the axions $b^a=0$ were set to zero again such that
\begin{align}
	\M_{ab} = -i\frac{1}{2}e^{-K}G_{ab}.
\end{align}
The calculation goes through as before. We have
\begin{align}
\nabla_a S = \partial_a S + \frac{1}{2}(\partial_a K)S,\qquad\text{where}\quad\partial_a S = \frac{1}{2}(\partial_a K))S + e^{\frac{1}{2}K}\partial_a|p^a\F_a|,
\end{align}
and thus
\begin{align}
\nabla_a S\bar\nabla_bS = \frac{1}{4}e^K\left(\frac{-3\K_a}{2\K}p^c\M_{cd}v^d+\M_{ac}p^c\right)\left(\frac{-3\K_b}{2\K}p^c\bar\M_{cd}v^d+\bar\M_{bd}p^d\right).
\end{align}
Hence,
\begin{align}
4e^KG^{ab}\nabla_aS\bar\nabla_b S &= G^{ab}\left(G_{ac}-\frac{3\K_a}{2\K}G_{cd}v^c\right)\left(G_{be}-\frac{3\K_b}{\K}G_{ef}v^f\right)p^cp^e\nl
&=G_{ab}p^ap^b-2\cdot\frac{3\K_a}{2\K}G^{ab}G_{be}G_{cd}v^dp^cp^e + \frac{9}{4}\frac{4}{3}(G_{ab}p^av^b)^2\nl
&= G_{ab}p^ap^b - 3\cdot\frac{4}{3}(G_{ab}p^av^b)^2 + 3(G_{ab}p^av^b)^2,
\end{align}
that is
\begin{align}
G^{ab}\nabla_aS\bar\nabla_b S = \frac{1}{4}e^{-K}\left(G_{ab}p^ap^b -(G_{ab}p^av^b)^2\right).
\label{Eq:E2Brane:GSSMitte}
\end{align}
In the last steps, we used
\begin{align}
G^{ab}\K_a=\frac{4}{3}\K v^b,\quad G^{ab}\K_a\K_b = \frac{4}{3}\K^2,
\end{align}
several times. The second term on the right-hand side of \eqref{Eq:E2Brane:GSSMitte} is equal to minus $S^2$:
\begin{align}
 S^2 &= e^K\left|p^a\left(-\frac{i}{2}e^{-K}G_{ab}iv^b\right)\right|^2\nl
 &= e^{-K}\frac{1}{4}(p^aG_{ab}v^b)^2.
\end{align}
Putting all terms together,
\begin{align}
S^2 + G^{ab}\nabla_a S\bar\nabla_b S = \frac{1}{4}e^{-K} p^aG_{ab}p^b.
\label{Eq:E2Brane:S2GSS}
\end{align}
Remember that we found a kinetic term\footnote{Note the appearance of the inverse $(\IM M)^{-1}$ in contrast to what we had in the last section.}
\begin{align}
	\int\left[ (\IM\M)^{-1}\right]^{\ha\hat b}\dif\tilde{\xi}_\ha\wedge *\dif\tilde{\xi}_{\hat b}
\end{align}
for the axions $\tilde\xi$ that couple to the instanton with charges $p$. Hence, the corresponding charge $\Q^2$ is
\begin{align}
	\Q^2 &=-\frac{1}{2}p^a(\IM\M) _{ab}p^b\nl
	&=\frac{1}{4}e^{-K}p^aG_{ab}p^b
\end{align}
This is equal to the right-hand side of \eqref{Eq:E2Brane:S2GSS} such that we finally arrive at the identity
\begin{empheq}[box={\mymath}]{equation}
 \Q^2 = S^2 + G^{ab}\nabla_a S\bar\nabla_b S.
 \label{Eq:E2Brane:ScalarWGC}
\end{empheq}
Note that we established \eqref{Eq:E2Brane:ScalarWGC} as a statement about the axions $\tilde\xi$ by turning off the charges $q$\footnote{This corresponds to the lower right part of $\mathscr M$ in \eqref{Eq:BPS:MMatrix}.}. However, it is sensible to assume that it holds also for the axions $\xi$. For these we found (see eq. \eqref{Eq:IIA:AxionAction}) a kinetic term
\begin{align}
\left(\R\I^{-1}\R +\I\right)_{\ha\hat b}\dif\xi^{\hat a}\wedge*\dif\xi^{\hat b}
\end{align}
which makes the calculation leading to \eqref{Eq:E2Brane:ScalarWGC} pretty awful. We will not carry it out but instead propose that \eqref{Eq:E2Brane:ScalarWGC} also holds if the instanton couples to $\xi$ or both types of axions, corresponding to having $q$ or both, $q$ and $p$ charges.

\subsubsection{Result}
Equation \eqref{Eq:E2Brane:ScalarWGC} is an extension of the Weak Gravity Conjecture \eqref{Eq:WGC:Axions} for axions to a situation with scalar fields present. For simplicity, we will now consider the case of a single scalar field $\phi$  and a single axion with decay constant $f$ coupled to the instanton. With $\Q = 1/f$ and still working in units where $M_p=1$, we then have
\begin{align}
	1 = f^2(S^2 + \partial_\phi S\bar \partial_\phi S).
\end{align}
In particular - since the second term is positive definite - we recover the axion WGC
\begin{align}
	1 \ge f S,
\end{align}
that is, $f < 1$ if we assume $S>1$.

We have established \eqref{Eq:E2Brane:ScalarWGC} for branes wrapping supersymmetric cycles. This is similar to the situation discussed in the last section where we saw that a D3-brane wrapping a supersymmetric three-cycle gives rise to a BPS particle. We argued that for two such particles, gauge, gravitational and scalar interaction would cancel, while for non-BPS states, the former would exceed the combined gravitational and scalar forces. Hence, it seems natural to propose that without supersymmetry, equation\eqref{Eq:E2Brane:ScalarWGC} becomes a bound
\begin{empheq}[box={\mymath}]{equation}
 \Q^2 \ge S^2 + G^{ab}\nabla_a S\bar\nabla_b S.
\end{empheq}

Motivated by this evidence, we propose that there is a general extension of the axion-instanton Weak Gravity Conjecture to situations with scalar fields which is analogous to the gauge-scalar Weak Gravity Conjecture and (reinstalling the Planck mass) takes the form
\begin{empheq}[box={\mymath}]{equation}
	M_p^2/f^2 \ge S^2+|\partial_\phi S|^2M_p^2.
\end{empheq}
Since the second term on the left-hand side of this relation is positive-definite, this lowers the bound $f < M_p$ for the axion decay constant which follows from demanding $e^{-S} < 1$ for an instanton and we have the new bound 
\begin{align}
f< M_p/\sqrt{1 +| \partial_\phi S|^2 M_p^2}\le M_p.
\end{align}
\newpage

\section{Conclusion}

We have discussed the Weak Gravity Conjecture in the presence of scalar fields and presented evidence from IIB supergravity where three-branes wrapping supersymmetric Lagrangian three-cycles look like (BPS) particles after compactification to four-dimensional spacetime. Such a particle is charged under the gauge fields arising from the expansion of the Type IIB four-form and saturates the Gauge-Scalar Weak Gravity Conjecture. If the particle carries charge under a single gauge field and with only one scalar field present, the bound takes the form 
\[m^2+|\partial_\phi m|^2M_p^2\le g^2M_p^2.\]
For two BPS particles, this translates to the statement that the combined gravitational, gauge and scalar forces between them vanishes.

Motivated by this, we studied compactification of Type IIA string theory, where a Euclidean E2-brane wrapping a supersymmetric Lagrangian three-cycle looks like an instanton in the four-dimensional theory. Expansion of the Type IIA three-form gave rise to certain axions and we found a relation for these which is similar to  the Gauge-Scalar Weak Gravity Conjecture:
\[\Q^2 = S^2 + G^{ab}\nabla_a S\bar\nabla_b S\]
where $S$ is the instanton action and $\Q^2$ describes the coupling of these axions to the instanton. More precisely,
\[\Q^2 = \frac{1}{2}\begin{pmatrix}
p && q
\end{pmatrix}\begin{pmatrix}
-(\I + \R\I^{-1}R) && \R\I^{-1}\\
\I^{-1}\R && -\I^{-1}
\end{pmatrix}\begin{pmatrix}
p \\ q
\end{pmatrix}\]
where the matrices $\I$ and $\R$ are defined in terms of the prepotential for the Calabi-Yau moduli.

We therefore argued that the axion-instanton Weak Gravity Conjecture might be extended to situations where scalar fields are present. More precisely, we proposed a bound of the form 
\[S^2+|\partial_\phi S|^2M_p^2\le  M_p^2/f^2\]
relating instanton action $S$, axion decay constant $f$ and scalar field $\phi$. While we gave a physical interpretation of the Gauge-Scalar Weak Gravity Conjecture in terms of forces, it is not clear to us how the additional $\partial_\phi S\bar\partial_\phi S$ should be understood in the case of axions and instantons. Since this term is positive definite, we argued that scalar fields actually lower the bound $f<M_p$ that follows from the axion-instanton Weak Gravity Conjecture when demanding $e^{-S} < 1$ for the instanton action.\newpage

\appendix 

\leavevmode\thispagestyle{empty}
\newpage
\section{Mathematical Preliminaries} 
\label{Appendix:Maths} 
While the reader is expected to be familiar with basic differential geometry \cite{Books:GaugeKnotsGravity}, we review some of the mathematical definitions and results on complex manifolds that are used in the thesis. More detailed discussions on this can be found in \cite{Nakahara, ComplexManifoldsLectureNotes, ComplexManifoldsLectureNotes2} while a brief treatment of K\"ahler and Calabi-Yau manifolds is given in \cite{Greene}.
\subsection{Complex manifolds}
Simply put, a complex manifold is a manifold $M$ that allows us to define \emph{holomorphic} functions $f:M\rightarrow\mathbb{C}$. To do so in a patch independent way, one needs the transition functions to be holomorphic. We define
\begin{definition}[Complex manifold]
	A \emph{complex manifold} is a differentiable manifold with a holomorphic atlas.
\end{definition}
Note that a complex manifold necessarily has even real dimension since the coordinate functions are maps onto $\mathbb{C}^m$. Instead of pondering on complex manifolds, we will begin with differential manifolds that are not necessarily complex but admit a so-called almost complex structure  which resembles the one that complex manifolds carry: While they might not even locally look like $\mathbb{C}^m$, their tangent spaces are complex. Eventually, we will think of complex manifolds as differentiable manifolds with almost complex structure and vanishing Nijenhuis tensor field. The reason we are choosing this approach is that it is the one that comes naturally when we discussing string compactification.

\begin{definition}[Almost complex structure]
A $(1,1)$-tensor field $\mathcal J$ on a differentiable manifold $M$ satisfying
\begin{align}
	\mathcal J^2= -\operatorname{id}
	\label{Maths:AlmostComplexStructure}
\end{align}
is called \emph{almost complex structure}. In that case, $(M,\mathcal{J})$ is called \emph{almost complex manifold}.
\end{definition}
Note that since $TM\otimes T^*M= \operatorname{End}(TM)$, the field $\mathcal{J}$ defines an endomorphism $\mathcal{J}_p$ of every fiber $T_pM$. Writing $\mathcal{J}_p$ in local coordinates,
\begin{align}
	\mathcal{J}_p = {\mathcal{J}_\mu}^\nu(p) \dif x^\mu\otimes \frac{\partial}{\partial x^\nu},
	\label{Maths:AlmostComplexLocalCoordinates}
\end{align}
condition \eqref{Maths:AlmostComplexStructure} becomes
\begin{align}
	{\mathcal{J}_\mu}^\sigma(p){\mathcal{J}_\sigma}^\nu(p) = -\delta_\mu^\nu.
	\label{Maths:AlmostComplexStructureLocal}
\end{align}
Taking the determinant of \eqref{Maths:AlmostComplexStructureLocal}, we find that almost complex manifolds need to have even real dimension. Not every $2m$-dimensional differentiable manifolds admits an almost complex structure, though. For example, the only spheres that do are $S^2$ and $S^6$ \cite{ComplexManifoldsLectureNotes}.

The almost complex structure now allows us to define (anti-)holomorphic vector fields as follows:
We extend $\mathcal{J}_p$ to the complexification\footnote{By the \emph{complexification} $V^\mathbb{C}$ of a real vector space $V$ we mean the complex space spanned by the linear combinations $a + i b$ with $a,b\in V$ and all vector space operations defined in the obvious way. It has complex dimension $\dim_\mathbb{C}V^\mathbb{C} = \dim_\mathbb{R}V$.} $T_pM^\mathbb{C}$. Since $\mathcal{J}_p^2 = -\operatorname{id}_{T_pM}$, the almost complex structure $\mathcal{J}_p$  has eigenvalues $\pm i$ and we can therefore decompose
	\begin{align}
	T_pM^\mathbb{C} = T_pM^+\oplus T_pM^-
	\end{align}
	where the two disjoint spaces $T_pM^\pm$ are spanned by the eigenvectors with eigenvalue $\pm i$. This can be extended to the whole tangent bundle and we write
	\begin{align}
		TM^{\mathbb{C}} = TM^+ \oplus TM^-.
		\label{Maths:ComplexifiedTangentBundle}
	\end{align}
	
	The complexification $\Gamma(TM)^\mathbb{C}$ consists of the vector fields $Z = X + iY$ with $X,Y\in \Gamma(TM)$ and we define the complex conjugate $\bar{Z}\coloneqq X-iY$. With a projection
	\begin{align}
		\mathcal{P}^\pm\coloneqq \frac{1}{2}\left(\operatorname{id}\mp i\mathcal{J}\right),
	\end{align}
	we can then decompose any vector field $W\in\Gamma(TM)^{\mathbb{C}}$ as
	\begin{align}
		W =Z_1+\bar{Z}_2
		\label{Maths:VecorFieldDecomposition}
	\end{align}
	with $Z_1 \coloneqq \mathcal{P}^+W$ and $\bar{Z}_2 \coloneqq \mathcal{P}^-W$.
	
	\begin{definition}[(Anti-)holomorphic vector field]
	A vector field $Z=\mathcal{P}^+Z$ is called \emph{holomorphic} and a vector field $\bar{Z}=\mathcal{P}^-\bar{Z}$ \emph{antiholomorphic}.
	\end{definition}
	
	The decomposition of the complexified tangent bundle \eqref{Maths:ComplexifiedTangentBundle} also defines the \emph{complexified cotangent bundle}
	\begin{align}
		T^*M^{\mathbb{C}} \coloneqq T^*M^+\oplus T^*M^-.
	\end{align}
	This gives us a way to define the space $\Omega^{r,s}(M)$ of $(r,s)$-forms on $M$:
	The exterior product $\Lambda^kT^*M^{\mathbb{C}}$ decomposes as
	\begin{align}
		\Lambda^kT^*M^{\mathbb{C}} = \Lambda^{0,k}M \oplus \Lambda^{1,k-1}M\oplus\cdots\oplus\Lambda^{k,0}M
	\end{align}
	where
	\begin{align}
		\Lambda^{r,s}M\coloneqq \Lambda^rT^*M^+\otimes\Lambda^sT^*M^-.
	\end{align}
	\begin{definition}[$(r,s)$-forms]
		A section of $\Lambda^{r,s}M$ is called a \emph{$(r,s)$-form} on $M$. We denote the set of $(r,s)$-forms with $\Omega^{r,s}(M)$.
	\end{definition}
	For a complex manifold, the exterior derivative of a $(r,s)$ form can be decomposed \cite{ComplexManifoldsLectureNotes2} as
	\begin{align}
		\dif \omega^{r,a} = \alpha^{r+1,s}+\beta^{r,s+1}
	\end{align}
	which amounts to a decomposition $\dif = \partial + \bar\partial$. We define:
	\begin{definition}[Dolbeault operators]
		The operators\begin{align}
			\partial: \Omega^{r,s}\rightarrow\Omega^{r+1,s},\quad\bar{\partial}:\Omega^{r,s}\rightarrow\Omega^{r,s+1}
		\end{align}
		from the decomposition $\dif = \partial + \bar{\partial}$ are called \emph{Dolbeault operators}. A $p$-form $\omega\in\Omega^{p,0}$ for which $\bar{\partial}\omega=0$ is called \emph{holomorphic}.
	\end{definition}

	We are now ready to give a criterion to determine whether an almost complex structure actually is a complex structure.
	\begin{satz}
	An almost complex manifold $(M,\J)$ admits a \emph{complex structure} if and only if the \emph{Nijenhuis tensor field} $N:\Gamma(TM)\times\Gamma(TM)\rightarrow\Gamma(TM)$ defined by
	\begin{align}
		N(X,Y)\coloneqq [X,Y]+\J[\J X,Y]+\J[X,\J Y]-[\J X,\J Y]
		\label{Maths:Nihenuis}
	\end{align}
	vanishes \cite{Nakahara}.
	\end{satz}
	In local coordinates \eqref{Maths:AlmostComplexLocalCoordinates}, the components of $N$ are
	\begin{align}
		{N^\mu}_{\nu\delta} = {\J_\nu}^{\sigma}\partial_{[\sigma}{\J_{\delta]}}^{\mu}-{\J_\delta}^{\sigma}\partial_{[\sigma}{\J_{\nu]}}^\mu.
	\end{align}
	If this condition is fulfilled, one can define complex coordinates in every patch in terms of which the almost complex structure $\J$ can be written as
	\begin{align}
		\J = i\dif z^j\otimes\frac{\partial}{\partial z^j} - i\dif\bar z^{\bar j}\otimes\frac{\partial}{\partial\bar z^{\bar j}}.
		\label{Maths:ComplexCoordinates}
	\end{align}
\begin{definition}[Hermitian metric]
	A Riemannian metric $g$ on a complex manifold $(M,\mathcal{J})$ is called \emph{Hermitian}, if it satisfies
	\begin{align}
		g(X,Y) = g(\mathcal{J}X,\mathcal{J}Y)
		\label{Maths:HermitianMetric}
	\end{align}
	for all vector fields $X,Y$ on $M$. In that case, $(M,g)$ is called \emph{Hermitian manifold}.
\end{definition}
It is easy to show that every complex manifold admits a Hermitian metric \cite{ComplexManifoldsLectureNotes}. In local coordinates, $g = g_{\mu\nu}\dif x^\mu \otimes \dif x^\nu$ and condition \eqref{Maths:HermitianMetric} translates as
\begin{align}
	g_{\mu\nu} = {\mathcal{J}_\mu}^\alpha{\mathcal{J}_\nu}^\beta g_{\alpha\beta}.
\end{align}
We note the following property \cite{Nakahara}: The components of a Hermitian metric with respect to a \emph{complex} basis,
\begin{align}
		g_{ij}(p) = g_p\left(\frac{\partial}{\partial z^i},\frac{\partial}{\partial z^j}\right),~ g_{\bar{i}j}(p) = g_p\left(\frac{\partial}{\partial \bar{z}^i},\frac{\partial}{\partial z^j}\right),~...,
	\end{align}
	satisfy
	\begin{align}
		g_{ij} = 0 = g_{\bar{i}\bar{j}}.
		\label{Maths:HermMetricPure}
	\end{align}
Thus, one can write
\begin{align}
	g = g_{i\bar{j}}\left(\dif z^i\otimes \dif \bar z^{\bar j}+\dif \bar z^{\bar j} \otimes \dif z^i\right).
\end{align}
\subsection{K\"ahler geometry}
\label{Appendix:Kähler}
\begin{definition}[K\"ahler form]
	On a Hermitian manifold $(M,g,\mathcal J)$, the tensor field $J$ defined by
	\begin{align}
		J(X,Y)\coloneqq g(\mathcal{J}X,Y)\qquad\text{for all }X,Y\in\Gamma(TM)
	\end{align}
	is antisymmetric\footnote{We have $g(\mathcal{J}X,Y)=g(\mathcal{J}^2X,\mathcal{J}Y)=-g(X,\mathcal{J}Y)=-g(\mathcal{J}Y,X)$.} and thus a two-form. Extending $J$ from $\Gamma(TM)$ to $\Gamma(TM)^\mathbb{C}$, it becomes a two-form of type $(1,1)$ and is called \emph{K\"ahler form} of the Hermitian metric.
\end{definition}
It's easy top see that the K\"ahler form is invariant under $\mathcal{J}$, i.e. $J(\mathcal{J}X,\mathcal{J}Y)=J(X,Y)$ for all vector fields $X,Y$. If we write $J$ in a complex basis,
\begin{align}
	J = J_{ij}\dif z^i\otimes\dif z^j + J_{i\bar{j}}\dif z^i\otimes\dif z^{\bar{j}}+\cdots
\end{align}
with
\begin{align}
J_{ij} = J\left(\frac{\partial}{\partial z^i},~\frac{\partial}{\partial z^j}\right) = ig_{ij}, J_{i\bar j} = ig_{i\bar{j},~\cdots},
\end{align}
we find that due to \eqref{Maths:HermMetricPure} the pure index components vanish and we can write
\begin{align}
	J = i g_{i\bar{j}}\dif z^i\wedge \dif \bar{z}^{\bar j}.
	\label{Maths:KählerComplexCoord}
\end{align}
Note that the K\"ahler form is real, $\bar{J} = J$. The K\"ahler form of a Hermitian manifold $M$ with complex dimension $m$ allows us to define a $2m$-form
\begin{align}
	J\wedge\cdots\wedge J.
	\label{Maths:VolumeForm}
\end{align}
Writing $z^i=x^i+iy^i$ and using $\sqrt{g}=2^m\det g_{i\bar j}$, we find
\begin{align}
	\frac{1}{m!}\int_Y J\wedge\cdots\wedge J = \int_Y \sqrt{g}\dif x^1\wedge\cdots\wedge\dif x^m\wedge\dif y^1\wedge\cdots\wedge\dif y^m = \operatorname{Vol}(Y)
\end{align}
Hence, \eqref{Maths:VolumeForm} serves as a \emph{volume form} on the manifold.
\begin{definition}[K\"ahler manifold]
	A Hermitian manifold $(M,g)$ with closed K\"ahler form $\dif J=0$ is called \emph{K\"ahler manifold} and its metric \emph{K\"ahler metric}.
\end{definition}
In local complex coordinates,
\begin{align*}
\dif J = (\partial+\bar{\partial})ig_{i\bar j}\dif z^i\wedge \dif z^{\bar j}=0
\end{align*}
from which we conclude
\begin{align*}
\frac{\partial g_{i\bar j}}{\partial z^l} = \frac{\partial g_{l\bar j}}{\partial z^i},\quad\frac{\partial g_{i\bar j}}{\partial \bar z^{\bar l}} = \frac{\partial g_{i\bar k}}{\partial \bar z^{\bar j}}.
\end{align*}
Thus, we can write $J$ in terms of a so-called \emph{K\"ahler potential} $K$:
\begin{align}
g_{i\bar j} = \frac{\partial^2K}{\partial z^i\partial z^{\bar j}},\quad J=\partial\bar{\partial} K.
\end{align}
Note that the metric can be expressed this way in a given patch $U_i$. Given two charts $(U_i,\varphi_i)$ and $(U_j,\varphi_j)$ with $U_i\cap U_j\neq\emptyset$ and coordinates $z=\varphi_i(p), w=\varphi_j(p)$, the K\"ahler potentials $K_i$ and $K_j$ do in general not coincide but are related by a \emph{K\"ahler transformation}
\begin{align}
	K_j(w,\bar{w}) = K_i(z,\bar{z})+f_{ij}(z)+g_{ij}(\bar{z})
\end{align}
with holomorphic (antiholomorphic) functions $f$ and $g$ \cite{Nakahara}.

With help of the Dolbeault operator $\bar{\partial}$, we can generalize de-Rham cohomology and define
\begin{definition}[Dolbeault cohomology group]
	For a complex manifold $M$, we call the quotient space of $\bar\partial$-closed modulo $\bar{\partial}$-exact $(r,s)$-forms,
	\begin{align}
		H^{r,s}(M)\coloneqq \frac{\ker{(\bar{\partial}:\Omega^{r,s}(M)\rightarrow\Omega^{r,s+1}(M))}}{\operatorname{im}(\bar\partial:\Omega^{r,s-1}(M)\rightarrow\Omega^{r,s}(M))}
	\end{align}
	the X{$(r,s)$-th Dolbeault cohomology group}. Its complex dimension
	\begin{align}
		h^{r,s}\coloneqq\dim_{\mathbb C}H^{r,s}(M)
	\end{align}
	is called \emph{Hodge number}.
\end{definition}
Note that since the K\"ahler form $J$ is closed, $J\in H^{r,s}(M)$. The corresponding equivalence class $[J]$ is called \emph{K\"ahler class}.

We further define the adjoint operators $\partial^\dagger\coloneqq -*\bar{\partial}*$ and $\bar{\partial}^\dagger\coloneqq -*\partial*$ with the Hodge-$*$ operator and the Laplacians $\Delta_{\partial}\coloneqq(\partial+\partial^\dagger)^2$ and $\Delta_{\bar{\partial}}\coloneqq(\bar{\partial}+\bar{\partial}^\dagger)^2$.

We quote the following properties of a K\"ahler manifold with complex dimension $m$ \cite{Nakahara}:
\begin{enumerate}[i)]
\item The Hodge numbers $h^{r,s}$ are related to the Betti numbers $b^p$ via
\begin{align}
	b^p = \sum_{r+s=p}h^{r,s},
\end{align}
\item they are symmetric, $h^{r,s}=h^{s,r}$ and $h^{r,s} = h^{m-r,s-r}$ and
\item the Laplacians defined above coincide with $\Delta = (\dif+\dif^\dagger)^2$,
 \begin{align}
 \Delta = 2\Delta_\partial = 2\Delta_{\bar{\partial}}.
 \end{align} 
\end{enumerate}
 \begin{definition}[Harmonic form]
 	A form satisfying $\Delta_{\bar{\partial}}=0$ is called \emph{harmonic} and we denote the set of harmonic $(r,s)$-forms by $\operatorname{Harm}^{r,s}(M)$.
 	\label{Maths:HarmonicForm:Definition}
 \end{definition}
The K\"ahler form is harmonic. Forms of this type are of great importance to us since massless forms in the type II supergravity actions are expanded in terms of harmonic forms on the compactification manifold. This is possible due to \emph{Hodge's theorem}: For $\Omega^{r,s}(M)$ there is a unique decomposition
\begin{align}
	\Omega^{r,s}(M) = \bar\partial\Omega^{r,s-1}(M)\oplus\bar{\partial}\Omega^{r,s+1}(M)\oplus\operatorname{Harm}^{r,s}(M).
\end{align}
From this follows in particular that
\begin{align}
	\operatorname{Harm}^{r,s}(M)\cong H^{r,s}(M).
	\label{HodgeIsomorphism}
\end{align}
\subsection{Calabi-Yau manifolds}
There are many different ways to define what a Calabi-Yau is. We cite from \cite{ComplexManifoldsLectureNotes2}:
\begin{satz}
	Let $(Y,J,g)$ be a compact K\"ahler manifold with complex dimension $\dim_{\mathbb C}(Y)=n$. The following statements are equivalent:
	\begin{enumerate}[i)]
		\item $Y$ is Ricci-flat, that is it has vanishing Ricci-form $\mathcal R=0$.
		\item It admits a globally defined and nowhere vanishing holomorphic $n$-form.
		\item It has holonomy $\operatorname{Hol}(g)\subset SU(n)$.
		\item Its first Chern class vanishes.
	\end{enumerate}
	\label{Maths:CalabiYau:Theorem}
\end{satz}
\begin{definition}[Calabi-Yau n-fold]
	A compact K\"ahler $n$-fold with one of the properties listed in \ref{Maths:CalabiYau:Theorem} is called \emph{Calabi-Yau} manifold.
	\label{Maths:CalabiYau:Definition}
\end{definition}
From now on, we denote the coordinates on a Calabi-Yau $Y$ by $\{y^i,\bar{y}^{\bar j}\}$ and only consider those of complex dimension $\dim_{\mathbb C}Y=3$. For a Calabi-Yau threefold, the Hodge numbers are depicted in the following figure:
\newline
\begin{figure}[h]
{\small 
	\begin{align*}
	\begin{array}{ccccccccc}
	&&&&h^{0,0}&&&& \\[3mm]
	&&&h^{1,0}&&h^{0,1}&&& \\[3mm]
	\ \ \ \ &&h^{2,0}&&h^{1,1}&&h^{0,2}&& \ \ \ \ \\[3mm]
	&h^{3,0}&&h^{2,1} &&h^{1,2}&&h^{0,3}\\[3mm]
	&&h^{3,1}&&h^{2,2}&&h^{1,2}&& \\[3mm]
	&&&h^{3,2}&&h^{2,3}&&&\\[3mm]
	&&&&h^{3,3}&&&& \\[3mm]
	\end{array} =
	\begin{array}{ccccccccc}
	&&&&1&&&& \\[3mm]
	&&&0&&0&&& \\[3mm]
	\ \ \ \ &&0&&h^{1,1}&&0&& \ \ \ \ \\[3mm]
	&1&&h^{2,1} &&h^{1,2}&&1\\[3mm]
	&&0&&h^{1,1}&&0&& \\[3mm]
	&&&0&&0&&&\\[3mm]
	&&&&1&&&& \\[3mm]
	\end{array} 
	\end{align*}
}
\caption[Hodge Diamond of Calabi-Yau threefold]{Hodge Diamond of a Calabi-Yau threefold.\label{Maths:CalabiYau:HodgeDiamond}}
\end{figure}
In particular, there is only one $(3,0)$-form which we denote by $\Omega$ (with $\bar\Omega$ the conjugate $(0,3)$-form), while there are no one- or five-forms. It is worth noticing that since $h^{3,3}=1$, the $(3,3)$-form $\Omega\wedge\bar\Omega$ must be proportional to $J\wedge J\wedge J$ and thus to the volume form.

As presented in the main text, for a Calabi-Yau $Y$  the cohomology groups $H^{1,1}(Y)$ and $H^{2,1}(Y)$ are associated with K\"ahler and complex structure moduli respectively. Thus, we introduce
\begin{enumerate}[i)]
\item a basis $\{\omega_A\}$ for $H^{1,1}(Y)$ with $A=1,...,h^{1,1}$ as well as a dual basis $\{\tilde{\omega}^A\}$ for $H^{2,2}(Y)$ normalized such that
\begin{align}
\frac{1}{\V_0}\int_Y \omega_A\wedge \tilde\omega_B = \delta^{A}_B,
\label{Maths:CalabiYau:H11Basis}
\end{align}
where we introduced a six-dimensional fixed Calabi-Yau references volume $\V_0$.
\item a basis $\{\eta_a\}$ for $H^{2,1}(Y)$ and $\{\bar{\eta}_a\}$ for $H^{1,2}(Y)$ with $a=1,...,h^{2,1}$.
\end{enumerate}

\subsection{Some integrals on Calabi-Yau threefolds}
\label{Maths:IntegralsOnCalabiYauThreefolds}
We collect a couple of integrals used in the thesis for the compactification of Type II supergravity. First, we need to know how deal with $\alpha_\ha \wedge *\alpha_{\hat b}$ and similar terms.
The real three-forms $\alpha_{\ha}$ and $\beta^{\ha}$ were defined to be orthogonal as stated in \eqref{Maths:CalabiYau:RealBasis}. Since their Hodge duals are again three-forms, we can expand
\begin{align}
*\alpha_{\ha} = {A_{\ha}}^{\hat b}\alpha_{\hat b}+B_{ab}\beta^{\hat b},\quad*\beta^{\ha} = C^{\ha\hat b}\alpha_{\hat b}+{D^{\ha}}_{\hat b}\beta^{\hat b}
\end{align}
in terms of some coefficient matrices $A,B,C$ and $D$. Thus,
\begin{align}
\frac{1}{\V_0}\int_Y \alpha_{\ha}\wedge*\alpha_{\hat b} = \frac{1}{\V_0}\int_Y\alpha_{\ha}\wedge (B_{b\hat c}\beta^{\hat c}) = B_{a\hat c}\delta_{\ha}^{\hat c} = B_{\hat b\ha}=B_{\ha\hat b}
\end{align}
and analogously
\begin{align}
\frac{1}{\V_0}\int_Y \beta^{\ha}\wedge*\beta^{\hat b} = -C^{\hat b\ha}=-C^{\ha\hat b},\qquad
\frac{1}{\V_0}\int_Y \alpha_{\ha}\wedge*\beta^{\hat b} &= {D^{\hat b}}_{\ha} = -{A_{\ha}}^{\hat b}.
\end{align}
These matrices can be written in terms of a matrix $M$ via
\begin{align}
A &= (\RE \M)(\IM \M)^{-1},\nonumber\\
B &= -(\IM \M) - (\RE \M)(\IM \M)^{-1}(\RE \M),\nonumber\\
C &= (\IM \M)^{-1},
\end{align}
which again is determined by the \emph{moduli}. This is derived in \ref{Appendix:MatrixM} and we note at this point:
\begin{align}
\frac{1}{\V_0}\int_Y \alpha_{\ha}\wedge*\alpha_{\hat b} &= -[\IM \M+(\RE\M)(\IM\M)^{-1}(\RE\M)]_{ab},\nl
\frac{1}{\V_0}\int_Y \beta^{\ha}\wedge*\beta^{\hat b} &= -[(\IM\M)^{-1}]^{\ha\hat b},\nl
\frac{1}{\V_0}\int_Y \alpha_{\ha}\wedge*\beta^{\hat b} &= -[(\RE\M)(\IM\M)^{-1}]_{\ha}^{\hat b},\nl
\frac{1}{\V_0}\int_Y\alpha_\ha\wedge\beta^{\hat b} &= \delta_{\ha}^{\hat b}
\label{Maths:AlphaIntegrals}
\end{align}
If we integrate one of the four-dimensional forms $\Lambda$ over the internal Calabi-Yau $Y$, we just get the form times the volume,
\begin{align}
	\int_Y \Lambda\wedge *\Lambda = \Lambda\wedge *\Lambda\int_Y*\mathds{1}=\Lambda\wedge*\Lambda\V= \frac{1}{6}\K\Lambda\wedge*\Lambda.
\end{align}
\leavevmode\thispagestyle{empty}
\newpage
\section{Type II Supergravity} 
\label{Appendix:String} 
In the thesis, we are dealing with the low energy effective actions of Type IIA and IIB string theory, namely Type IIA and Type IIB supergravity (\emph{SUGRA}). This appendix reviews some facts about Type II supergravity that are important for the thesis.

In supersymmetry (\emph{SUSY}), the Poincaré and internal symmetries of a quantum theory are extended to include $\mathcal{N}$ charges which are \emph{spinors}. The Poincaré and SUSY charges form a so-called superalgebra that necessarily includes bosonic and fermionic elements. Supergravity is supersymmetry where the SUSY parameters $\varepsilon=\varepsilon(x)$ are local. The name hints at the fact that local supersymmetry necessarily includes gravity and conversely, a consistent supersymmetric theory of gravity must be local. This is standard textbook material and a thorough treatment can be found in the literature. See \cite{Polchinski1,Polchinski2}, \cite{Books:BeckerBecker} or \cite{Ibanez} for general string theory and \cite{Supersymmetry}, \cite{Books:Supergravity} or the appendix of \cite{Polchinski2} for supersymmetry and supergravity. Both here and in the rest of the thesis, a basic knowledge of these topics is presumed.

Supergravity restricts the number of spacetime dimensions in which it can be consistently formulated to eleven as a higher-dimensional theory leads to more than eight gravitinos upon toroidal compactification to four dimensions. These can only be  embedded in a representation that also has spins $\ge 5/2$ for which - as is well-known - no consistent interactions exist \cite{Books:Supergravity}.
\subsection{Type IIA SUGRA from compactification of $11$-dim. SUGRA}
\label{Appendix:String:IIA}
We begin with SUGRA in the largest allowed dimension $d=11$. This is not only the low-energy effective action of M-theory but we will construct Type IIA supergravity by compactifying $d=11$-SUGRA on a circle. It contains:
\begin{itemize}
\item The graviton $g_{MN}$ transforming in the symmetric traceless of $SO(d-2)$ which has dimension $d(d-3)/2 = 44$. 
\item Its superpartner, the gravitino $\psi_M$: A Majorana spinor in the vector-spinor representation of $SO(d-2)$ which has dimension $(d-3)2^{(d-2)/2}=128$.
\item A three-form $C_3$ that accounts for the remaining $128-44=84=\binom{d-2}{3}$ bosonic states.
\end{itemize}
The unique action keeping only the bosonic fields is
\begin{align}
 S ^{(11)}= \frac{1}{2\kappa_{11}^2}\int *\mathcal{R}-\frac{1}{2\kappa_{11}^2}\int F_4\wedge*F_4 -\frac{1}{6\kappa_{11}^2}\int F_4\wedge F_4\wedge C_3
\end{align}
with  $F_4=\dif C_3$ and the gravitational coupling $\kappa_{11}$. In $d$ dimensions, the latter relates to the $d$-dimensional Newton's constant $G_{d}$ via
\begin{align}
	\kappa_{d}^2 = 8\pi G_{d}.
\end{align}
As mentioned before, Type IIA SUGRA is obtained upon dimensional reduction of $S^{(11)}$, i.e. compactification on $S^1$ keeping only the massless modes\footnote{This is due to the fact that M-theory compactified on a circle with radius $R$ is corresponding to Type IIA string theory with string coupling $g_s=R/\sqrt{\alpha'}$}. In the following, we will sketch the usual Kaluza-Klein procedure as presented for instance in \cite{Polchinski1}.

We take one of the spatial dimensions and call it $y$, demanding that $y\cong y+2\pi R$, where $R$ is the radius of the compactification circle. This corresponds to $\mathbb{R}^{1,10}\rightarrow\mathbb{R}^{1,9}\times S^1$ and gives rise to a scalar $\sigma$ and a gauge field $A_1$ from the lower-dimensional perspective. To see this, consider the general ansatz
\begin{align}
	\hat G_{MN}(x,y)\dif x^M\dif x^N = G_{\mu\nu}(x)+e^{2\sigma(x)}(\dif y+A_\mu(x)\dif x^\nu)^2,\quad \mu,\nu = 0,...,9,
\end{align}
or
\begin{align}
\hat G_{MN}=\begin{pmatrix}
G_{\mu\nu}+e^{2\sigma}A_\mu A_\nu && e^{2\sigma} A_\mu \\
e^{2\sigma} A_\nu && e^{2\sigma}
\end{pmatrix},
\end{align}
where the hats denote $11$-dim. quantities. Likewise, the three-form $\hat C_3$ splits into a three-form $C_3$ and a two-form $B_2$. The precise procedure is discussed in section \ref{Section:Compactification} and we won't carry out the explicit compactification but rather quote the resulting ten-dimensional action from \cite{Polchinski2}:
\begin{align}
S^{(10)} =& \frac{1}{2\kappa_{10}^2}\int \left(e^\sigma *\mathcal R-\frac{1}{2}e^{3\sigma}F_2\wedge *F_2\right)\nl
	&- \frac{1}{4\kappa_{10}^2}\int \left(e^{-\sigma} H_3\wedge *H_3 + e^\sigma F_4\wedge *F_4\right)\nl
	&-\frac{1}{4\kappa_{10}^2}\int B_2\wedge F_4\wedge F_4,
\end{align}
where we defined the ten-dimensional coupling $\kappa^2=2\pi R \kappa_{10}^2$ and
\begin{align}
H_3=\dif B_2,\quad F_4=\dif C_3-A_1\wedge H_3.
\end{align} 
To make contact with string theory, we perform a Weyl-rescaling \cite{Grimm2,PaltiPhD}
\begin{align}
	G_{\mu\nu}&\rightarrow \Omega^{-2}G_{\mu\nu}
	\label{WeylRescaling}
\end{align}
under which
\begin{align}
	G^{\mu\nu}&\rightarrow\Omega^2G^{\mu\nu},\nl
	\sqrt{-G}&\rightarrow\Omega^{-d}\sqrt{-G},\nl
	\int\dif x^d\sqrt{-G}\mathcal{R}&\rightarrow \int\dif x^d\sqrt{-G}\Omega^{2-d}\left(\mathcal R+(d-1)(d-2)\Omega^{-2}\partial_\mu\Omega\partial^\mu\Omega \right).
\end{align}
Note that wedge products of the form $C_p\wedge * C_p$ contain $\sqrt{-G}$ and $p$ times the metric, such that
\begin{align}
C_p\wedge * C_p\rightarrow \Omega^{2p-d}C_p\wedge *C_p.
\end{align}
With $\Omega = e^{\frac{\sigma}{2}}$ and $d=10$, the action is
\begin{align}
S^{(10)} =& \frac{1}{2\kappa_{10}^2}\int \left[e^{-3\sigma}  \left(*\mathcal R + 9\cdot 2\dif\sigma\wedge*\dif\sigma\right) -\frac{1}{2}F_2\wedge *F_2\right]\nl
&- \frac{1}{4\kappa_{10}^2}\int \left(e^{-3\sigma} H_3\wedge *H_3 + F_4\wedge *F_4\right)\nl
&-\frac{1}{4\kappa_{10}^2}\int B_2\wedge F_4\wedge F_4.
\end{align} Defining $\sigma = \frac{2}{3}\phi$ and rearranging the terms, we finally get
\begin{align}
S^{(10)}_{\text{IIA}} = &\frac{1}{2\kappa_{10}^2}\int e^{-2\phi}\left(*\mathcal R+ 4\dif\phi\wedge*\dif\phi- \frac{1}{2}H_3\wedge *H_3\right)\nl
&-\frac{1}{4\kappa_{10}^2}\int \left(F_2\wedge* F_2 + F_4\wedge* F_4\right) - \frac{1}{4\kappa_{10}^2}\int B_2\wedge F_4\wedge F_4
\label{Appendix:IIAAction}
\end{align}
which has the structure
\begin{align}
	S_{\text{IIA}}^{(10)} = S_{\text{NS}}+S_{\text{R}}^{\text{
		IIA}}+S_{\text{CS}}^{\text{IIA}}
\end{align}
with fields in the NS-NS and R-R sector of Type IIA string theory and a Chern-Simons term.
\subsection{Type IIB SUGRA}
Since we only consider massless fields, the difference between the field content of Type IIA and Type IIB string theory lies in the R-R sector where the former contains the odd $p$-form gauge fields $A_1$ and $C_3$ while the latter has the even fields $C_0$, $C_2$ and $C_4$. The form of $S_{\text{IIA}}^{(10)}$ thus suggests to define similarly
\begin{align}
	S_{\text{IIB}}^{(10)} &= S_{\text{NS}}+S_{\text{R}}^{\text{IB}}+S_{\text{CS}}^{\text{IIB}},
\end{align}
explicitly,
\begin{align}
S_{\text{IIB}}^{(10)}=& \frac{1}{2\kappa_{10}^2}\int e^{-2\phi}\left(*\mathcal R+ 4\dif\phi\wedge*\dif\phi- \frac{1}{2}H_3\wedge *H_3\right)\nl
&-\frac{1}{4\kappa_{10}^2}\int \left(F_1\wedge* F_1 + F_3\wedge* F_3 + F_5\wedge*F_5\right)-\frac{1}{4\kappa_{10}^2}\int C_4\wedge H_3\wedge F_3
\label{Appendix:IIBAction}
\end{align}
with
\begin{align}
	H_3&=\dif B_2,\nl
	F_1&=\dif C_0,\nl
	F_3&=\dif C_2-C_0H_3,\nl
	F_5&=\dif C_4+B_2\wedge \dif C_2.
\end{align}
There is an obstacle though, since self-duality $F_5=*F_5$ is not implied by \eqref{Appendix:IIBAction} and thus must be imposed by an additional constraint on the solution. As a consequence, the above action does not have the same number of bosonic and fermionic degrees of freedom in thus is not supersymmetric. The equations of motion, though, are supersymmetric if the constraint is imposed.
\subsection{$\N=2$ supergravity in $d=4$ and special geometry}
\label{Appendix:String:Sugra}
As discussed in section \ref{Section:Compactification}, the four-dimensional theory resulting from compactification of Type II string theory on a Calabi-Yau manifold has $\N=2$ supersymmetry. Hence, the actions obtained from the Type II supergravities are $\N=2$ supergravities in four dimensions. We briefly review some of their features needed in the thesis - a comprehensive treatment can be found e.g. in \cite{Books:Supergravity}
\subsubsection{Multiplets of $\N=2$ supergravity}
An extensive treatment on this can be found in \cite{Books:Supergravity}. The (massless) field content of $\N=2$ SUGRA in $d=4$ is given by $n_V$ one-forms $V^A$ and complex scalars $z^A$, the metric $g$ and a one-form $V^0$ called the \emph{graviphoton} as well as $2n_H$ additional complex scalars $t^a,\xi^a$. These form three multiplets that are listed in table \ref{Table:Strings:SugraMultiplets}.
\begin{table}
	\begin{center}
		\begin{tabular}{|c|c|c|}
			\hline 
			\rowcolor{\BColor}
			\rule[-1.5ex]{0pt}{4.5ex} \textbf{Multiplet}& \textbf{(Massless) field content}  & \textbf{Number} \\
			\hline 
			\rule[-1.5ex]{0pt}{4.5ex} Gravity multiplet& $(g_{\mu\nu},A^0)$  & 1 \\
			\hline 
			\rule[-1.5ex]{0pt}{4.5ex} Gauge multiplets & $(V^A,z^A)$ & $n_V$ \\
			\hline
			\rule[-1.5ex]{0pt}{4.5ex} Hypermultiplets &$(t^a,\xi^a)$ & $n_H$\\
			\hline
		\end{tabular} 
	\end{center}
	\caption[$\N=2$ supergravity multiplets]{$\N=2$ supergravity multiplets in $d=4$.\label{Table:Strings:SugraMultiplets}}
\end{table}
The $\sigma$-model describing their self-interactions factorizes in a product of a \emph{special K\"ahler manifold} for the scalars $z^A$ in the gauge multiplets and a quaterionic manifold for the scalars in the hypermultiplets \cite{BlackHoles}. It is the former scalars that parameterize the matrices $\I$ and $\R$ in \ref{Eq:BPS:Action}. Since there are $n_V+1$ vectors, the special K\"ahler manifold is \emph{projective} and we can use $n_v+1$ holomorphic sections $Z^{\hat A}(z^A)$ as projective coordinates on the scalar manifold as well as $n_V+1$ dual vector field $\F_{\hat A}$ such that the sections $(Z^{\hat A},\F_{\hat A})$ form symplectic vectors. We will now elaborate on this in more detail. First, we will discuss \emph{special geometry} for global SUSY since it is a little easier before turning to SUGRA. For further discussion on this topic see \cite{PaltiPhD, Davidse} and especially \cite{Books:Supergravity}.
\subsubsection{Rigid special K\"ahler manifolds}
\label{Appendix:String:SpecialGeometry}
\begin{definition}[Rigid special K\"ahler geometry]
Let $(M,g)$ be a K\"ahler manifold with $\dim_{\mathbb C}M=n$ on which complex coordinates $z^a$ and $2n$ fields $X^A(z)$, $F_A(z)$ are defined where the latter transform as a vector of the symplecic group $\operatorname{Sp}(2n,\mathbb{R})$. 
Writing $V=(X^A,\F_A)$, we can define an inner product
\begin{align}
\langle V,W \rangle = V^T\begin{pmatrix}
0 && \1 \\
-\1 && 0
\end{pmatrix}W
\end{align}
which obviously is invariant under symplectic transformations. We call $M$ a \emph{rigid special K\"ahler manifold} if
\begin{align}
	g_{i\bar j} = i\langle\partial_i V,\partial_{\bar j}\bar V\rangle,\qquad 	\langle \partial_i V,\partial_j V\rangle =0.
	\label{Eq:String:SUGRA:RigidSpecialKähler}
\end{align}
\end{definition}
Note that $i\langle\partial_i V,\partial_{\bar j}\bar V\rangle = \partial_i\partial_{\bar j} i\langle V,\bar V\rangle$, i.e. the symplectic invariant $i\langle V,\bar V\rangle$ serves as a K\"ahler potential for the metric. Since both indices $i$ and $I$ range from $1,...,n$, the expression
\begin{align}
\partial_i X^I = \frac{\partial X^I}{\partial z^i}
\end{align}
is an $n\times n$ matrix. If it is invertible - which is the case if $g$ is positive definite \cite{Books:Supergravity} - then we can choose  the $X^I$ as new coordinates and multiply the second equation in \eqref{Eq:String:SUGRA:RigidSpecialKähler} with the inverse $\partial z^i/\partial X^K$ to get
\begin{align}
	\frac{\partial \F_K}{\partial z^j} = \frac{\partial \F_I}{\partial X^K}\frac{\partial X^I}{\partial z^j}.
	\label{String:SUGRA:RigidPrepotentialCondition}
\end{align}
This is the condition for $\F_I$ to possess a \emph{prepotential} $\F$ which is a holomorphic function of the coordinates $X$ with
\begin{align}
\F_I(X) =\frac{\partial\F(X)}{\partial X^I}.
\label{String:SUGRA:RigidPrepotential}
\end{align}
Since we are eventually interested in $\N=2, d=4$ SUGRA obtained from Type II string theory, we now turn to so-called \emph{projective special K\"ahler geometry}. Henceforth, we will drop the word ``projective''.
\subsubsection{Special K\"ahler manifolds}
Special geometry in its symplectic formulation is suited to describe the moduli spaces of Calabi-Yau threefolds. In \eqref{HolomThreeFormExp}, the fields $Z^\ha$ and $\F_\ha$, which are defined as periods of the holomorphic three-form\footnote{Note that unlike the main text, we are not careful about a Calabi-Yau reference volume factor $\V_0$ in this appendix. It can always be reinstalled by dimensional analysis, though.},
\begin{align}
Z^{\ha} = \int_{\mathcal A_{\ha}}\Omega,\qquad \F_{\ha}=\int_{\mathcal B^\ha}\Omega,
\end{align}
form a symplectic vector
\begin{align}
	v \coloneqq \binom{Z^\ha(z)}{\F_\ha(z)}
	\label{Eq:String:SUGRA:Vectorv}
\end{align}
with $\ha = 0,1,...,h^{2,1}$. Note that unlike the case of rigid special geometry, this is one degree of freedom more than the number of coordinates $z^a$. However, since a rescaling of $\Omega$ amounts to a rescaling of the periods $Z^\ha$ while the holomorphic three-form is defined only up to a complex rescaling, the $Z^\ha$ are \emph{projective},
\begin{align}
(Z^0,Z^1,...,)\cong (\lambda Z^0,\lambda Z^1,...).
\end{align}
Defining $h^{2,1}$ inhomogeneous coordinates\footnote{In a chart where $Z^0=0$, we can chose a $Z^b\neq 0$, since the $Z^\ha$ do not simultaneously vanish, and rename the indices.} $Z^\ha/Z^0$ - provided the matrix
\begin{align}
	\partial_a\left(\frac{Z^b}{Z^0}\right)
\end{align}
is invertible - we have the right number to chose
\begin{align}
z^a=\frac{Z^a}{Z^0}
\end{align}
or $Z^\ha = (1,z^a)$. Hence, we can write the periods $\F_\ha$ as functions\footnote{By a small abuse of notation, we use the same name for $F_\ha(z)$ and $F_\ha(Z)$ again.} of the coordinates $Z$. We arrive at a condition similar to the second equation of \eqref{Eq:String:SUGRA:RigidSpecialKähler} from
\begin{align}
	\int \partial_\ha \Omega\wedge\Omega &= \int(\alpha_\ha-\partial_\ha\F_{\hat b}\beta^{\hat b})\wedge (Z^\ha\alpha_\ha -\F_\ha\beta^\ha)\nl
	&= -\F_\ha +(\partial_\ha\F_{\hat b})Z^{\hat b},
\end{align}
where $\partial_\ha = \partial/\partial Z^\ha$, since this integral vanishes. This follows directly from the expansion
\begin{align}
\partial_{a}\Omega = k_a\Omega + i\eta_a
\end{align}
derived in \eqref{Calculations:Kodaira} and we conclude
\begin{align}
	\F_\ha = (\partial_\ha\F_{\hat b})Z^{\hat b}=\frac{1}{2}\partial_\ha\left(\F_{\hat b}Z^{\hat b}\right).
\end{align}
Thus, similar to \eqref{String:SUGRA:RigidPrepotential}, the periods $\F_I$ can be written in terms of a prepotential via
\begin{align}
	\F_\ha = \frac{\partial\F(Z)}{\partial Z^\ha},\quad\text{where}\quad\F(Z) = \frac{1}{2}\F_\ha Z^\ha.
\end{align}
Unlike the rigid case, the symplectic product $i\langle v,\bar v\rangle$ does not serve as a K\"ahler potential. Rather, we define
\begin{align}
e^{-K}\coloneqq -i\langle v,\bar v\rangle
\end{align}
and we will see in a minute, that $K$ is a K\"ahler potential for the metric on the complex structure moduli space. As mentioned several times, the holomorphic three-form is not uniquely defined but be can consider redefinitions
\begin{align}
	\Omega\rightarrow e^{f(Z)}\Omega
	\label{String:SUGRA:OmegaRescaling}
\end{align}
that should not change the physics. We see that under \eqref{String:SUGRA:OmegaRescaling}, the quantity $K$  transform as
\begin{align}
	K\rightarrow K - f(Z)-\bar{f}(\bar Z)
\end{align}
which is a K\"ahler transformation. Now, we are almost in place to write down an expression for the K\"ahler metric similar to the rigid case \eqref{Eq:String:SUGRA:RigidSpecialKähler}. First, we define the vectors
\begin{align}
	V\coloneqq e^{\frac{1}{2}K} v
\end{align}
as well as \emph{K\"ahler covariant derivatives}
\begin{align}
	\nabla_a Z^{\hat b} &\coloneqq \partial_a Z^{\hat b} + (\partial_aK)Z^{\hat b},\nl
	\bar \nabla_{\bar{a}} \bar Z^{\hat b} &\coloneqq \partial_{\bar a} \bar Z^{\hat b} + (\partial_{\bar a}K)\bar Z^{\hat b}.
	\label{Eq:String:SUGRA:KählerCovariant}
\end{align}
Note that these transform as
\begin{align}
\nabla_aZ^{\hat b}\rightarrow \nabla_a Z^{\hat b} e^{-f},\quad\bar\nabla_{\bar a}\bar Z^{\hat b}\rightarrow\bar\nabla_{\bar a}\bar Z^{\hat b} e^{-\bar f}
\end{align}
under a combined K\"ahler transformation
\begin{align}
	Z^{\hat a}\rightarrow Z^{\hat a}e^{-f},\quad \bar Z^{\hat a}\rightarrow \bar Z^{\hat a}e^{-\bar f},\quad K\rightarrow K+f+\bar f.
	\label{String:SUGRA:KählerTransformation}
\end{align}
Then, the expression
\begin{align}
	i\langle\nabla_a V,\bar \nabla_{\bar b}\bar V\rangle
	\label{String:SUGRA:KählerMetric}
\end{align}
where $\nabla_a V = e^{K}\nabla v$ is K\"ahler and symplectic covariant. Explicitly,
\begin{align}
		i\langle\nabla_a V,\bar \nabla_{\bar b}\bar V\rangle &= ie^{K}\left(\partial_a\partial_{\bar b}\langle v,\bar v\rangle + \partial_a\langle v,\bar v\rangle\partial_{\bar b}K + \partial_a K\partial_{\bar b}\langle v,\bar v\rangle + \partial_a K\partial_{\bar b}K\langle v,\bar v\rangle\right)\nl
		&= (\partial_a\partial_{\bar b}K + \partial_a K\partial_{\bar b}K) + (-\partial_a K\partial_{\bar b}K) +(-\partial_a K\partial_{\bar b}K) +(\partial_a K\partial_{\bar b}K)\nl
		&= \partial_a\partial_{\bar b} K,
		\label{Eq:String:SUGRA:KählerMetricSymplectic}
\end{align}
which means that it is the K\"ahler metric corresponding to the K\"ahler potential $K$. In the discussion of the complex structure moduli space in section \ref{Section:Compactification:ComplexStructureModuliSpace} we see that this is the K\"ahler metric $G_{a\hat b}$.

\section{Calculations}
\label{Appendix:Calculations}
\subsection{Compactification}
\subsubsection{Expansion of $\partial_{z^a}\Omega$}
\label{Calculations:Kodaira}
We derive eq. \eqref{Compactification:Kodaira}. First, we show that $\partial_{z^a}\Omega\in H^{3,0}+H^{2,1}$:
\begin{align}
	\partial_{z^a}\Omega &= \frac{1}{3!}(\partial_{z^a}\Omega_{ijk})\dif y^i\wedge\dif y^j\wedge\dif y^k + \frac{1}{2}\Omega_{ijk}\partial_{z^a}(\dif y^i)\wedge \dif y^j\wedge\dif y^k.
\end{align}
The first part is a $(3,0)$-form, while the second term is the wedge product of the derivative $\partial_{z^a}\dif y^v$ and a $(2,0)$-form. We will take a closer look at this derivative. Expanding
\begin{align}
	y^i(z^a+\delta z^a) = y^i(z^a) + \Lambda^i_a\delta z^a,
\end{align}
we find
\begin{align}
	\partial_{z^a}(\dif y^i)&= \dif\Lambda_a^i = \frac{\partial\Lambda_a^i}{\partial y^j}\dif y^j+\frac{\partial\Lambda_a^i}{\dif\bar y^{\bar j}}\dif\bar y^{\bar j}.
\end{align}
Thus, $\partial_{z^a}\dif y^v$ is composed of a $(1,0)$-form and a $(0,1)$-form, i.e. $\partial_{z^a}\Omega\in H^{3,0}+H^{2,1}$. We calculate the $(2,1)$-part
\begin{align}
	\frac{1}{2}\Omega_{ijk}\frac{\partial\Lambda_a^i}{\partial \bar y^{\bar l}}\dif \bar y^{\bar l}\wedge\dif y^j\wedge\dif y^k
	\label{Eq:Calculations:Omega}
\end{align}
by differentiating the Calabi-Yau metric:
\begin{align}
	0 = \partial_{z^a}(2g_{i\bar j}\dif y^i\dif\bar y^{\bar j}) =2\frac{\partial g_{i\bar j}}{\partial z^a}\dif y^i\dif \bar y^{\bar j} + 4g_{k\bar j}\frac{\partial\Lambda_a^k}{\partial y^i}\dif y^i\dif\bar y^{\bar j} 
\end{align}
and find
\begin{align}
g^{\bar i k}g_{k i}\frac{\partial g_{\bar i\bar j}}{\partial z^a} &= 	\frac{\partial g_{i\bar j}}{\partial z^a} = -2g_{k\bar j}\partial_i\Lambda_a^k.
\end{align}
From \eqref{Maths:KählerModuli:MetricExpansion}, we have
\begin{align}
	\frac{\partial g_{\bar i\bar j}}{\partial z^a}=-\frac{i}{\|\Omega\|^2}(\eta_a)_{\bar i kl}{\bar\Omega^{kl}}_{\bar j}
\end{align}
with $\partial_i \coloneqq \partial /\partial y^i$, that is,
\begin{align}
	\partial_i\Lambda_a^k = \frac{i}{2\|\Omega\|^2}(\eta_a)_{\bar ilm} \bar{\Omega}^{lmk}.
\end{align}
Thus, we confirm that \eqref{Eq:Calculations:Omega} is given by $\eta_a$, i.e.
\begin{align}
	\frac{\partial}{\partial z^a}\Omega = k_a\Omega + i\eta_a.
\end{align}
\subsubsection{K\"ahler moduli metric}
\label{Calculations:Compactification:KählerModuliMetric}
We derive the metric \eqref{KählerModuliMetric} from the Kähler moduli part of \eqref{Eq:Compactification:ModuliSpaceMetric}:
\begin{align}
2G_{AB}t^A\bar t^B &= -\frac{1}{2\V}\int \dif^6x \sqrt{g}g^{i\bar k}g^{j\bar l}\left(\delta g_{i\bar l}\delta g_{j\bar k}-\delta B_{i\bar l}\delta B_{j\bar k}\right)\nl
&= -\frac{1}{2\V}\int\dif^6x\sqrt{g}g^{i\bar k}g^{j\bar l}\left[(-i)^2v^A(\omega_A)_{i\bar l}v^B(\omega_B)_{j\bar k}-b^A(\omega_A)_{i\bar l}b^B(\omega_B)_{j\bar k}\right]\nl
&=  \frac{1}{2\V}\int\dif^6x\sqrt{g}g^{i\bar k}g^{j\bar l}(v^Av^B+b^Ab^B)(\omega_A)_{i\bar l}(\omega_B)_{j\bar k}\nl
&= \frac{1}{2\V}\int \omega_A\wedge*\omega_B t^A\bar t^B,
\end{align}
that is,
\begin{align}
G_{AB} = \frac{1}{4\V}\int \omega_A\wedge *\omega_B.
\end{align}
\subsubsection{K\"ahler moduli space prepotential}
\label{Calculations:Compactification:KählerModuliPrepotential}
It is shown that the cubic prepotential indeed is a prepotential for the K\"ahler moduli:
\begin{align}
i(\bar{X}^{\hat A}\mathcal{F}_{\hat{A}}-X^{\hat A}\bar{\mathcal F}_{\hat A}) &=\frac{-i}{3!}\K_{ABC}\left(-\bar X^0\frac{X^AX^BX^C}{{X^0}^2}+3\bar{X}^A\frac{X^BX^C}{X^0}-\text{c.c.}\right)\nonumber\\
&= \frac{i}{3!}\K_{ABC}\left(t^At^Bt^C + 3\bar t^At^Bt^C-\bar t^A\bar t^B \bar t^C - 3t^A\bar t^B\bar t^C\right)\nonumber\\
&= \frac{i}{3!}\K_{ABC}(t^A-\bar{t}^A)(t^B-\bar t^B)(t^C-\bar t^C)\nonumber\\ 
&=\frac{i}{3!}\K_{ABC}(2i)^3v^Av^Bv^C\nonumber\\
&=\frac{4}{3}\K
\end{align}
\subsection{Supersymmetric black holes}
\subsubsection{Matrix $\M$ in relation for central charge}
\label{Calculations:SUSYBlackHoles:MatrixM}
We show that
\begin{align}\frac{1}{2}
	\begin{pmatrix}
	p && q
	\end{pmatrix}\begin{pmatrix}
	-\left(\I+\R\I^{-1}\R\right)&&\R\I^{-1}\\
	\I^{-1}\R&& -\I^{-1}
	\end{pmatrix}\binom{p}{q} = -\frac{1}{2}(q-\N p)^T\I^{-1}(q-\bar\N p)
\end{align}
in order to prove \eqref{Eq:SUSYBlackHoles:CentralChargeRelation2}. After multiplying both sides by $\frac{1}{2}$, the left-hand side reads
\begin{align}
	-p^A(\I+\R\I^{-1}\R)_{AB}p^B + p^A{(\R\I^{-1})_A}^Bq_B+q_A{(\I^{-1}\R)^A}_Bp^B-q_A(\I^{-1})^{AB}q_B.
\end{align}
We find the same $q^T\I^{-1} q$-term on both sides  and the remaining terms on the right-hand side are
\begin{align}
	-\N_{AC}p^C(\I^{-1})^{AB}q_B &= -p^C(\R+i\I)_{AC}(\I^{-1})^{AB}q_B\nl
	&= -p^A{(\R\I^{-1})_A}^{B}q_B - ip^Aq_A,\nl
	-q_A(\I^{-1})^{AB}\bar\N_{BC}p^C &= -q_A(\I^{-1})^{AB}(\R-i\I)_{BC}p^C\nl
	&= -q_A{(\I^{-1}\R)^A}_Bp^B + ip^Aq_A,\nl
	\N_{AC}p^C(\I^{-1})^{AB}\bar\N_{BD}p^D &= p^C(\R+i\I)_{AC}(\I^{-1})^{AB}(\R-i\I)_{BD}p^D\nl
	&= -p^A(\I+\R\I^{-1}\R)_{AB}p^B,
\end{align}
which adds up to precisely what we found for the left-hand side.
\subsection{Gauge-coupling matrix}
\label{Appendix:MatrixM}
The matrices defined by the expansions
\begin{align}
*\alpha_{\hat a} = {A_{\hat a}}^{\hat b}\alpha_{\hat b}+B_{\hat a\hat b}\beta^{\hat b},\quad*\beta^{\hat a} = C^{\hat a\hat b}\alpha_{\hat b}+{D^{\hat a}}_{\hat b}\beta^{\hat b}
\end{align}
can be expressed in terms of the moduli. To do so, notice that the Hodge star acts on a $(3,0)$-form as $*\Omega = -i\Omega$ which lets us write \eqref{HolomThreeFormExp} as
\begin{align}
*\Omega &= Z^{\hat a}{*\alpha_{\hat a}}-\mathcal{F}_{\hat b}{*\beta^{\hat b}} \nonumber\\
&= Z^{\hat a}({A_{\hat a}}^{\hat b}\alpha_{\hat b} + B_{\hat a\hat b}\beta^{\hat b}) - \mathcal{F}_{\hat b}(C^{\hat b\hat c}\alpha_{\hat c}-{A_{\hat c}}^{\hat b}\beta^{\hat c})\nonumber\\
&= -i\Omega\nonumber\\
&= -i(Z^{\hat a}\alpha_{\hat a} - \mathcal{F}_{\hat b}\beta^{\hat b}),
\end{align}
and by equating coefficients
\begin{align}
Z^{\hat a}{A_{\hat a}}^{\hat b}-\mathcal{F}_{\hat a}C^{\hat a\hat b} &= -iZ^{\hat b},\nonumber\\
Z^{\hat a} B_{\hat a\hat b}+{A_{\hat b}}^{\hat a} &= i\mathcal{F}_{\hat b}.
\label{OmegaAC}
\end{align}
We found the expression \eqref{Eq:Compactification:Kodaira3} for $k_a$ in the expansion of $\Omega$, which we now calculate explicitly with help of the prepotential:
\begin{align}
k_a &=\partial_{z^a} \ln \left(\bar{Z}^{\hat b}\mathcal{F}_{\hat b}-Z^{\hat b}\bar{\mathcal{F}}_{\hat b}\right)\nl
    &=\frac{\bar{Z}^{\hat b}\mathcal{F}_{a\hat b}-\bar{\mathcal{F}}_{a}}{i\left(\bar{Z}^{\hat b}\mathcal{F}_{\hat b}-Z^{\hat b}\bar{\mathcal{F}}_{\hat b}\right)},\quad \mathcal{F}_{\hat a} = {\mathcal{F}}_{\hat a\hat b}Z^{\hat b}\nl
    &= \frac{Z^{\hat b}\mathcal{F}_{a\hat b} - \bar{\mathcal{F}}_{a\hat b}\bar{Z}^{\hat b}}{\bar{Z}^{\hat b}Z^{\hat c}\mathcal{F}_{\hat{b}\hat c}-Z^{\hat b}\bar{Z}^{\hat c}\bar{\mathcal{F}}_{\hat b\hat c}}\nl
    &= \frac{2\IM \mathcal{F}_{a\hat b}\bar Z^{\hat b}}{2\IM \mathcal{F}_{\hat b\hat c}\bar{Z}^{\hat{b}}Z^{\hat c}}.
\end{align}
Thus,
\begin{align}
\partial_{z^a}\Omega= \frac{1}{\IM \mathcal{F}_{\hat b\hat c}\bar{Z}^{\hat{b}}Z^{\hat c}}\IM \mathcal{F}_{a\hat b}\bar Z^{\hat b}\Omega + i\eta_a.
\end{align}
Since also $\partial_{z^0}\Omega\in H^{3,0}+H^{2,1}$, one can define $\eta_0$ via
\begin{align}
\partial_{Z^0}\Omega= \frac{1}{\IM \mathcal{F}_{\hat b\hat c}\bar{Z}^{\hat{b}}Z^{\hat c}}\IM \mathcal{F}_{0\hat b}\bar Z^{\hat b}\Omega + i\eta_0.
\end{align}
Using equation \eqref{OmegaAC}, it follows that
\begin{align}
\alpha_{\hat a}-\mathcal{F}_{\hat a\hat b}\beta^{\hat b} &=\partial_{Z^{\hat a}}\Omega =\frac{1}{\IM \mathcal{F}_{\hat b\hat c}\bar{Z}^{\hat{b}}Z^{\hat c}}\IM \mathcal{F}_{\hat a\hat b}\bar Z^{\hat b}\Omega + i\eta_{\hat a}\nl
&= \frac{1}{\IM \mathcal{F}_{\hat b\hat c}\bar{Z}^{\hat{b}}Z^{\hat c}}\IM \mathcal{F}_{a\hat b}\bar Z^{\hat b}(Z^{\hat c}\alpha_{\hat c}-\mathcal{F}_{\hat c}\beta^{\hat c})+i\eta_{\hat a},
\end{align}
i.e.
\begin{align}
i\eta_{\hat a} =& \left(\delta^{\hat c}_{\hat a}- \frac{\IM\mathcal{F}_{\hat a\hat b}\bar Z^{\hat b} Z^{\hat c}}{\IM \mathcal{F}_{\hat b\hat d}\bar{Z}^{\hat{b}}Z^{\hat d}}\right)\alpha_{\hat c} - \left(\mathcal{F}_{\hat a\hat c}-\frac{\IM\mathcal{F}_{\hat a\hat b}\bar{Z}^{\hat b}\mathcal{F}_{\hat c}}{\IM\mathcal{F}_{\hat b\hat d}\bar{Z}^{\hat b}Z^{\hat d}}\right)\beta^{\hat c}\nl
=& *\eta_{\hat a}\nl
=& -i\left(\delta^{\hat c}_{\hat a}- \frac{\IM\mathcal{F}_{\hat a\hat b}\bar Z^{\hat b} Z^{\hat c}}{\IM \mathcal{F}_{\hat b\hat d}\bar{Z}^{\hat{b}}Z^{\hat d}}\right)(A_{\hat c}^{\hat e}\alpha_{\hat e}+B_{\hat c\hat e}\beta^{\hat e})\nl
&+i\left(\mathcal{F}_{\hat a\hat c}-\frac{\IM\mathcal{F}_{\hat a\hat b}\bar{Z}^{\hat b}\mathcal{F}_{\hat c}}{\IM\mathcal{F}_{\hat b\hat d}\bar{Z}^{\hat b}Z^{\hat d}}\right)(C^{\hat c\hat e}\alpha_{\hat e}-A_{\hat e}^{\hat c}\beta^{\hat e})
\end{align}
from which we find
\begin{align}
\left(\delta^{\hat c}_{\hat a}- \frac{\IM\mathcal{F}_{\hat a\hat b}\bar Z^{\hat b} Z^{\hat c}}{\IM \mathcal{F}_{\hat b\hat d}\bar{Z}^{\hat{b}}Z^{\hat d}}\right)(\delta^{\hat e}_{\hat c} +iA_{\hat c}^{\hat e})&= i\left(\mathcal{F}_{\hat a\hat c}-\frac{\IM\mathcal{F}_{\hat a\hat b}\bar{Z}^{\hat b}\mathcal{F}_{\hat c}}{\IM\mathcal{F}_{\hat b\hat d}\bar{Z}^{\hat b}Z^{\hat d}}\right)C^{\hat c\hat e}\nl
i(-A_{\hat c}^{\hat e}Z^{\hat c}+C^{\hat c\hat e}\mathcal{F}_{\hat c})\frac{\IM\mathcal{F}_{\hat a\hat b}\bar{Z}^{\hat b}}{\IM\mathcal{F}_{\hat b\hat d}\bar{Z}^{\hat b}Z^{\hat d}} + \delta_{\hat a}^{\hat e} - \frac{\IM\mathcal{F}_{\hat a\hat b}\bar{Z}^{\hat b}Z^{\hat e}_{\hat e}}{\IM\mathcal{F}_{\hat b\hat d}\bar{Z}^{\hat b}Z^{\hat d}}&=-i(A_{\hat a}^{\hat e} -\mathcal{F}_{\hat a\hat c}C^{\hat c\hat e}).
\end{align}
Using \eqref{OmegaAC} again, this is
\begin{align}
-Z^{\hat e}\frac{\IM\mathcal{F}_{\hat a\hat b}\bar{Z}^{\hat b}}{\IM\mathcal{F}_{\hat b\hat d}\bar{Z}^{\hat b}Z^{\hat d}} + \delta_{\hat a}^{\hat e} - \frac{\IM\mathcal{F}_{\hat a\hat b}\bar{Z}^{\hat b}Z^{\hat e}}{\IM\mathcal{F}_{\hat b\hat d}\bar{Z}^{\hat b}Z^{\hat d}}&=-i(A_{\hat a}^{\hat e} -\mathcal{F}_{\hat a\hat c}C^{\hat c\hat e}),
\end{align}
i.e.
\begin{align}
A_{\hat a}^{\hat b}-\mathcal{F}_{\hat a\hat c}C^{\hat c\hat b} = i\delta_{\hat a}^{\hat b} - \frac{2i}{\IM\mathcal{F}_{\hat c\hat d}\bar{Z}^{\hat c}Z^{\hat d}}\IM\mathcal F_{\hat a\hat c}\bar{Z}^{\hat c}Z^{\hat b}
\end{align}
and separating real and imaginary part
\begin{align}
A_{\hat a}^{\hat b} = \RE\mathcal{F}_{\hat a\hat c}C^{\hat c\hat b}+\delta_{\hat a}^{\hat b}+ i\IM\mathcal{F}_{\hat a\hat c}\left(C^{\hat c\hat b}-\frac{2}{\IM\mathcal{F}_{\hat c\hat d}\bar{Z}^{\hat c}Z^{\hat d}}\bar{Z}^{\hat c}Z^{\hat b}\right).
\end{align}
Likewise, one finds
\begin{align}
B_{\hat a\hat b} + \mathcal{F}_{\hat a\hat c}A_{\hat b}^{\hat c} &= -i\mathcal{F}_{\hat a\hat b}+ \frac{2i}{\IM\mathcal F_{\hat c\hat d}\bar Z^{\hat c}Z^{\hat d}}\IM\mathcal{F}_{\hat a\hat c}\hat{Z}^{\hat c}\mathcal{F}_{\hat b}
\end{align}
Introducing a matrix
\begin{empheq}[box=\mymath]{equation}
\mathcal M_{\hat a\hat b}:= \bar{\mathcal{F}}_{\hat a\hat b}+\frac{2i}{Z^{\hat a}\IM\mathcal{F}_{\hat a\hat b}Z^{\hat b}}\IM\mathcal{F}_{\hat a\hat c}Z^{\hat c}\IM\mathcal{F}_{\hat b\hat d}Z^{\hat d},
\label{Eq:Calculations:MatrixM}
\end{empheq}
we can write
\begin{align}
A &= (\RE \M)(\IM \M)^{-1},\nonumber\\
B &= -(\IM \M) - (\RE \M)(\IM \M)^{-1}(\RE \M),\nonumber\\
C &= (\IM \M)^{-1}.
\end{align}
We derive another identity which we use in the thesis:
\begin{align}
	\M_{\hat a\hat b}Z^{\hat c} &= \bar\F_{\ha\hat b}Z^{\hat b}+2i\IM\F_{\hat a\hat c}Z^{\hat c}\nl
	&= \F_{\ha\hat b}Z^{\hat b},
\end{align}
i.e.
\begin{empheq}[box=\mymath]{equation}
\F_\ha =\M_{\hat a\hat b}Z^{\hat b}.
\label{Eq:Calculations:MZ}
\end{empheq}
For a prepotential of the form
\begin{align}
\mathcal{F} \coloneqq -\frac{1}{3!}\K_{abc}\frac{Z^aZ^bZ^c}{Z^0},\qquad \mathcal{F}_{\ha}\coloneqq\partial_{Z^{\ha}}\mathcal F,\qquad Z = (1,t^A=b^A+iv^A),
\end{align}
like the one for the K\"ahler moduli space in section \ref{Section:KählerModuliSpace}, the matrix $\M$ can be brought to an explicit form which we will now to. We have
\begin{align}
\mathcal{F}_{00} =& -\frac{1}{3!}\partial_0\K_{abc}\frac{Z^aZ^bZ^c}{-(Z^0)^2}\nl
=& -\frac{1}{3}\K_{abc}Z^aZ^bZ^c,\nl
\mathcal F_{a0} =& \frac{1}{3!}\K_{bcd}\partial_a \frac{Z^bZ^cZ^d}{(Z^0)^2}\nl
=& \frac{1}{2}\K_{abc}Z^bZ^c,\nl
\mathcal{F}_{ab} =& -\K_{abc}Z^c
\end{align}
and
\begin{align}
\IM\mathcal{F}_{ab} =& -\K_{abc}v^c = -\K_{ab},\nl
\IM\mathcal{F}_{00} =& -\frac{1}{3}\IM\left[(b^a+iv^a)(b^b+iv^b)(b^c+iv^c)\right]\nl
=& -\frac{1}{3}\K_{abc}(v^ab^bb^c+v^bb^ab^c+v^cb^ab^b -v^av^bv^c)\nl
=& - \K_{ab}b^ab^b + \frac{1}{3}\K\quad\text{with }\K = \K_av^a = \K_{ab}v^av^b=\K_{abc}v^av^bv^c,\nl
\IM\mathcal{F}_{a0} =& \frac{1}{2}\K_{abc}\IM\left[(b^b+iv^b)(b^c+iv^c)\right]\nl
=& \frac{1}{2}\K_{abc}(b^bv^c+v^bb^c)\nl
=& \K_{ab}b^b.
\end{align}
Thus,
\begin{align}
Z^{\hat a}\IM\mathcal{F}_{\hat a\hat b}Z^{\hat b} =& \left(-\K_{ab}b^ab^b + \frac{1}{3}\K\right) + 2\left(\K_{ab}b^a\right)Z^b+ \left(-\K_{ab}Z^aZ^b \right).
\end{align}
With
\begin{align}
\K_{ab}Z^aZ^b =& \K_{ab}(b^a+iv^a)(b^b+iv^b)\nl
=& \K_{ab}b^a b^b + 2i\K_{a}b^a-\K,\nl
\K_{ab}b^aZ^b =& \K_{ab}b^ab^b+i\K_ab^a,
\end{align}
we have
\begin{align}
Z^{\hat a}\IM\mathcal{F}_{\hat a\hat b}Z^{\hat b} =& \K_{ab}b^ab^b+ \frac{1}{3}\K+2i\K_ab^a -\K_{ab}b^ab^b-2i\K_{a}b^a+\K\nl
=& \frac{4}{3}\K.
\end{align}
Thus,
\begin{align}
\M_{00} =& -\frac{1}{3}\K_{abc}(b^a-iv^a)(b^b-iv^b)(b^c-iv^c) + \frac{2i}{\frac{4}{3}\K}(\IM\mathcal{F}_{0\hat b}Z^{\hat b})^2
\label{M00}
\end{align}
where
\begin{align}
i(\IM\mathcal{F}_{0\hat b}Z^{\hat b})^2 =& i(-\K_{ab}b^ab^b+\frac{1}{3}\K + \K_{ab}b^aZ^b)^2 \nl
=& i\left(-\K_{ab}b^ab^b+\K_{ab}b^ab^b+\frac{1}{3}\K+i\K_{ab}b^av^b\right)^2\nl
=& i\left( \frac{1}{3}\K+i\K_ab^a\right)^2\nl
=& \left(\frac{1}{3}\K+i\K_ab^a\right)\left(i\frac{1}{3}\K-\K_ab^a\right).
\end{align}
The imaginary part is
\begin{align}
\IM\M_{00} =& -\frac{1}{3}\K_{abc}(-3v^ab^bb^c+v^av^bv^c) + \frac{3}{2\K}\left[\left(\frac{1}{3}\K\right)^2-\K_ab^a\K_bb^b\right]\nl
=& \K_{ab}b^ab^b-\frac{1}{3}\K+\frac{\K}{6}-\frac{3}{2\K}\K_ab^a\K_bb^b\nl
=&-\frac{\K}{6}\left(1-\frac{6}{\K}\K_{ab}+\frac{9}{\K^2}\K_a\K_b\right)b^ab^b\nl
=&-\frac{\K}{6}\left(1+4G_{ab}b^ab^b\right).
\end{align}
The $0a$-term is
\begin{align}
\M_{0a} = \bar{\mathcal{F}}_{0a} + \frac{3i}{2\K}(\IM\mathcal{F}_{a\hat b}Z^{\hat b})(\IM\mathcal{F}_{0\hat c}Z^{\hat c})
\label{M0a}
\end{align}
with
\begin{align}
(\IM\mathcal{F}_{a\hat b}Z^{\hat b}) =& \IM\mathcal{F}_{a0} + \IM\mathcal{F}_{ab}Z^b\nl
=& \K_{ab}b^b - \K_{ab}(b^b+iv^b)\nl
=& -i\K_a,
\end{align}
i.e.
\begin{align}
\IM\M_{0a} &= -\K_{ab}b^b +\IM\left[\frac{3i}{2\K}(-i\K_a)\left(\frac{1}{3}\K+i\K_bb^b\right)\right]\nl
&= -\K_{ab}b^b+\frac{3}{2\K}\K_a\K_bb^b\nl
&= -\frac{\K}{6}\left(-4\frac{-3}{2\K}\right)\left(\K_{ab}-\frac{3}{2\K}\K_a\K_b\right)b^b\nl
&= -\frac{\K}{6}(-4G_{ab}b^b).
\end{align}
The $ab$-term is
\begin{align}
\M_{ab} &= \overline{\mathcal{F}}_{ab} + \frac{3i}{2\K}\IM\mathcal{F}_{a\hat c}Z^{\hat c}\IM\mathcal{F}_{b\hat d}Z^{\hat d}\nl
&= -\K_{abc}(b^c-iv^c) + \frac{3i}{2\K}(-i\K_a)(-i\K_b)
\end{align}
and has imaginary part
\begin{align}
\IM\M_{ab} &= \K_{ab}-\frac{3}{2\K}\K_a\K_b\nl
&= -\frac{\K}{6}4\left(-\frac{3}{2\K}\right)\left(\K_{ab}-\frac{3}{2\K}\K_a\K_b\right)\nl
&= -\frac{\K}{6}4G_{ab}.
\label{Mab}
\end{align}
Gathering the terms \eqref{M00}, \eqref{M0a} and \eqref{Mab}, this can be written in matrix notation as
\begin{align}
\IM\M_{\hat a\hat b} = -\frac{\K}{6}\begin{pmatrix}
1+4G_{ab}b^ab^b && -4G_{ab}b^b\\
-4G_{ab}b^b && 4G_{ab}
\end{pmatrix}.
\end{align}
Now, we'll compute
\begin{align}
\RE \mathcal M_{\hat a\hat b}= \RE\left(\bar{\mathcal{F}}_{\hat a\hat b}+\frac{2i}{Z^{\hat a}\IM\mathcal{F}_{\hat a\hat b}Z^{\hat b}}\IM\mathcal{F}_{\hat a\hat c}Z^{\hat c}\IM\mathcal{F}_{\hat b\hat d}Z^{\hat d}\right).
\end{align}
To do so, recall that we already found
\begin{align}
\M_{00} =& -\frac{1}{3}\K_{abc}(b^a-iv^a)(b^b-iv^b)(b^c-iv^c) + \frac{2i}{\frac{4}{3}\K}(\IM\mathcal{F}_{0\hat b}Z^{\hat b})^2
\label{M00}
\end{align}
where
\begin{align}
i(\IM\mathcal{F}_{0\hat b}Z^{\hat b})^2 =\left(\frac{1}{3}\K+i\K_ab^a\right)\left(i\frac{1}{3}\K-\K_ab^a\right).
\end{align}
Thus,
\begin{align}
\RE \M_{00} &=-\frac{1}{3}\K_{abc}b^ab^bb^c+\K_ab^a+\frac{3}{2\K}\left(-\frac{1}{3}\K\K_ab^a-\frac{1}{3}\K_ab^a\K\right)\nl
&= -\frac{1}{3}\K_{abc}b^ab^bb^c + \K_ab^a-\K_ab^a\nl
&= -\frac{1}{3}\K_{abc}b^ab^bb^c.
\end{align}
In order to compute the real part of
\begin{align}
\M_{0a} = \bar{\mathcal{F}}_{0a} + \frac{3i}{2\K}(\IM\mathcal{F}_{a\hat b}Z^{\hat b})(\IM\mathcal{F}_{0\hat c}Z^{\hat c}),
\end{align}
we note that
\begin{align}
\RE\mathcal F_{a0} &= \frac{1}{2}\RE\left[\K_{abc}(b^b+iv^b)(b^c+iv^c)\right]\nl
&= \frac{1}{2}\K_{abc}(b^bb^c-v^bv^c)\nl
&= \frac{1}{2}\K_{abc}b^bb^c-\frac{1}{2}\K_a
\end{align}
and recall
\begin{align}
(\IM\mathcal{F}_{a\hat b}Z^{\hat b}) = -i\K_a.
\end{align}
Thus,
\begin{align}
\RE\M_{a0} &= \frac{1}{2}\K_{abc}b^bb^c-\frac{1}{2}\K_a+\frac{3}{2\K}\RE\left[i(-i\K_a)\left(\frac{1}{3}\K+i\K_bb^b\right) \right]\nl
&= \frac{1}{2}\K_{abc}b^bb^c-\frac{1}{2}\K_a+\frac{3}{2\K}\K_a\frac{1}{3}\K\nl
&= \frac{1}{2}\K_{abc}b^bb^c.
\end{align}
The $ab$-term is
\begin{align}
\M_{ab} &= \overline{\mathcal{F}}_{ab} + \frac{3i}{2\K}\IM\mathcal{F}_{a\hat c}Z^{\hat c}\IM\mathcal{F}_{b\hat d}Z^{\hat d}\nl
&= -\K_{abc}(b^c-iv^c) + \frac{3i}{2\K}(-i\K_a)(-i\K_b)
\end{align}
and has real part
\begin{align}
\RE\M_{ab} = -\K_{abc}b^c.
\end{align}
Gathering all terms, we have
\begin{align}
\RE\M_{\hat a\hat b} = \begin{pmatrix}
-\frac{1}{3}\K_{cde}b^cb^db^e && \frac{1}{2}\K_{acd}b^cb^d \\ \frac{1}{2}\K_{acd}b^cb^d && -\K_{abc}b^c
\end{pmatrix}.
\end{align}
After this lengthy calculations, we finally arrive at the explicit expression
\begin{empheq}[box={\mymath}]{equation}
\M = \begin{pmatrix}
-\frac{1}{3}\K_{cde}b^cb^db^e && \frac{1}{2}\K_{acd}b^cb^d \\ \frac{1}{2}\K_{acd}b^cb^d && -\K_{abc}b^c
\end{pmatrix} + i\frac{-\K}{6}\begin{pmatrix}
1+4G_{ab}b^ab^b && -4G_{ab}b^b\\
-4G_{ab}b^b && 4G_{ab}
\end{pmatrix}.
\label{Eq:MatrixMExplicit}
\end{empheq}

\leavevmode\thispagestyle{empty}
\newpage



\printbibliography[heading=bibintoc]

@book{Books:QFT,
	Author = {Matthew D. Schwartz},
	Publisher = {Cambridge Univ. Press},
	Title = {Quantum Field Theory and the Standard Model},
	Year = {2015}}

@phdthesis{Gurrieri,
	Archiveprefix = {arXiv},
	Author = {Gurrieri, Sebastien},
	Eprint = {hep-th/0408044},
	Primaryclass = {hep-th},
	School = {Marseille, CPT},
	Slaccitation = {%%CITATION = HEP-TH/0408044;%%},
	Title = {{N=2 and N=4 supergravities as compactifications from string theories in 10 dimensions}},
	Year = {2003}}

@article{Grimm,
	Archiveprefix = {arXiv},
	Author = {Grimm, Thomas W.},
	Doi = {10.1002/prop.200510253},
	Eprint = {hep-th/0507153},
	Journal = {Fortsch. Phys.},
	Pages = {1179-1271},
	Primaryclass = {hep-th},
	Reportnumber = {DESY-THESIS-2005-020, ZMP-HH-05-14},
	Slaccitation = {%%CITATION = HEP-TH/0507153;%%},
	Title = {{The Effective action of type II Calabi-Yau orientifolds}},
	Volume = {53},
	Year = {2005}}

@article{Grimm2,
	Archiveprefix = {arXiv},
	Author = {Grimm, Thomas W. and Louis, Jan},
	Doi = {10.1016/j.nuclphysb.2004.08.005},
	Eprint = {hep-th/0403067},
	Journal = {Nucl. Phys.},
	Pages = {387-426},
	Primaryclass = {hep-th},
	Reportnumber = {LPTENS-04-14},
	Slaccitation = {%%CITATION = HEP-TH/0403067;%%},
	Title = {{The Effective action of N = 1 Calabi-Yau orientifolds}},
	Volume = {B699},
	Year = {2004}}

@article{Micu,
	Archiveprefix = {arXiv},
	Author = {Louis, Jan and Micu, Andrei},
	Doi = {10.1016/S0550-3213(02)00338-3},
	Eprint = {hep-th/0202168},
	Journal = {Nucl. Phys.},
	Pages = {395-431},
	Primaryclass = {hep-th},
	Slaccitation = {%%CITATION = HEP-TH/0202168;%%},
	Title = {{Type 2 theories compactified on Calabi-Yau threefolds in the presence of background fluxes}},
	Volume = {B635},
	Year = {2002}}

@article{Grana,
	Archiveprefix = {arXiv},
	Author = {Grana, Mariana},
	Doi = {10.1016/j.physrep.2005.10.008},
	Eprint = {hep-th/0509003},
	Journal = {Phys. Rept.},
	Pages = {91-158},
	Primaryclass = {hep-th},
	Reportnumber = {LPTENS-05-26, CPHT-RR-049-0805},
	Slaccitation = {%%CITATION = HEP-TH/0509003;%%},
	Title = {{Flux compactifications in string theory: A Comprehensive review}},
	Volume = {423},
	Year = {2006}}

@article{Ceresole,
	Archiveprefix = {arXiv},
	Author = {Ceresole, Anna and D'Auria, R. and Ferrara, S.},
	Booktitle = {{S duality and mirror symmetry. Proceedings, Conference, Trieste, Italy, June 5-9, 1995}},
	Doi = {10.1016/0920-5632(96)00008-4},
	Eprint = {hep-th/9509160},
	Journal = {Nucl. Phys. Proc. Suppl.},
	Pages = {67-74},
	Primaryclass = {hep-th},
	Reportnumber = {POLFIS-TH-10-95, CERN-TH-95-244},
	Slaccitation = {%%CITATION = HEP-TH/9509160;%%},
	Title = {{The Symplectic structure of N=2 supergravity and its central extension}},
	Volume = {46},
	Year = {1996}}

@phdthesis{Davidse,
	Archiveprefix = {arXiv},
	Author = {Davidse, Marijn},
	Eprint = {hep-th/0603073},
	Primaryclass = {hep-th},
	Reportnumber = {ITP-UU-06-10, SPIN-06-08},
	School = {Utrecht U.},
	Slaccitation = {%%CITATION = HEP-TH/0603073;%%},
	Title = {{Membrane and fivebrane instantons and quaternionic geometry}},
	Url = {https://inspirehep.net/record/712090/files/arXiv:hep-th_0603073.pdf},
	Year = {2006}}

@article{MirrorSymmetry,
	Archiveprefix = {arXiv},
	Author = {Hosono, S. and Klemm, A. and Theisen, S.},
	Booktitle = {{Proceedings, 3rd Baltic Rim Student Seminar: Integrable Models and Strings: Helsinki, Finland, September 13-17, 1993}},
	Doi = {10.1007/3-540-58453-6_13},
	Eprint = {hep-th/9403096},
	Journal = {Lect. Notes Phys.},
	Pages = {235-280},
	Primaryclass = {hep-th},
	Reportnumber = {HUTMP-94-01, LMU-TPW-94-02},
	Slaccitation = {%%CITATION = HEP-TH/9403096;%%},
	Title = {{Lectures on mirror symmetry}},
	Volume = {436},
	Year = {1994}}

@article{WGC,
	Archiveprefix = {arXiv},
	Author = {Arkani-Hamed, Nima and Motl, Lubo\v{s} and Nicolis, Alberto and Vafa, Cumrun},
	Doi = {10.1088/1126-6708/2007/06/060},
	Eprint = {hep-th/0601001},
	Journal = {JHEP},
	Pages = {060},
	Primaryclass = {hep-th},
	Reportnumber = {HUTP-05-A0057},
	Slaccitation = {%%CITATION = HEP-TH/0601001;%%},
	Title = {{The String landscape, black holes and gravity as the weakest force}},
	Volume = {06},
	Year = {2007}}

@article{WGC:Palti1,
	Archiveprefix = {arXiv},
	Author = {Palti, Eran},
	Doi = {10.1007/JHEP08(2017)034},
	Eprint = {1705.04328},
	Journal = {JHEP},
	Pages = {034},
	Primaryclass = {hep-th},
	Slaccitation = {%%CITATION = ARXIV:1705.04328;%%},
	Title = {{The Weak Gravity Conjecture and Scalar Fields}},
	Volume = {08},
	Year = {2017}}

@article{WGC:GaugeGravity,
	Archiveprefix = {arXiv},
	Author = {Banks, Tom and Johnson, Matt and Shomer, Assaf},
	Doi = {10.1088/1126-6708/2006/09/049},
	Eprint = {hep-th/0606277},
	Journal = {JHEP},
	Pages = {049},
	Primaryclass = {hep-th},
	Reportnumber = {RU-06-08, SCIPP-06-07},
	Slaccitation = {%%CITATION = HEP-TH/0606277;%%},
	Title = {{A Note on Gauge Theories Coupled to Gravity}},
	Volume = {09},
	Year = {2006}}

@article{WGC:GlobalSymmetries,
	Archiveprefix = {arXiv},
	Author = {Banks, Tom and Seiberg, Nathan},
	Doi = {10.1103/PhysRevD.83.084019},
	Eprint = {1011.5120},
	Journal = {Phys. Rev.},
	Pages = {084019},
	Primaryclass = {hep-th},
	Slaccitation = {%%CITATION = ARXIV:1011.5120;%%},
	Title = {{Symmetries and Strings in Field Theory and Gravity}},
	Volume = {D83},
	Year = {2011}}

@article{WGC:Fencing,
	Archiveprefix = {arXiv},
	Author = {Brown, Jon and Cottrell, William and Shiu, Gary and Soler, Pablo},
	Doi = {10.1007/JHEP10(2015)023},
	Eprint = {1503.04783},
	Journal = {JHEP},
	Pages = {023},
	Primaryclass = {hep-th},
	Reportnumber = {MAD-TH-15-04},
	Slaccitation = {%%CITATION = ARXIV:1503.04783;%%},
	Title = {{Fencing in the Swampland: Quantum Gravity Constraints on Large Field Inflation}},
	Volume = {10},
	Year = {2015}}

@article{WGC:Vafa,
	Archiveprefix = {arXiv},
	Author = {Vafa, Cumrun},
	Eprint = {hep-th/0509212},
	Primaryclass = {hep-th},
	Reportnumber = {HUTP-05-A043},
	Slaccitation = {%%CITATION = HEP-TH/0509212;%%},
	Title = {{The String landscape and the swampland}},
	Year = {2005}}

@article{WGC:Vafa2,
	Archiveprefix = {arXiv},
	Author = {Ooguri, Hirosi and Vafa, Cumrun},
	Doi = {10.1016/j.nuclphysb.2006.10.033},
	Eprint = {hep-th/0605264},
	Journal = {Nucl. Phys.},
	Pages = {21-33},
	Primaryclass = {hep-th},
	Reportnumber = {CALT-68-2600, HUTP-06-A017},
	Slaccitation = {%%CITATION = HEP-TH/0605264;%%},
	Title = {{On the Geometry of the String Landscape and the Swampland}},
	Volume = {B766},
	Year = {2007}}

@article{WGC:Palti2,
	Archiveprefix = {arXiv},
	Author = {Grimm, Thomas W. and Palti, Eran and Valenzuela, Irene},
	Eprint = {1802.08264},
	Primaryclass = {hep-th},
	Slaccitation = {%%CITATION = ARXIV:1802.08264;%%},
	Title = {{Infinite Distances in Field Space and Massless Towers of States}},
	Year = {2018}}

@article{WGC:Palti3,
	Archiveprefix = {arXiv},
	Author = {L{\"u}st, Dieter and Palti, Eran},
	Doi = {10.1007/JHEP02(2018)040},
	Eprint = {1709.01790},
	Journal = {JHEP},
	Pages = {040},
	Primaryclass = {hep-th},
	Slaccitation = {%%CITATION = ARXIV:1709.01790;%%},
	Title = {{Scalar Fields, Hierarchical UV/IR Mixing and The Weak Gravity Conjecture}},
	Volume = {02},
	Year = {2018}}

@article{WGC:Pablo,
	Archiveprefix = {arXiv},
	Author = {Shiu, Gary and Soler, Pablo and Cottrell, William},
	Eprint = {1611.06270},
	Primaryclass = {hep-th},
	Reportnumber = {MAD-TH-16-07},
	Slaccitation = {%%CITATION = ARXIV:1611.06270;%%},
	Title = {{Weak Gravity Conjecture and Extremal Black Holes}},
	Year = {2016}}

@article{WGC:Naturalness,
	Archiveprefix = {arXiv},
	Author = {Cheung, Clifford and Remmen, Grant N.},
	Doi = {10.1103/PhysRevLett.113.051601},
	Eprint = {1402.2287},
	Journal = {Phys. Rev. Lett.},
	Pages = {051601},
	Primaryclass = {hep-ph},
	Reportnumber = {CALT-68-2879},
	Slaccitation = {%%CITATION = ARXIV:1402.2287;%%},
	Title = {{Naturalness and the Weak Gravity Conjecture}},
	Volume = {113},
	Year = {2014}}

@article{WGC:Inflation,
	Archiveprefix = {arXiv},
	Author = {Rudelius, Tom},
	Doi = {10.1088/1475-7516/2015/09/020, 10.1088/1475-7516/2015/9/020},
	Eprint = {1503.00795},
	Journal = {JCAP},
	Number = {09},
	Pages = {020},
	Primaryclass = {hep-th},
	Slaccitation = {%%CITATION = ARXIV:1503.00795;%%},
	Title = {{Constraints on Axion Inflation from the Weak Gravity Conjecture}},
	Volume = {1509},
	Year = {2015}}

@article{WGC:Inflation2,
	Archiveprefix = {arXiv},
	Author = {Junghans, Daniel},
	Doi = {10.1007/JHEP02(2016)128},
	Eprint = {1504.03566},
	Journal = {JHEP},
	Pages = {128},
	Primaryclass = {hep-th},
	Reportnumber = {LMU-ASC-21-15},
	Slaccitation = {%%CITATION = ARXIV:1504.03566;%%},
	Title = {{Large-Field Inflation with Multiple Axions and the Weak Gravity Conjecture}},
	Volume = {02},
	Year = {2016}}

@article{Heidenreich,
	Archiveprefix = {arXiv},
	Author = {Heidenreich, Ben and Reece, Matthew and Rudelius, Tom},
	Doi = {10.1007/JHEP02(2016)140},
	Eprint = {1509.06374},
	Journal = {JHEP},
	Pages = {140},
	Primaryclass = {hep-th},
	Slaccitation = {%%CITATION = ARXIV:1509.06374;%%},
	Title = {{Sharpening the Weak Gravity Conjecture with Dimensional Reduction}},
	Volume = {02},
	Year = {2016}}

@article{BeckerBecker,
	Archiveprefix = {arXiv},
	Author = {Becker, Katrin and Becker, Melanie and Strominger, Andrew},
	Doi = {10.1016/0550-3213(95)00487-1},
	Eprint = {hep-th/9507158},
	Journal = {Nucl. Phys.},
	Pages = {130-152},
	Primaryclass = {hep-th},
	Reportnumber = {NSF-ITP-95-62},
	Slaccitation = {%%CITATION = HEP-TH/9507158;%%},
	Title = {{Five-branes, membranes and nonperturbative string theory}},
	Volume = {B456},
	Year = {1995}}

@book{Polchinski1,
	Author = {Polchinski, J.},
	Isbn = {9780521633031},
	Publisher = {Cambridge University Press},
	Slaccitation = {%%CITATION = INSPIRE-487240;%%},
	Title = {{String theory. Vol. 1: An introduction to the bosonic string}},
	Year = {2007}}

@book{Polchinski2,
	Author = {Polchinski, J.},
	Isbn = {9780521672283},
	Publisher = {Cambridge University Press},
	Slaccitation = {%%CITATION = INSPIRE-487241;%%},
	Title = {{String theory. Vol. 2: Superstring theory and beyond}},
	Year = {2007}}

@book{Ibanez,
	Author = {Ibanez, Luis E. and Uranga, Angel M.},
	Isbn = {9781139227421},
	Publisher = {Cambridge University Press},
	Slaccitation = {%%CITATION = INSPIRE-1112474;%%},
	Title = {{String theory and particle physics: An introduction to string phenomenology}},
	Year = {2012}}

@article{Supersymmetry,
	Archiveprefix = {arXiv},
	Author = {Bilal, Adel},
	Eprint = {hep-th/0101055},
	Primaryclass = {hep-th},
	Reportnumber = {NEIP-01-001},
	Slaccitation = {%%CITATION = HEP-TH/0101055;%%},
	Title = {{Introduction to supersymmetry}},
	Year = {2001}}

@book{Nakahara,
	Author = {Nakahara, M.},
	Journal = {Boca Raton, USA: Taylor \& Francis (2003) 573 p},
	Slaccitation = {%%CITATION = INSPIRE-640779;%%},
	Title = {{Geometry, topology and physics}},
	Year = {2003}}

@book{ZeeGroups,
	Author = {Zee, A.},
	Isbn = {9780691162690},
	Lccn = {2015037408},
	Publisher = {Princeton University Press},
	Series = {In a nutshell},
	Title = {Group Theory in a Nutshell for Physicists},
	Year = {2016}}

@inproceedings{ComplexManifoldsLectureNotes,
	Author = {Stefan Vandoren},
	Title = {Lectures on Riemannian Geometry, Part II: Complex Manifolds},
	Url = {https://www.semanticscholar.org/paper/Lectures-on-Riemannian-Geometry%2C-Part-II%3A-Complex-Vandoren/9eef42d482293387ffc37ca54d8979b43e651258},
	Year = {2009}}

@article{ComplexManifoldsLectureNotes2,
	Archiveprefix = {arXiv},
	Author = {Bouchard, Vincent},
	Eprint = {hep-th/0702063},
	Primaryclass = {HEP-TH},
	Slaccitation = {%%CITATION = HEP-TH/0702063;%%},
	Title = {{Lectures on complex geometry, Calabi-Yau manifolds and toric geometry}},
	Year = {2007}}

@inproceedings{Greene,
	Archiveprefix = {arXiv},
	Author = {Greene, Brian R.},
	Booktitle = {{Fields, strings and duality. Proceedings, Summer School, Theoretical Advanced Study Institute in Elementary Particle Physics, TASI'96, Boulder, USA, June 2-28, 1996}},
	Eprint = {hep-th/9702155},
	Pages = {543-726},
	Primaryclass = {hep-th},
	Reportnumber = {CU-TP-812},
	Slaccitation = {%%CITATION = HEP-TH/9702155;%%},
	Title = {{String theory on Calabi-Yau manifolds}},
	Year = {1996}}

@book{Books:GaugeKnotsGravity,
	Author = {Baez, J. and Muniain, J. P.},
	Journal = {Singapore, Singapore: World Scientific (1994) 465 p. (Series on knots and everything, 4)},
	Slaccitation = {%%CITATION = INSPIRE-397363;%%},
	Title = {{Gauge fields, knots and gravity}},
	Year = {1995}}

@book{Books:Supergravity,
	Address = {Cambridge, UK},
	Author = {Freedman, Daniel Z. and Van Proeyen, Antoine},
	Isbn = {9781139368063, 9780521194013},
	Publisher = {Cambridge Univ. Press},
	Slaccitation = {%%CITATION = INSPIRE-1123253;%%},
	Title = {{Supergravity}},
	Year = {2012}}

@book{Books:BeckerBecker,
	Author = {Becker, K. and Becker, M. and Schwarz, J. H.},
	Isbn = {9780511254864, 9780521860697},
	Publisher = {Cambridge University Press},
	Slaccitation = {%%CITATION = INSPIRE-744404;%%},
	Title = {{String theory and M-theory: A modern introduction}},
	Year = {2006}}

@phdthesis{PaltiPhD,
	Archiveprefix = {arXiv},
	Author = {Palti, Eran},
	Eprint = {hep-th/0608033},
	Primaryclass = {hep-th},
	School = {Sussex U.},
	Slaccitation = {%%CITATION = HEP-TH/0608033;%%},
	Title = {{Aspects of moduli stabilisation in string and M-theory}},
	Year = {2006}}

@article{Susskind,
	Archiveprefix = {arXiv},
	Author = {Susskind, Leonard},
	Eprint = {hep-th/9501106},
	Primaryclass = {hep-th},
	Reportnumber = {SU-ITP-95-1},
	Slaccitation = {%%CITATION = HEP-TH/9501106;%%},
	Title = {{Trouble for remnants}},
	Year = {1995}}

@article{Candelas:ModuliSpaces,
	Author = {Candelas, Philip and de la Ossa, Xenia},
	Doi = {10.1016/0550-3213(91)90122-E},
	Journal = {Nucl. Phys.},
	Pages = {455-481},
	Reportnumber = {UTTG-07-90},
	Slaccitation = {%%CITATION = NUPHA,B355,455;%%},
	Title = {{Moduli Space of {Calabi-Yau} Manifolds}},
	Volume = {B355},
	Year = {1991}}

@article{StromingerIdentity,
	Author = {Strominger, Andrew},
	Doi = {10.1103/PhysRevLett.55.2547},
	Issue = {23},
	Journal = {Physical Review Letters},
	Month = {Dec},
	Numpages = {0},
	Pages = {2547--2550},
	Publisher = {American Physical Society},
	Title = {Yukawa Couplings in Superstring Compactification},
	Url = {https://link.aps.org/doi/10.1103/PhysRevLett.55.2547},
	Volume = {55},
	Year = {1985}}

@article{CosmicCensor,
	Archiveprefix = {arXiv},
	Author = {Kallosh, Renata and Linde, Andrei D. and Ortin, Tomas and Peet, Amanda W. and Van Proeyen, Antoine},
	Doi = {10.1103/PhysRevD.46.5278},
	Eprint = {hep-th/9205027},
	Journal = {Phys. Rev.},
	Pages = {5278-5302},
	Primaryclass = {hep-th},
	Reportnumber = {SU-ITP-92-13},
	Slaccitation = {%%CITATION = HEP-TH/9205027;%%},
	Title = {{Supersymmetry as a cosmic censor}},
	Volume = {D46},
	Year = {1992}}

@article{GlobalSymmetriesInST,
	Author = {Banks, Tom and Dixon, Lance J.},
	Doi = {10.1016/0550-3213(88)90523-8},
	Journal = {Nucl. Phys.},
	Pages = {93-108},
	Reportnumber = {PUPT-1086, SCIPP-8805},
	Slaccitation = {%%CITATION = NUPHA,B307,93;%%},
	Title = {{Constraints on String Vacua with Space-Time Supersymmetry}},
	Volume = {B307},
	Year = {1988}}

@article{BlackHoles,
	Archiveprefix = {arXiv},
	Author = {Dall'Agata, Gianguido},
	Booktitle = {{Proceedings, INFN-Laboratori Nazionali di Frascati School on the Attractor Mechanism on Supersymmetric Gravity and Black Holes (SAM 2009): Frascati, Italy, 29 Jun - 3 Jul, 2009}},
	Doi = {10.1007/978-3-642-31380-6_1},
	Eprint = {1106.2611},
	Journal = {Springer Proc. Phys.},
	Pages = {1-45},
	Primaryclass = {hep-th},
	Reportnumber = {DFPD-11-TH-07},
	Slaccitation = {%%CITATION = ARXIV:1106.2611;%%},
	Title = {{Black holes in supergravity: flow equations and duality}},
	Volume = {142},
	Year = {2013}}
\newpage

\newdateformat{deutsch}{\THEDAY{. }\monthname[\THEMONTH], \THEYEAR}
\setlength{\parindent}{0em}
\thispagestyle{empty}
\section*{Declaration of Authorship}
\vspace{2\baselineskip}
{\large\textbf{Erkl\"arung:}\par}
\vspace{2\baselineskip}
Ich versichere, dass ich diese Arbeit selbstst\"{a}ndig verfasst und keine
anderen als die angegebenen Quellen und Hilfsmittel benutzt habe.\par
\vspace{5\baselineskip}
Heidelberg, den 4. August, 2018

\end{document}